\documentclass[a4paper,11pt]{article}
\pdfoutput=1 
\usepackage{jheppub} 
\usepackage[T1]{fontenc} 
\usepackage{amsmath,amssymb,amsfonts,cases}
\usepackage{amsthm}
\usepackage{enumitem}
\usepackage{graphicx,caption,subcaption}
\usepackage[T1]{fontenc} 
\usepackage{enumitem}  
\usepackage{empheq}
\usepackage{dsfont}
\usepackage{hyperref}
\usepackage[mathscr]{euscript}
\usepackage{dcolumn}
\usepackage{bm}
\usepackage{bbm}
\usepackage{slashed}
\usepackage{wrapfig}
\usepackage{mathtools}
\usepackage[utf8]{inputenc}
\usepackage{sectsty}
\allsectionsfont{\boldmath}

\DeclarePairedDelimiter{\ceil}{\lceil}{\rceil}

\newcommand{\beq}{\begin{equation}}
\newcommand{\eeq}{\end{equation}}
\newcommand{\bea}{\begin{eqnarray}}
\newcommand{\eea}{\end{eqnarray}}
\newcommand{\bi}{\begin{itemize}}
\newcommand{\ei}{\end{itemize}}
\newcommand{\ben}{\begin{enumerate}}
\newcommand{\een}{\end{enumerate}}

\def\Tr{{\rm Tr}}

\newcommand{\mds}[1]{\mathds{#1}}

\newcommand{\N}{{\mathcal N}}

\newcommand{\braketbis}[1]{\left< #1 \right>}

\newcommand{\e}{\epsilon}

\newcommand{\eps}{\varepsilon}

\newcommand\eq[1]{eq.~(\ref{eq:#1})}

\newcommand\mf[1]{{\mathfrak{#1}}}
\newcommand\mc[1]{{\mathcal{#1}}}

\newcommand\CM{{\mc{M}}}
\newcommand\CN{{\mc{N}}}
\newcommand\CO{{\mc{O}}}

\newcommand\vrho{\varrho}

\newcommand\II{\rm I\hspace{-0.02cm} I}

\title{\LARGE Central Charges of 2d Superconformal Defects}
\author[1]{Adam Chalabi,}
\author[1]{Andy O'Bannon,}
\author[2]{Brandon Robinson,}
\author[1]{Jacopo Sisti}

\affiliation[1]{STAG Research Centre, Physics and Astronomy, University of Southampton, Highfield, Southampton SO17 1BJ, UK}
\affiliation[2]{Instituut voor Theoretische Fysica, K.U. Leuven, Celestijnenlaan 200D, BE-3001 Leuven, Belgium}
\emailAdd{a.chalabi@soton.ac.uk}
\emailAdd{a.obannon@soton.ac.uk}
\emailAdd{brandon.robinson@kuleuven.be}
\emailAdd{J.Sisti@soton.ac.uk}

\abstract{In conformal field theories (CFTs) of dimension $d>3$, two-dimensional (2d) conformal defects are characterised in part by central charges defined via the defect's contribution to the trace anomaly. However, in general for interacting CFTs these central charges are difficult to calculate. For superconformal 2d defects in supersymmetric (SUSY) CFTs (SCFTs), we show how to compute these defect central charges from the SUSY partition function either on $S^d$ with defect along $S^2$, or on $S^1 \times S^{d-1}$ with defect along $S^1 \times S^1$. In the latter case we propose that defect central charges appear in an overall normalisation factor, as part of the  SUSY Casimir energy. For 2d half-BPS defects in 4d $\N=2$ SCFTs and in the 6d $\N=(2,0)$ SCFT we obtain novel, exact results for defect central charges using existing results for partition functions computed using SUSY localisation, SUSY indices, and correspondences to 2d Liouville, Toda, and $q$-deformed Yang-Mills theories. Some of our results for defect central charges agree with those obtained previously via holography, showing that the latter are not just large-$N$ and/or strong-coupling limits, but are exact. Our methods can be straightforwardly extended to other superconformal defects, of various codimension, as we demonstrate for a 4d defect in the 6d $\N=(2,0)$ SCFT.}


\begin{document}
\maketitle
\flushbottom


\section{Introduction}
\label{sec:intro}

Defect operators are essential for classifying Quantum Field Theories (QFTs)~\cite{Douglas:2010ic,Aharony:2013hda,Gukov:2013zka,Gukov:2014gja}. For example, two gauge theories with the same gauge algebra but different gauge groups can have identical correlators of all local operators, but different spectra of 1d (line) operators, such as Wilson and 't Hooft lines~\cite{Aharony:2013hda}. Furthermore, such 1d operators are the order parameters classifying vacua as confining, Higgs, Coulomb, etc. Similarly, higher-dimensional defect operators are order parameters classifying phases in which higher-dimensional objects, such as strings, condense~\cite{Gukov:2013zka,Gukov:2014gja}.

A fundamental question in QFT is therefore how to characterise and classify defect operators. A formidable obstacle to answering this question is in dealing with (strongly) interacting degrees of freedom of the ambient QFT and/or on the defect. To overcome this obstacle we will employ a common strategy: impose highly restrictive symmetries. Specifically, we will require both conformal symmetry and SUperSYmmetry (SUSY). Furthermore, we will focus exclusively on 2d defects, which in 4d are also called surface operators. In short, we will focus on 2d superconformal defects in SuperConformal Field Theories (SCFTs).

We use conformal symmetry to provide order in the space of QFTs. In particular, Conformal Field Theories (CFTs) occupy privileged places in the space of QFTs, as fixed points of Renormalisation Group (RG) flows, and $c$-theorems then imply irreversibility along those RG flows, providing a hierarchy among QFTs. More specifically, $c$-theorems state that certain central charges must decrease monotonically along RG flows. As a result, these central charges can count degrees of freedom, which we expect to decrease along RG flows as the UltraViolet (UV) physics becomes more coarse grained and massive modes decouple. Proofs exist for $c$-theorems in 2d~\cite{Zamolodchikov:1986gt,Cappelli:1990yc,Casini:2004bw,Komargodski:2011xv}, 3d ($F$-theorem)~\cite{Casini:2012ei,Casini:2017vbe}, and 4d ($a$-theorem)~\cite{Cardy:1988cwa,Osborn:1989td,Jack:1990eb,Osborn:1991gm,Komargodski:2011vj,Komargodski:2011xv,Casini:2017vbe}.

How to extend $c$-theorems to defect CFTs remains an open question. Currently only two defect $c$-theorems have been proven. The first is for RG flows along 1d interfaces separating 2d CFTs. In these systems the ``$g$-theorem"~\cite{Affleck:1991tk,Friedan:2003yc,Casini:2016fgb} requires the interface entropy, denoted $\ln(g)$, to decrease monotonically along the RG flow. Intuitively, $\ln(g)$ measures the ground state degeneracy of the 1d quantum system and thus counts the number of degrees of freedom localised at the interface. Often $\ln(g)$ can be computed, even in (strongly) interacting systems, using powerful methods available to 2d CFT.\footnote{The $g$-theorem also applies to a 2d CFT with a boundary, interpreted as an interface with an ``empty'' CFT on one side, and to a 2d CFT with a point-like impurity.}

The second defect $c$-theorem is for 2d defects in CFTs of dimension $d\geq 3$, with an RG flow on the defect~\cite{Jensen:2015swa,Casini:2018nym}. To be precise, let $\cal{M}$ denote the background manifold for a $d$-dimensional Euclidean CFT with coordinates $\{x^\mu\}$, with $\mu =1,2,\ldots,d$, and let $\Sigma \hookrightarrow \cal{M}$ with coordinates $\{\xi^a\}$, with $a=1,2$, be the 2d submanifold on which the defect has support. In these cases the trace anomaly includes the usual contribution from the ambient CFT, which can be non-zero only if $d$ is even, plus a contribution delta-function localised at $\Sigma$~\cite{Graham:1999pm,Henningson:1999xi,Asnin:2008ak,Schwimmer:2008yh},
\beq
\label{eq:defecttrace}
T^{\mu}_{~\mu}\Big|_{\Sigma} = - \frac{1}{24\pi} \left(b \, E_2 + d_1\,\mathring{\mathrm{I\!I}}^\mu_{ab}\mathring{\mathrm{I\!I}}_\mu^{ab} - d_2 \, W_{ab}{}^{ab} \right)\,,
\eeq
where we use the sign and normalisation conventions of refs.~\cite{Estes:2018tnu,Jensen:2018rxu}.\footnote{For a surface defect in $d=4$ that breaks parity, additional parity-odd terms may also appear in the trace anomaly~\cite{Cvitan:2015ana,Jensen:2018rxu}. These will play no role in this paper.} In eq.~\eqref{eq:defecttrace}, $E_2$ and $\mathring{\mathrm{I\!I}}^\mu_{ab}$ are the Euler density and traceless second fundamental form of $\Sigma $, respectively, and $W_{abcd}$ is the pullback of the ambient Weyl tensor to $\Sigma$. In general, a 2d defect thus has three possible central charges, $b$, $d_1$, and $d_2$. In the classification of ref.~\cite{Deser:1993yx}, $b$ is type A while $d_1$ and $d_2$ are both type B. Among other things, the classification means $b$ cannot depend on defect marginal couplings, while $d_1$ and $d_2$ can. In general, all three can depend on marginal couplings of the ambient CFT~\cite{Herzog:2017xha,Herzog:2019rke}, unless the defect preserves 2d $\N=(2,0)$ SUSY, in which case $b$ cannot depend on ambient marginal couplings either~\cite{Bianchi:2019umv}.

The central charge $b$ obeys a $c$-theorem for defect RG flows~\cite{Jensen:2015swa,Casini:2018nym}, sometimes called the ``$b$-theorem.'' Whether $d_1$ and $d_2$ obey $c$-theorems still remains unknown. However, proofs exist that in reflection-positive theories $d_1\geq 0$~\cite{Herzog:2017xha,Herzog:2017kkj}, and if the average null energy condition is obeyed in the presence of the defect then $d_2 \geq 0$~\cite{Jensen:2018rxu}. Free field computations show that $b$ can be negative, but whether $b$ obeys a lower bound has not been proven. For $d=3$, a lower bound $b \geq - \frac{2}{3} d_1$ was conjectured in ref.~\cite{Herzog:2017kkj}.

In general, $b$, $d_1$, and $d_2$ are difficult to calculate, and indeed to our knowledge they have been calculated only in free-field CFTs~\cite{Henningson:1999xi,Gustavsson:2003hn,Gustavsson:2004gj,Asnin:2008ak} and holographic CFTs~\cite{Berenstein:1998ij,Graham:1999pm,Drukker:2008wr,Garriga:2008ks,Garriga:2009hy,Fiol:2010un,Fiol:2010wf,Jensen:2013lxa,Korovin:2013gha,Estes:2014hka,Seminara:2017hhh,Seminara:2018pmr,Estes:2018tnu,Jensen:2018rxu}. This is why we turn to SUSY, to provide tractable examples of interacting theories in which we can calculate $b$, $d_1$, and $d_2$ without using holography.

In particular, we will consider 1/2-BPS 2d superconformal defects in 4d and 6d SCFTs. In 4d we focus on $\N=2$ SCFTs of class $\cal{S}$, which are obtained generically by wrapping a stack of $N$ coincident M5-branes on a genus-$g$ $n$-punctured Riemann surface $\mc{C}_{g,n}$~\cite{Gaiotto:2009hg}. The torus, $\mc{C}_{1,0}$, is a special case where the 4d SUSY is enhanced, producing the maximally SUSY SCFT in 4d, $\N=4$ $SU(N)$ super Yang-Mills (SYM) theory.  We consider superconformal defects preserving 2d $\N=(2,2)$ SUSY (enhanced to $\N=(4,4)$ in $\N=4$ SYM theory) arising as either M2-branes ending on the M5-branes, which sit at a point on $\mc{C}_{g,n}$, or a second stack of M5-branes intersecting the first stack over a codimension 2 surface, and which wrap all of $\mc{C}_{g,n}$~\cite{Alday:2009fs,Gaiotto:2009fs,Kozcaz:2010af,Alday:2010vg}.

In 6d we focus on the worldvolume theory of $N$ coincident M5-branes (not wrapping a Riemann surface), namely a 6d $A_{N-1}$ $\N=(2,0)$ SCFT. Our 2d superconformal defects will arise from M2-branes ending on the M5-branes, producing a so-called ``Wilson surface'' operator in the 6d SCFT, which preserves 2d $\N=(4,4)$ SUSY~\cite{Ganor:1996nf}.

Our goal is to use SUSY methods to compute $b$, $d_1$, and $d_2$ for these classes of 2d superconformal defects. In fact, in a 4d SCFT with a 2d $\N=(2,0)$ superconformal defect, ref.~\cite{Bianchi:2019sxz} proved that the SUSY algebra requires $d_1=d_2$. Ref.~\cite{Bianchi:2019sxz} further provided compelling evidence for the conjecture that $d_1 \propto d_2$ for 2d $\N=(2,2)$ superconformal defects in any $d$. All of these results extend equally well to 2d $\CN=(4,4)$ superconformal defects. We will thus only explicitly compute $b$ and $d_2$.

We will compute $b$ only for 2d superconformal defects in 4d $\CN\geq 2$ SCFTs. To do so, we will compute the partition function of the Euclidean CFT on ${\cal{M}}=S^4$, with the defect wrapping an equatorial sphere $\Sigma=S^2$, and then perform a re-scaling of the $S^4$ radius, with all other scales held fixed. The trace anomaly is the statement that the partition function changes by an overall power of the sphere radius under Weyl re-scaling, fixed by the central charge of the SCFT. We obtain $b$ by calculating that power and subtracting any contribution from the ambient SCFT's type A central charge.

Luckily, many methods exist to compute the partition function of an $\N=2$ SCFT on ${\cal{M}}=S^4$ with a superconformal defect along $\Sigma = S^2$: holography~\cite{Maldacena:1997re,Gubser:1998bc,Witten:1998qj}, SUSY localisation~\cite{Nekrasov:2002qd,Pestun:2007rz,Drukker:2010jp,Alday:2010vg,Kanno:2011fw,Benini:2012ui,Doroud:2012xw,Gomis:2014eya,Nawata:2014nca,Lamy-Poirier:2014sea,Gomis:2016ljm,Pan:2016fbl,Gorsky:2017hro}, the AGT correspondence~\cite{Alday:2009aq,Alday:2009fs,Wyllard:2009hg,Kozcaz:2010af}, geometric engineering \cite{Kozcaz:2010af,Dimofte:2010tz}, and many others. We will use existing results from SUSY localisation and the AGT correspondence (to 2d Liouville/Toda CFTs) to extract new results for $b$ in several examples, some of which provide non-trivial tests of the $b$-theorem.

We will compute $d_2$ only for 2d $\N=(4,4)$ superconformal defects in $\N=4$ SYM, and for Wilson surfaces in the M5-brane theory. To do so, we will compute the SCFT's SUSY partition function on ${\cal{M}}= S^1_R \times S^3$ or $S^1_R \times S^5$, where $S^1_R$ is a circle of radius $R$, with the defect wrapping $\Sigma = S^1_R \times S^1$, where the latter $S^1$ is equatorial inside $S^3$ or $S^5$. Here again, many methods exist to compute the $S^1 \times S^{d-1}$ SUSY partition function: holography~\cite{Maldacena:1997re,Gubser:1998bc,Witten:1998qj}, SUSY localisation~\cite{Kim:2012ava, Assel:2014paa,Bullimore:2014upa,Bobev:2015kza}, correspondences to 2d topological QFTs~\cite{Gadde:2011ik,Bullimore:2014nla}, characters of modules in vertex operator algebras (VOAs)~\cite{Beem:2013sza,Beem:2014kka,Beem:2014rza,Cordova:2017mhb}, and many others.

We will make a general argument for how to extract $d_2$ from the SUSY partition function on ${\cal{M}}=S^1_R \times S^{d-1}$. These partition functions turn out to be a product of two factors. One factor is the Schur index~\cite{Kinney:2005ej,Gadde:2011uv}. By appealing to a growing body of evidence from various perspectives~\cite{Kim:2012ava,DiPietro:2014bca,Assel:2015nca,Bobev:2015kza,Closset:2019ucb,Bianchi:2019sxz}, we claim that the other factor is $e^{-R E_c }$, where $E_c$ is the SUSY Casimir Energy (SCE). We propose that introducing the defect shifts $E_c$ by a term $\propto d_2$, and provide compelling evidence from our two examples.

Our first example is $\N=4$ $SU(N)$ SYM, where we will use the fact that the Schur limit of the 4d SUSY partition function is equivalent to the partition function of 2d $SU(N)$ $q$-deformed Yang-Mills (qYM) theory on a Riemann surface ($\mc{C}_{g,n}$) in the zero area limit~\cite{Gadde:2011ik}. In this correspondence, the insertion of $p$ 2d superconformal defects labelled by representations $\mc{R}_i$ of $SU(N)$ deforms the 4d SUSY partition function~\cite{Gaiotto:2012xa} in a way that is captured in 2d qYM as a $p$-point correlation function $\langle \CO_{\mc{R}_1}\ldots\CO_{\mc{R}_p}\rangle$~\cite{Gaiotto:2012xa,Alday:2013kda}. Our second example is the Wilson surface in the M5-brane theory, for which a form for the SUSY partition function on $S^1_R \times S^5$ was proposed in ref.~\cite{Bullimore:2014upa}. In fact, we will obtain a more general result: we will compute the shift in $E_c$ due to \textit{two} intersecting Wilson surfaces, which turns out to be more than just a sum of the contributions from two individual Wilson surfaces, possibly because of additional degrees of freedom arising at the 1d intersections along $S^1_R$.

Holographic results for $b$ and $d_2$ exist for the 2d $\N=(4,4)$ superconformal defects in $\N=4$ SYM theory, in the 't Hooft large-$N$ limit with large 't Hooft coupling ~\cite{Graham:1999pm,Drukker:2008wr,Gentle:2015ruo,Jensen:2018rxu}, and for Wilson surfaces in the M5-brane SCFT, in the large-$N$ limit~\cite{Berenstein:1998ij,Graham:1999pm,Estes:2018tnu,Jensen:2018rxu}. Our results using SUSY methods agree perfectly with the holographic results, whenever they overlap. However, the SUSY methods involve no approximations and are valid at all couplings and for any $N$: they provide \textit{exact} results for $b$ and $d_2$. We thus find that the holographic results are exact, and not merely large-$N$ or strong coupling limits. Furthermore, the agreement with holography provides compelling evidence that $E_c \propto d_2$, especially in the Wilson surface case, as we will discuss.

Ultimately, our main message is the methods themselves. For the classes of superconformal defects that we study, we find practical ways to obtain exact results for $b$ and $d_2$.  The various methods that we present also provide different perspectives on what $b$ and $d_2$ are counting. Furthermore, these methods can be straightforwardly generalised to superconformal defects of other (co-)dimension, such as SUSY interfaces or domain walls between SCFTs~\cite{DHoker:2006qeo,Drukker:2010jp}. In fact, we further apply the method of computing $d_2$ via the change in SCE to compute a putative central charge for 4d superconformal defects as the character of a semi-degenerate module in a $W_N$-algebra~\cite{Beem:2014kka,Bullimore:2014upa}. In this case the form of the trace anomaly remains unknown, so we cannot say exactly which (linear combination of) defect central charges we obtain. Nevertheless, we believe the methods we develop in this paper can play a crucial role in characterising and classifying defects quite generally.

This paper is organised as follows. In section~\ref{sec:review} we review key facts we will need about 2d superconformal defects of SCFTs of class ${\cal{S}}$. In section~\ref{sec:s4} we present our calculations of $b$ using SUSY localisation and the AGT correspondence. In section~\ref{sec:Casimirs} we present our calculations of $d_2$ for superconformal defects in $\N=4$ SYM theory, using $q$-deformed YM, and for Wilson surfaces in the M5-brane theory, using the $S^1 \times S^5$ partition function. In section~\ref{sec:discussion} we conclude with a summary, and discuss possible directions for future research. We collect in two appendices various technical results that we will use along the way.

\section{Review: 2d Superconformal Defects}
\label{sec:review}
\addtocontents{toc}{\protect\setcounter{tocdepth}{1}}

In this section, we will provide a short overview of the relevant features of 2d superconformal defects that will be useful for our computations below. In particular, all of our work will focus on deforming 4d and 6d SCFTs by the addition of such defects. Even for these narrowly focused applications, there is a vast amount of extant literature, the surface of which we will only scratch.

In general, surface operators in a 4d theory can be characterised in two distinct but sometimes related ways~\cite{Gukov:2006jk}. One may either 
\begin{enumerate}[label=(\roman*)]
	\item assign singular behaviour to ambient 4d fields at the 2d submanifold $\Sigma$, or 
	\item introduce an auxiliary 2d theory at $\Sigma$ and couple it to the ambient 4d theory.
\end{enumerate}
This is a broad partitioning of defects according to whether we are using only the behaviour of ambient fields to describe a surface defect, or adding new degrees of freedom supported only on $\Sigma$. They are sometimes related, for example, integrating out degrees of freedom on $\Sigma$ may produce singular behaviour of the ambient fields at $\Sigma$, or sometimes the two descriptions can be related by dualities~\cite{Gukov:2006jk,Gukov:2008sn,Gomis:2014eya,Frenkel:2015rda,Ashok:2017odt,Ashok:2019rwa}.

\subsection{2d Levi Type-$\mds{L}$ Defects}

Approach (i) to describing surface defects is quite powerful and, by now, well-studied \cite{Gukov:2006jk,Gukov:2008sn,Gukov:2014gja}. Here we briefly review the construction for $\CN=2$ and $\CN=4$ gauge theories.  

Consider $\CN=4$ SYM theory on $\CM=\mathds{R}^4$ with coordinates $\{x^\mu\}$, with $\mu=1,\ldots,4$, and a $1/2$-BPS surface operator supported on $\Sigma = \mathds{R}^2$ with coordinates $\{x^1, x^2\}$. Let us write the coordinates on the normal bundle $N\Sigma = \mds{C}$ as $x^3+ix^4 = z = r e^{i\theta}$. To define the surface defect~\cite{Kapustin:2006pk}, one needs to prescribe a singularity in the normal component of the gauge field $A=A_z \, dz$ and the 1-form scalar in the adjoint $\CN=2$ hypermultiplet $\varphi= (\varphi_1 +i \varphi_2)\, dz$. In preserving 2d $\CN=(4,4)$ SUSY along $\Sigma$, $A$ and $\varphi$ have to satisfy a set of simple BPS conditions, which in the GL twist of $\N=4$ SYM take the form of Hitchin's equations \cite{Hitchin:1987mz, Gukov:2006jk}. We will not consider the GL twist of $\N=4$ SYM in the following, except for this sub-section and the next, where the Hitchin moduli space, $\mathscr{M}_H$, provides an informative perspective of the defect. The leading singular behaviour of the BPS solutions that additionally preserve defect conformal symmetry is given by
\begin{align}\label{eq:Levi-defect}
A &= \alpha \, d\theta\,,\qquad \varphi= \frac{1}{2z}(\beta + i\gamma)\,dz\,,
\end{align}
for constants $(\alpha,\,\beta,\,\gamma)$.  Relaxing the constraint that the defect preserves conformal symmetry would allow for non-trivial dependence on the radial coordinate $r$.  

If the ambient 4d theory has only $\CN=2$ SUSY rather than $\CN=4$, one constructs a $1/2$-BPS surface defect by prescribing a singularity in the 4d gauge field only. In these cases the BPS conditions do not allow for a singularity in any scalar fields, so no analogue of $\beta$ or $\gamma$ exists --- those are special to $\CN=4$ SYM.

The data $(\alpha,\,\beta,\,\gamma)$ describing the $1/2$-BPS defect are valued in $\mds{T} \times \mf{t} \times \mf{t}$, where $\mds{T}$ is the maximal torus of the gauge group $G$ and $\mf{t}$ is the associated Cartan subalgebra.  Thus, quantisation of the 2d-4d system requires the preserved gauge symmetry consistent with solving the BPS equations to be a subgroup of $G$ containing $\mds{T}$, called the \textit{Levi subgroup} $\mds{L}\subset G$. There are a number of ways to construct $\mds{L}$, and choosing a particular $\mds{L}\subset G$ is part of defining the defect. Thus, 2d defects of the types that we have been describing are called Levi type-$\mds{L}$. Unless otherwise specified, we will only consider $G= U(N)$ or $SU(N)$ and Levi subgroups $\mds{L} = \left[\prod_{i=1}^{n+1} U(N_i)\right]$ or $S\left[\prod_{i=1}^{n+1} U(N_i)\right]$, respectively, with the constraint $\sum_{i=1}^{n+1} N_i=N$.

There are two types of 2d Levi type-$\mds{L}$ defects commonly encountered in the literature that are given special names and will be considered below.  For gauge group $G=SU(N)$, if $\mds{L} = S[U(N-1)\times U(1)]$ then the surface defect is called \textit{simple}, and if $\mds{L}=\mds{T} = U(1)^{N-1}$ then the surface defect is called \textit{full}.

Lastly, in addition to $\mds{L}$ and $(\alpha,\,\beta,\,\gamma)$, one can turn on a quantum 2d theta angle parameter, $\eta$, along the defect. The importance of $\eta$ can be seen in studying the behaviour of Levi type-$\mds{L}$ defects under dualities. Under, say, S-duality $(\alpha,\eta) \rightarrow (\eta,-\alpha)$, and so for a generic 2d $\CN=(4,4)$ superconformal defect in 4d $\CN=4$ SYM theory, specifying $(\mds{L}; \alpha,\,\beta,\,\gamma,\,\eta)$ completely describes the defect. The parameters $(\beta,\,\gamma)$ are together valued in the $\mds{L}$-invariant part of $\mf{t}$, while $\alpha$ is valued in $\mds{T}$ and $\eta$ is valued in the maximal torus ${}^L\mds{T}$ of the Langlands dual ${}^LG$ of $G$. All of the parameters grouped together transform in the part of $(\mds{T}\times\mf{t}\times\mf{t}\times {}^L\mds{T})$ invariant under the Weyl group of $\mds{L}$~\cite{Gukov:2006jk,Gukov:2008sn}. 

Unless otherwise stated, the Levi type-$\mds{L}$ surface defect examples considered below will have $\beta=\gamma=0$.  This is particularly relevant for the computation of superconformal indices or twisted partition functions for $\CN\geq(2,2)$ defects.  The parameter $\beta+i\gamma$ being non-zero is generally incompatible with the necessary symmetries for computing the defect index. In particular, non-zero $\beta$ and/or $\gamma$ breaks rotational symmetry in the plane normal to $\Sigma$.\footnote{We thank L.~Bianchi and M.~Lemos for pointing this out to us.}

Having set the basis to describe Levi type-$\mds{L}$ defects, it is useful to understand the physical meaning of the parameters $(\alpha,\,\beta,\,\gamma,\,\eta)$. 2d $\CN=(4,4)$ superconformal defects defined by \eq{Levi-defect} are elements of Hitchin's moduli space, $\mathscr{M}_H$, arising in the GL-twist of $\CN=4$ SYM theory. As mentioned, the $\eta$ parameter is a 2d theta angle, but the other ``classical'' parameters $(\alpha,\,\beta,\,\gamma)$ encode geometric information about $\mathscr{M}_H$.  $\mathscr{M}_H$ is constructed by a hyper-K\"ahler quotient \cite{Hitchin:1986ea}, and as such there are both complex structure --- one of three labelled $I,\,J,\,K$ --- and K\"ahler parameters that describe the local geometry.  By making a choice of which parameters go into the solution for $\varphi$ in \eq{Levi-defect}, we are in effect picking a complex structure, while the parameter controlling the singular behaviour of the 4d gauge field $A$ is the K\"ahler parameter. In ref.~\cite{Gukov:2006jk}, the combination $\beta+i\gamma$ was identified with complex structure $I$, and in this complex structure $\alpha$ was the K\"ahler parameter. Cyclicly permuting the roles of the parameters, one may identify $\gamma+i\alpha$ and $\alpha+i\beta$ with complex structures $J$ and $K$ with K\"ahler parameters $\beta$ and $\gamma$, respectively

\subsection{2d Defects from 2d QFTs}
\label{sec:review2dqfts}

In approach (ii), we begin with a 4d theory and add 2d degrees of freedom localised on $\Sigma$. We will consider cases where the latter are a Gauged Linear Sigma Model (GLSM) or a Non-Linear Sigma Model (NLSM). The 4d and 2d degrees of freedom can be coupled in various ways, for example by superpotential couplings and/or by gauging a shared symmetry group~\cite{Gukov:2006jk, Gukov:2008sn, Gadde:2013ftv, Gomis:2014eya, Gomis:2016ljm}. 

\begin{figure}[t] 
	\begin{center}
		\includegraphics[scale=0.7]{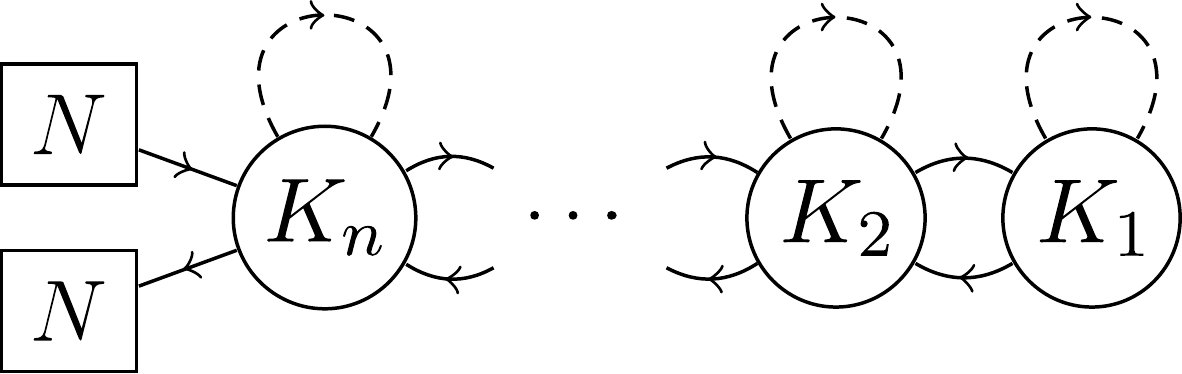}
	\end{center}
	\vspace{-.5cm}
	\caption{\label{quiver1}
		\small
		Linear quiver diagram corresponding to a 2d $\CN=(2,2)$ GLSM. Its field content consists of $U(K_i)$ 2d vector multiplets for $i=1, \ldots, n$, $N$ fundamental $\phi^{\text{\tiny fund}}_n$ and $N$ anti-fundamental $\tilde \phi^{\text{\tiny anti-fund}}_n$ chiral multiplets coupled to the $U(K_n)$ vector, one chiral multiplet $\phi^{\text{\tiny bif}}_{i(i+1)}$ in the bifundamental representation of $({\bf{K_i}} ,{\bf{\overline K_{i+1}}})$, and one chiral multiplet $\phi^{\text{\tiny bif}}_{(i+1)i}$ in the bifundamental representation $({\bf{\overline K_i}}, \bf{K_{i+1}})$ for each $1\le i\le n-1$.  Additionally, depending on the particular details of the 2d $\CN=(2,2)$ gauge theory, there can be one adjoint chiral $X_i$ of $U(K_i)$ for each node. This quiver diagram can be used to construct a surface operator whenever the 4d $\mathcal{N}=2$ gauge theory has at least an $S[U(N)\times U(N)]$ flavour or gauge symmetry group.
	}
\end{figure}

We will consider 4d SCFTs that enjoy at least $\CN \geq 2$ SUSY. To engineer a $1/2$-BPS surface defect, consider the GLSM with $\CN \geq (2,2)$ SUSY and gauge group $G_\text{2d}$ described by the quiver in figure~\ref{quiver1}. The $i^{\textrm{th}}$ circular node denotes a 2d gauge multiplet with gauge group $U(K_i)$, the arrows connecting the $i^{\textrm{th}}$ and $(i+1)^{\textrm{th}}$ node represent chiral multiplets in the bifundamental representation $({\textbf{K}}_i, \overline{{\textbf{K}}}_{i+1})$ or $(\overline{{\textbf{K}}}_i, {{\textbf{K}}}_{i+1})$  of $U(K_i)\times U(K_{i+1})$, depending on the direction of the arrow. We collectively denote the fields in the bifundamental by $\phi^{\text{\tiny bif}}_{i(i+1)}$ and $\phi^{\text{\tiny bif}}_{(i+1)i}$, respectively. The dashed arrows starting and ending on the same node are adjoint chiral multiplets $X_i$. In what follows our quivers will always have the bifundamental fields, but may or may not have the adjoint chirals, depending on the type of defect we wish to study. For each gauge node, we may also turn on an FI parameter and a 2d theta angle for its $U(1)$ factor. The square nodes on the left indicate the number of flavours of the (anti-)fundamental chiral multiplets under the $U(K_n)$ gauge group. We denote these fundamental and anti-fundamental chirals by $\phi^{\text{\tiny fund}}_n$ and $\tilde \phi^{\text{\tiny anti-fund}}_n$, respectively.

If we set the (real) twisted masses of all matter fields to zero, and provided that the FI parameters do not run, then such a GLSM may flow to an interacting InfraRed (IR) fixed point~\cite{Benini:2016qnm}. To the best of our knowledge, it is unknown whether such an IR fixed point exists, but whenever it does the 2d SCFT has central charge $c_\text{2d}$ given by
\begin{equation}\label{eq:c2d}
\frac{c_\text{2d}}{3}=\sum_{\mathcal{R}}(1-q_\mathcal{R})\, \text{dim} \,\mathcal{R} - \text{dim}\, G_\text{2d}\,,
\end{equation}
where $\mathcal{R}$ are the 2d fields' representations of $G_\text{2d}$ and $q_{\mathcal{R}}$ are their R-charges. The representation data and dim $G_\text{2d}$ can be expressed in terms of the ranks of the gauge groups, $K_i$. We will compute several explicit examples in section~\ref{sec:s4}, but an important illustrative example is $1/2$-BPS surface defects in $\CN=4$ $SU(N)$ SYM. In this case $c_\text{2d}$ can be written more usefully in terms of the difference of adjacent ranks, $N_i=K_i - K_{i-1}$, with $K_0\equiv 0$ and $K_{n+1} \equiv N$. In particular, for an $\CN=(4,4)$ GLSM,
\begin{align}\label{eq:c2dSYM}
\frac{c_\text{2d}}{3} = N^2 - \sum_{i=1}^{n+1} N_i^2 \,,
\end{align}
a result that we will find again in several different ways in the following.

To obtain a $1/2$-BPS superconformal defect, one couples the ambient 4d theory to a GLSM, and flows to the IR fixed point, if it exists. Typically, the Vacuum Expectation Values (VEVs) of the 4d fields enter as twisted mass parameters in the 2d partition function~\cite{Gomis:2016ljm, Pan:2016fbl, Gorsky:2017hro}. A planar $1/2$-BPS superconformal defect then breaks the 4d superconformal algebra to a subalgebra: $\mf{su}(2,2|2)\rightarrow \mf{su}(1,1|1) \oplus \mf{su}(1,1|1) \oplus \mf{u}(1)$ for an $\CN=2$ SCFT, or  $\mf{psu}(2,2|4)\rightarrow \mf{psu}(1,1|2) \oplus \mf{psu}(1,1|2) \oplus \mf{u}(1)$ for an $\CN=4$ SCFT.

An alternative description of a $1/2$-BPS surface defect can be obtained by coupling a 2d NLSM to the ambient field theory~\cite{Gukov:2006jk, Gukov:2008sn, Gadde:2013ftv, Gukov:2014gja}. The NLSM description is obtained from the GLSM above by a defect RG flow: the gauge group is generically Higgsed, and the 2d vector multiplets become massive. By taking the gauge coupling in the GLSM to be parametrically large, the massive modes decouple and one obtains the NLSM as an effective theory. The moduli space of the GLSM becomes the target space of the NLSM.

As mentioned above, under certain conditions the two ways (i) and (ii) of introducing a surface defect are equivalent. One may suspect that integrating out the 2d degrees of freedom produces the delta-function singularities in the 4d fields on the support of the defect. Indeed, this is the case for a $1/2$-BPS surface defect in 4d $\CN = 4$ SYM theory with gauge group $G=SU(N)$~\cite{Gukov:2006jk,Gukov:2008sn,Gadde:2013ftv}. In order to obtain a Levi type-$\mds{L}$ defect from a GLSM, the latter needs to have $\CN = (4,4)$ SUSY and global symmetry $G$, which we gauge to couple the GLSM to the 4d theory. The Levi subgroup $\mds{L}$ is captured by the gauge symmetry in the linear quiver: Consider the linear quiver of figure~\ref{quiver1} whose gauge group $U(K_1)\times \ldots \times U(K_n)$ is such that $K_i > K_{i-1}$ for all $i=2, \ldots, n$. Then, the Levi subgroup is $\mds{L} = S[\prod_{i=1}^{n+1} U(N_i)]$ where $N_i = K_i - K_{i-1}$ with $K_0\equiv 0$ and $K_{n+1} \equiv N$. The parameters $(\alpha, \, \beta, \, \gamma,\, \eta)$ are encoded in the GLSM as follows. The linear combination $\alpha_k+i\eta_k$, with $k=1, \ldots, n+1$, corresponds to the complexified FI parameters of the GLSM, and the complex structure moduli $\beta_k+i\gamma_k$ characterise the 2d superpotentials. 

In the NLSM description, the requirement of $\CN=(4,4)$ SUSY and global symmetry $G$ translate to requiring the target space to be hyper-K\"ahler and to admit a $G$-action. In terms of the NLSM description, $(\alpha,\, \beta,\,\gamma)$ are encoded in the moduli of the target space, whereas $\eta$ is associated with the 2-form $B$-field. The authors of ref.~\cite{Gadde:2013ftv} conjecture that the NLSM target space is $T^*(G/\mds{L})$, which agrees with the moduli space $\mathscr{M}_H$ of the Levi type-$\mds{L}$ defect. The complex dimension of the target space is 
\begin{align}\label{eq:Gadde-Gukov-cpx-dim}
\text{dim}_{\mathds{C}}\,T^*(G/\mds{L})= N^2 - \sum_{i=1}^{n+1} N_i^2 \,,
\end{align}
which holds for general values of the parameters $(\alpha,\,\beta,\,\gamma,\,\eta)$. Note that this agrees with $c_\text{2d}/3$ of the associated $\CN=(4,4)$ GLSM in~\eq{c2dSYM}.

In the case of an ambient $\CN=2$ theory, there is a similar but weaker statement. A 2d GLSM with $\CN = (2,2)$ SUSY --- or NLSM whose target space admits a K\"ahler structure --- coupled to a 4d $\CN=2$ theory, is equivalent in the IR to the $\CN=2$ theory with prescribed singularities in the gauge field on the support of the defect~\cite{Frenkel:2015rda}.

\subsection{2d Defects in Theories of Class $\mathcal{S}$}

A large class of 4d $\CN=2$ SCFTs are the theories of ref.~\cite{Gaiotto:2009we}, often called \textit{class} $\mathcal{S}$. These come from the $6d$ $\mathcal{N}=(2,0)$ SCFT of type $A_{N-1}$ on a product manifold $\mathcal{M}_4 \times \mathcal{C}_{g,n}$ where $\mathcal{M}_4$ is a four-manifold and $\mc{C}_{g,n}$ is a genus-$g$ Riemann surface with $n$ punctures. A SUSY twist makes the corresponding partition function independent of the size of both $\mathcal{M}_4$ and $\mc{C}_{g,n}$ (though still dependent on their shape). Thus, one can shrink either $\mathcal{M}_4$ or $\mc{C}_{g,n}$ to zero without affecting the value of the partition function on $\mathcal{M}_4 \times \mc{C}_{g,n}$. By definition, class $\mathcal{S}$ theories are those obtained by shrinking $\mc{C}_{g,n}$. When $\mathcal{M}_4=S^4$ and the class $\mathcal{S}$ theory has a known Lagrangian description, then its partition function can be computed exactly via SUSY localisation~\cite{Pestun:2007rz}. The AGT correspondence is the statement that the $S^4$ partition function is equivalent to a Liouville/Toda correlator on $\mc{C}_{g,n}$~\cite{Alday:2009aq, Wyllard:2009hg}.

Within M-theory, these 4d theories can be found by wrapping M5-branes on $\mathcal{M}_4 \times \mc{C}_{g,n}$ and shrinking $\mc{C}_{g,n}$ to zero size. SUSY defects in the 4d SCFT can then be engineered by introducing a stack of either M2- or M5-branes ending on or intersecting the initial stack of M5-branes. In the 6d SCFT these M2- or M5-branes describe a 2d or 4d defect, respectively. To obtain a 2d defect in the 4d SCFT obtained by reducing on $\mc{C}_{g,n}$, we must either place the M2-branes at a point on $\mc{C}_{g,n}$ or let the M5-branes wrap all of $\mc{C}_{g,n}$.

The M2-branes localised at a point on $\mathcal{C}_{g,n}$ were discussed in detail in ref.~\cite{Gomis:2014eya} and refined in ref.~\cite{Gomis:2016ljm}. In the 4d SCFT, the 2d defect arising from M2-branes is described by the $n$-node quiver GLSM of figure \ref{quiver1} with $K_i < K_{i+1}$, an adjoint chiral multiplet on every node except the $n^{\textrm{th}}$ one, and non-vanishing FI parameter only for the $n^{\textrm{th}}$ node. The information encoded in the quiver can be summarised by a Young tableau of width $n$ which labels a representation of $A_{N-1}$. The length of the $j^{\textrm{th}}$ column is the difference in the ranks of the $j^{\textrm{th}}$ and $(j-1)^{\textrm{th}}$ node, i.e. $K_j - K_{j-1}$. Furthermore, the authors of refs.~\cite{Gomis:2014eya,Gomis:2016ljm} show that in the AGT correspondence to Liouville/Toda theory on $\mc{C}_{g,n}$, this surface defect corresponds to the insertion of a degenerate Toda primary labelled by the Young tableau. Its position on $\mc{C}_{g,n}$ is specified by the FI parameter of the $n^{\textrm{th}}$ node. The FI parameters of all other nodes of the quiver are turned off.

The M5-branes wrapping $\mc{C}_{g,n}$ were studied in refs.~\cite{Alday:2010vg,Kozcaz:2010yp}. After compactifying on $\mathcal{C}_{g,n}$, one obtains a different type of surface defect in a 4d $\CN=2$ SCFT which can be described by a Wess-Zumino-Novikov-Witten (WZNW) model on $\mathcal{C}_{g,n}$.

In the 4d SCFT, the authors of refs.~\cite{Gukov:2006jk,Gukov:2008sn,Frenkel:2015rda,Ashok:2017odt} proposed a Seiberg-like duality between the 2d defects that arise from these M2- and M5-branes in 6d. More specifically, the duality is a particular type of integral transform between the partition functions of the corresponding Liouville/Toda and WZNW theories living on $\mc{C}_{g,n}$. We will not need any details of this duality except that, like any duality, it is a mapping between physical observables of the two cases. Of importance to us is the fact that under the duality the metric on $\mathcal{M}_4$ and the submanifold $\Sigma$ are invariant, and the stress tensor maps to itself. As a result, the Weyl anomaly is invariant under the duality, and hence the defect central charges are also. We will see that this is the case in our examples below.

\subsection{Holographic Results}

Refs.~\cite{Kobayashi:2018lil,Jensen:2018rxu} showed that for a 2d conformal defect in a higher-$d$ CFT, the entanglement entropy of a sphere centred on the defect includes a logarithmic term with a universal coefficient given by a linear combination of three central charges: the ambient CFT's type A central charge (when $d$ is even), $b$, and $d_2$. Furthermore, $d_2$ determines the stress tensor one-point function in the presence of the defect. By calculating this entanglement entropy and stress tensor one-point function, holographic calculations of $b$ and $d_2$ have been performed for Levi type-$\mds{L}$ defects in 4d $\N=4$ $SU(N)$ SYM~\cite{Gentle:2015jma, Gentle:2015ruo,Jensen:2018rxu} and for Wilson surfaces in the 6d $\N=(2,0)$ $A_{N-1}$ SCFT~\cite{Gentle:2015jma,Rodgers:2018mvq,Estes:2018tnu,Jensen:2018rxu}. One of our goals is to reproduce these results using purely field theory means, so let us review them in detail.

For the Levi type-$\mathds{L}$ surface defect in 4d $\CN=4$ $SU(N)$ SYM theory, the holographic results for $b$ and $d_2$ are
\begin{subequations}
\label{eq:4d-Levi-surface}
	\begin{align}\label{eq:4d-Levi-surface-b}
	b &=3\left( N^2 -\sum_{i=1}^{n+1} N_i^2\right)\,,\\\label{eq:4d-Levi-surface-d2}
	d_2&= 3\left( N^2 -\sum_{i=1}^{n+1} N_i^2\right) +\frac{24\pi^2 N}{\lambda} \sum_{i=1}^{n+1} N_i \left|\beta_i^2+\gamma_i^2\right|\,,
	\end{align}
\end{subequations}
where $\lambda$ is the 't Hooft coupling of $\N=4$ SYM.

As mentioned in section~\ref{sec:intro}, in this case ref.~\cite{Bianchi:2019sxz} proved that $d_2=d_1$, so in fact the holographic calculations provide all three defect central charges. As also mentioned in section~\ref{sec:intro}, these defects preserve enough SUSY that $b$ cannot depend on defect or ambient marginal couplings, while $d_1$ and $d_2$ can. The $b$ in eq.~\eqref{eq:4d-Levi-surface-b} indeed does not depend on defect or ambient marginal couplings, and in fact depends only on the choice of Levi subgroup $\mds{L}$. On the other hand, $d_2=d_1$ manifestly depends on the defect marginal parameters $\beta_i$ and $\gamma_i$ and on the ambient marginal coupling $\lambda$.

The S-duality of $\N=4$ SYM sends $N/\lambda \to \lambda/N$, under which $d_2=d_1$ appears to change. However, as mentioned at the end of the previous subsection, Weyl anomaly coefficients are invariant under any duality that leaves the metric on $\mathcal{M}_4$ and the submanifold $\Sigma$ invariant, and maps the stress tensor to itself. This includes the S-duality of 4d $\N=4$ SYM. Indeed, after accounting for the S-duality transformations of $\beta_i$ and $\gamma_i$, described in ref.~\cite{Gukov:2006jk}, the combination of $N$, $\lambda$, $\beta_i$, and $\gamma_i$ in eq.~\eqref{eq:4d-Levi-surface-d2} is invariant under S-duality.

Our first result is simply the observation that $b/3$ from eq.~\eqref{eq:4d-Levi-surface-b} agrees exactly with $c_{\text{2d}}/3$ from eq.~\eqref{eq:c2dSYM} for the GLSM construction of the 2d Levi type-$\mds{L}$ defect, and thus also with the complex dimension of the target space of the NLSM construction, eq.~\eqref{eq:Gadde-Gukov-cpx-dim}. Moreover, the expression in~\eq{Gadde-Gukov-cpx-dim} was conjectured to hold for arbitrary values of the parameters $(\alpha,\,\beta,\,\gamma,\,\eta)$, which strongly suggests that we can uniquely identify $c_{\text{2d}}/3=\text{dim}_{\mathds{C}}\,X$ with $b/3$ and not $d_2/3=d_1/3$, since the latter depend on $\beta_i$ and $\gamma_i$.

As mentioned above, however, in all that follows we will take $\beta_i=0$ and $\gamma_i=0$. In that case, the holographic results of eq.~\eqref{eq:4d-Levi-surface} have $b = d_2 = d_1$, so we will not be able to distinguish these three central charges from one another. This will be important in section~\ref{sec:Casimirs}, where we will make a proposal for how to extract a defect central charge from the SUSY partition function of $\N=4$ SYM on $S^1_R \times S^3$ with the Levi type-$\mathds{L}$ defect on $S^1_R \times S^1$. We will only perform an explicit calculation with $\beta_i=0$ and $\gamma_i=0$, so strictly speaking we will not be able to identify uniquely which central charge we calculate, although we will provide multiple arguments that we almost certainly compute $d_2=d_1$.

For Wilson surfaces in the 6d $\CN=(2,0)$ $A_{N-1}$ SCFT we will be able to distinguish $b$ from $d_2$, since in that case generically $b \neq d_2$. A Wilson surface defect is labelled by a Young tableau corresponding to a representation of $\mf{su}(N)$ with highest weight $\omega$. The holographic results for $b$ and $d_2$ for a Wilson surface are~\cite{Gentle:2015jma,Rodgers:2018mvq,Estes:2018tnu,Jensen:2018rxu}
\begin{subequations}
	\begin{align}\label{eq:6d-Wilson-surface-b}
	b&= 24(\rho,\omega)+3(\omega,\omega)\,,\\\label{eq:6d-Wilson-surface-d2}
	d_2 &= 24(\rho,\omega)+6(\omega,\omega)\,,
	\end{align}
\end{subequations}
where $\rho$ is the Weyl vector of $\mf{su}(N)$. Clearly in these cases $d_2 = b + 3(\omega,\omega)$, so that generically $b \neq d_2$, at least at large $N$. In section~\ref{sec:Casimirs} we will extract a defect central charge from the SUSY partition function of the 6d $\CN=(2,0)$ $A_{N-1}$ SCFT on $S^1_R \times S^5$ with a Wilson surface along $S^1_R \times S^1$. Since $b \neq d_2$, we can unambiguously say the defect central charge we obtain is $\propto d_2$. However, in this case ref.~\cite{Bianchi:2019sxz} provided compelling evidence, though not a rigorous proof, that $d_2 = d_1$, so the defect central charge we obtain could in fact be a linear combination of $d_2$ and $d_1$.

\addtocontents{toc}{\protect\setcounter{tocdepth}{2}}

\section{Partition Function on $S^4$}
\label{sec:s4}

In this section, we extract defect central charges from partition functions of $\CN\geq 2$ SCFTs on $\CM= S^4$ with $1/2$-BPS superconformal defects along an equatorial $\Sigma = S^2$. For arbitrary $\CM$ and $\Sigma$ an infinitesimal Weyl transformation $\delta g_{\mu\nu}= 2 g_{\mu\nu}\, \delta\omega $ of the partition function $Z$ gives rise to an integrated Weyl anomaly of the general form, including the defect contribution~\eq{defecttrace},
\begin{equation}
\begin{split}
\delta_\omega \ln Z = -\dfrac{1}{16\pi^2}&\int_{\mathcal{M}} d^4x \sqrt{g}\,\left(a_{\text{4d}}\,E_4  - c_{\text{4d}} \, W_{\mu\nu\rho\sigma}W^{\mu\nu\rho\sigma}\right)\, \delta\omega \\
+ \dfrac{1}{24\pi} &\int_\Sigma d^2x \sqrt{\gamma} \, \left(b \, E_2 + d_1\,\mathring{\mathrm{I\!I}}{}^\mu_{ab}\mathring{\mathrm{I\!I}}{}_\mu^{ab} - d_2 \, W_{ab}{}^{ab} \right)\, \delta\omega \, ,
\end{split}
\end{equation}
where $E_4$ and $E_2$ are the Euler densities for $\CM$ and $\Sigma$, respectively, $\gamma_{ab}$ is the induced metric on $\Sigma$, and $a_{\text{4d}}$ and $c_{\text{4d}}$ are the central charges of the 4d CFT. When $\CM = S^4$ and the defect wraps an equatorial $\Sigma = S^2$, all the type B terms above vanish. In particular, $\CM = S^4$ is conformally flat, so its Weyl tensor vanishes, hence $W_{\mu\nu\rho\sigma}W^{\mu\nu\rho\sigma}=0$ and $W_{ab}{}^{ab}=0$. The second fundamental form for $S^2\hookrightarrow S^4$ is pure trace, so $\mathring{\mathrm{I\!I}}{}^\mu_{ab}\mathring{\mathrm{I\!I}}{}_\mu^{ab}=0$ as well. Thus, the full integrated Weyl anomaly reduces to a linear combination of the A-type anomaly coefficients $a$ and $b$,
\begin{equation}
\label{eq:typeAccs}
\delta_\omega \ln Z = -4a_{\text{4d}} + \frac{b}{3} \, .
\end{equation}
In other words, under a global Weyl re-scaling of $\CM = S^4$, $Z \to e^{(-4a_{\text{4d}} + \frac{b}{3})\,\omega} Z$. Hence, we may extract the linear combination of central charges in eq.~\eqref{eq:typeAccs} from the transformation of the partition function $Z$ under a global Weyl re-scaling, and if we know $a_{\text{4d}}$ for the 4d SCFT, then we can identify $b$. Below we exploit two methods for computing $Z$ that make use of this idea to obtain $b$ if the CFT enjoys enough SUSY, namely SUSY localisation and the AGT correspondence.

\subsection{SUSY Localisation} \label{sec:SUSYloc}

In this subsection, we use existing results for $Z$ computed via SUSY localisation~\cite{Nekrasov:2002qd, Pestun:2007rz} to extract novel results for $b$. 

SUSY localisation is usually performed on the $\Omega$-background, $\mathds{R}^4_{\e_1,\e_2}$, or on an $S^4$ deformed by the ratio of equivariant parameters $\epsilon_{2}/\e_1\equiv\mf{b}^2$. The two ultimately give equivalent results, and we will follow the latter approach. The dimensionless parameter $\mf{b}$ determines how the sphere is ``squashed,'' which we denote as $S^4_{\mf{b}}$. Viewed as a hypersurface in $\mathds{R}^5$, $S^4_\mf{b}$ is defined by
\begin{equation}
x_0^2+ (r\epsilon_1)^2 \, (x_1^2 + x_2^2) + (r\epsilon_2)^2 \, (x_3^2 + x_4^2) = r^2 \, ,
\end{equation}
where $\{x_i\}$, with $i = 0, \ldots, 4$, are the Euclidean coordinates on $\mathds{R}^5$, and $r$ is the equatorial radius. Note that the mass dimensions of $\epsilon_{1,2}$ are 1, which we denote by $[\epsilon_{1,2}]=1$. The round $S^4$ of radius $r$ is recovered in the limit $\epsilon_1=\epsilon_2=\frac{1}{r}$. The deformation parameters $\epsilon_{1,2}$ break the isometry group of the 4-sphere to $U(1) \times U(1)$.  An $\mathcal{N}=2$ theory on this background preserves an $\mathfrak{su}(1|1)\subset\mathfrak{osp}(2|4)$ SUSY subalgebra of the round $S^4$. 

Generically, the localised partition function of a 4d $\mathcal{N}\geq2$ gauge theory without a defect factorises into three contributions~\cite{Pestun:2007rz}: a classical part $Z_{\text{class}}$, a 1-loop part $Z_\text{1-loop}$, and an instanton part $Z_{\text{inst}}$. Each of these is parametrised by the VEV of the adjoint scalar $\langle \Phi \rangle = a$ which is valued in the Cartan subalgebra $\mathfrak{h} \subset \mathfrak{g}$. The full partition function is obtained by integrating $a$ over $\mathfrak{h}$. Schematically,
\begin{equation}
Z_{S^4_\mathfrak{b}}=\int_{\mathfrak{h}}da \, Z_{\text{class}}Z_{\text{1-loop}}|Z_{\text{inst}}|^2 \, . \label{eq:Z4d}
\end{equation}

We implement global Weyl re-scalings by taking $\epsilon_1=\epsilon_2=\frac{1}{r}$ and then re-scaling $r$. If the theory is a SCFT, the 4d Weyl anomaly implies $Z_{S^4}\to r^{-4a_{\text{4d}}}Z_{S^4}$. The only contributions to $a_{\text{4d}}$ come from the integration measure $da$ and $Z_{\text{1-loop}}$, since the other factors are Weyl-invariant. More specifically, $Z_{\text{1-loop}}$ is a product of one-loop determinants of Laplacians for fields of different spins. Each such one-loop determinant is an infinite product of eigenvalues that diverges, and needs to be regulated. As explained in appendix~\ref{ap:special_fn}, we use zeta-function regularisation, which, crudely speaking, means that via analytic continuation we replace each infinite product with special functions, usually combinations of (multiple) Gamma functions. From that point of view, the ``quantum'' contribution to the Weyl anomaly of $Z_{S^4_\mathfrak{b}}$ comes from the ``anomalous'' scaling properties of these special functions, while the ``classical'' contribution comes from $da$. We provide more details of this in appendix~\ref{ap:special_fn}, and we will see explicit examples below.

Now consider a surface defect wrapping $\Sigma = S^2_{\e_1} \hookrightarrow S^4_\mathfrak{b}$ located at $x_3=x_4=0$ in $\mathds{R}^5$ such that it preserves the $U(1)\times U(1)$ isometry. Its embedding into $\mathds{R}^5$ is
\begin{equation}
x_0^2+ (r\epsilon_1)^2 \, (x_1^2 + x_2^2) = r^2 \, .
\end{equation}
A 2d $\mathcal{N}=(2,2)$ theory on $\Sigma$ preserves the same $\mathfrak{su}(1|1)\subset\mathfrak{osp}(2|2)$ SUSY subalgebra as above. Thus one can introduce couplings between the 2d $\mathcal{N}=(2,2)$ theory to the ambient 4d $\mathcal{N}=2$ theory on $\Sigma$ without breaking any further SUSY. As mentioned in section~\ref{sec:review2dqfts}, we can couple the ambient 4d fields to the 2d fields on $\Sigma$ by introducing superpotential couplings on $\Sigma$ to couple the 2d and 4d matter multiplets and/or by gauging a global symmetry on $\Sigma$ and identifying it with an ambient 4d global/gauge symmetry~\cite{Gomis:2014eya, Gomis:2016ljm}.

Some of the examples of 2d superconformal defects considered below are constructed from 2d $\CN=(2,2)$ GLSMs in the UV before flowing to the putative IR superconformal fixed point. The $S^2_{\e_1}$ partition function $Z_{S^2}$ of a purely 2d GLSM with gauge group $G_\text{2d}$ can also be computed through SUSY localisation. This is most conveniently done on the Coulomb branch of the moduli space~\cite{Benini:2012ui, Doroud:2012xw}. The field configurations on the locus are parametrised by a GNO-quantised 2d gauge flux $\mathfrak{m} = \frac{1}{2 \pi}\int F$ on $S^2_{\e_1}$ and the VEV of a real vector multiplet scalar $\sigma$. Schematically, combining the 1-loop and non-perturbative partition functions yields
\begin{equation}\label{eq:Localized-Z-S2}
Z_{S^2}=\dfrac{1}{|\mathcal{W}_{\mathfrak{g}_\text{2d}}|}\sum_{\mathfrak{m} \in \mathfrak{h}_\text{2d}^\mds{Z}} \int_{\mathfrak{h}_\text{2d}}d\sigma \, Z_{\text{class}}Z^{\text{gauge}}_{\text{1-loop}}Z^{\text{matter}}_{\text{1-loop}} \, ,
\end{equation}
where $\mathcal{W}_{\mathfrak{g}_\text{2d}}$ is the Weyl group of the associated gauge Lie algebra $\mathfrak{g}_\text{2d}$, and $\mathfrak{h}_\text{2d}^\mds{Z}$ is the GNO-lattice.\footnote{In practical terms, $\mathfrak{m} \in \mathfrak{h}_\text{2d}^\mds{Z}$ has integer eigenvalues on any representation of the gauge group $G_{\text{2d}}$.} Note that the kinetic term of $\sigma$ in the vector multiplet action is normalised such that $[\sigma]=1$. We emphasise that the localised partition function is independent of the 2d Yang-Mills coupling $g$, and only depends on the $S^2$ (or $S^2_{\e_1}$) through its equatorial radius.

We will again implement a global Weyl re-scaling by re-scaling $r$, in which case the 2d Weyl anomaly implies $Z_{S^2} \to r^{c_{\text{2d}}/3}Z_{S^2}$. Similarly to $Z_{S^4}$, the quantum contribution to the 2d Weyl anomaly comes from zeta-function regularisation of the infinite products in $Z^{\text{gauge}}_{\text{1-loop}}Z^{\text{matter}}_{\text{1-loop}}$, while the classical contribution comes from $d\sigma$. For more details, see appendix~\ref{ap:special_fn}. We will also see explicit examples in the following.

As explained in section~\ref{sec:intro} and above, our aim in this subsection is to extract $b$ from the Weyl anomaly of the localised partition function of 2d-4d coupled systems. For such systems many SUSY localised partition functions have been computed, but we will focus on cases where the 4d ambient theory is conformal, namely $N^2$ free massless hypermultiplets, $\CN=4$ $SU(N)$ SYM theory, and $\CN=2$ $SU(N)$ SQCD with $2N$ flavours with $1/2$-BPS $\CN=(2,2)$ surface operators, enhanced to $\CN=(4,4)$ for $\CN=4$ SYM. 

\subsubsection{Free Massless Hypermultiplets with a Generic Surface Defect}
\label{sec:fmh}

To start, we consider the theory of $N^2$ free massless hypermultiplets on $S^4_\mathfrak{b}$, which arises in the $A_{N-1}$ class $\mathcal{S}$ construction where $\mathcal{C}_{g,n}$ is an $S^2$ with two full punctures and one simple puncture. This theory enjoys global $USp(2N^2)$ flavour symmetry. To this ambient theory we couple the 2d GLSM in figure~\ref{quiver1}, which we put on $\Sigma=S^2_{\e_1}$. The GLSM enjoys an $SU(N) \times SU(N)$ symmetry acting on the (anti-)fundamental chirals, whereas the bifundamental and adjoint chirals enjoy a $U(1)$ symmetry. We couple the ambient free hypers to the GLSM via cubic and quintic superpotential couplings that identify the shared 2d-4d flavour symmetry $SU(N) \times SU(N) \times U(1) \subset USp(2N^2)$~\cite{Gomis:2014eya,Gomis:2016ljm}.

Absent an ambient 4d vector multiplet to couple to the GLSM, the saddle points of the 2d-4d theory are parametrised by independent contributions from decoupled 2d and 4d loci, and so the SUSY localised partition function of this theory factorises \cite{Gomis:2014eya,Gomis:2016ljm} 
\begin{equation}\label{eq:free-hyper-Z-generic}
Z_{\Sigma \hookrightarrow S^4_\mathfrak{b}} = Z_{S^4_\mathfrak{b}}^{\text{free}}\, Z_{\Sigma} \,.
\end{equation}
We denote by $Z_{\Sigma}$ the partition function of the GLSM on $S^2_{\e_1}$, and
\begin{equation}\label{eq:free_hypers}
Z_{S^4_\mathfrak{b}}^{\text{free}} = \left(\Upsilon\left(\frac{\epsilon_1+\epsilon_2}{2}\bigg|\epsilon_1, \epsilon_2\right)\right)^{-N^2}
\end{equation}
is the partition function of the $N^2$ free massless hypers in zeta-function regularisation. The Upsilon function is defined as
\begin{equation}
\Upsilon (z|a_1,a_2) \equiv \dfrac{1}{\Gamma_2(z|a_1,a_2) \Gamma_2(a_1+a_2-z|a_1,a_2)}\, , \label{eq:Upsilon}
\end{equation}
where $\Gamma_2(z|a_1,a_2)$ is the double Gamma function.  For more details about these special functions, see appendix~\ref{ap:special_fn}. However, the only information we currently need about the Upsilon function is its behaviour under re-scaling of its arguments, eq.~\eqref{eq:scaleUpsilon},
\beq
\label{eq:upsilonscaling}
\Upsilon\left(\frac{z}{r}\left|\frac{a_1}{r},\frac{a_2}{r}\right.\right)=r^{-2\zeta_2 (0;z|a_1,a_2)}\Upsilon(z|a_1,a_2)\,,
\eeq
where $\zeta_2(s;z|a_1,a_2)$ is the Barnes double zeta-function defined in eq.~\eqref{eq:Barnes}.

Since the 2d-4d partition function factorises, it is sufficient to just consider the scaling of $Z_{\Sigma}$ in order to compute $b$. Hence, $b$ is identified with $c_\text{2d}$.

However, to be clear, we hasten to add that this $c_{2d}$ is not (necessarily) the central charge of a 2d CFT, because the 2d stress tensor of our defect degrees of freedom is not necessarily conserved, due to the coupling to the ambient 4d fields. This implies various differences from a 2d CFT: no lower bound on our $b = c_{2d}$ is currently known, the usual 2d $c$-theorem does not necessarily apply (although the $b$-theorem does), and so on. In practical terms, however, the upshot is that we still compute $c_{2d}$ from eq.~\eqref{eq:c2d}, which in particular requires identifying the representations and R-charges of the 2d fields.

Let us consider the three contributions to \eq{Localized-Z-S2} separately and explicitly study their scaling behaviour. If the 2d gauge group $G_{\text{2d}}$ has $n$ $U(1)$ factors, the classical part of the localised partition function takes the form
\begin{equation}\label{eq:free-hyper-generic-classical}
Z_{\text{class}} = \prod_{j=1}^n z_j{}^{\text{Tr}_j(ir\sigma+\frac{\mathfrak{m}}{2})}\bar{z}_j{}^{\text{Tr}_j(ir\sigma-\frac{\mathfrak{m}}{2})} \, ,
\end{equation}
where $z_j \equiv e^{-2\pi \xi_j + i \theta_j}$, $\xi_j$ are FI parameters, $\theta_j$ are theta-angles, and $\text{Tr}_j$ denotes a projection to the $j^{\textrm{th}}$ $U(1)$ factor \cite{Benini:2016qnm}. Writing the measure in \eq{Localized-Z-S2} as $\int_{\mathfrak{h}_\text{2d}}d\sigma = r^{-\text{rank }G_\text{2d}} \int_{\mathfrak{h}_\text{2d}}d(r\sigma)$, it is clear that $r\sigma$ in \eq{free-hyper-generic-classical} is just a dimensionless dummy variable in the integration over the locus. Hence, the classical part $Z_{\text{class}}$ is trivial under Weyl re-scalings, as advertised. The measure, however, contributes factors of the $S^4$ radius $r$, which transform under Weyl re-sclaings with weight $-\text{rank }G_\text{2d}$.

The 1-loop contribution coming from the gauge sector takes the form
\begin{equation}\label{eq:free-hyper-1-loop-gauge-generic}
Z^{\text{gauge}}_{\text{1-loop}} = e^{2 \pi i \rho_{\rm{2d}} (\mathfrak{m})} \prod_{\alpha \in \Delta^+} \left[\dfrac{1}{r^2}\left(\dfrac{\alpha (\mathfrak{m})^2}{4} + \alpha (r\sigma)^2 \right) \right] \, ,
\end{equation}
where $\Delta^+$ is the set of positive roots and $\rho_{\rm{2d}}$ is the Weyl vector of $\mathfrak{g}_{\text{2d}}$, the Lie algebra associated to $G_{\text{2d}}$. Collecting the overall factors of $r$, we see that under Weyl re-scalings \eq{free-hyper-1-loop-gauge-generic} transforms with weight $-\text{dim }G_{\text{2d}}+\text{rank }G_{\text{2d}}$.

After zeta-function regularisation, the 1-loop partition function of the matter sector --- composed of massless chiral multiplets in the $\mc{R}$ representation of $G_{\rm 2d}$ --- becomes
\begin{equation}\label{eq:free-hyper-1-loop-matter-generic}
Z^{\text{matter}}_{\text{1-loop}}=\prod_{\mathcal{R}} \prod_{\{h_{\mathcal{R}}\}} \dfrac{\Gamma\left(\frac{q_\mathcal{R}}{2}-ih_{\mathcal{R}}(r\sigma)-\frac{h_{\mathcal{R}}(\mathfrak{m})}{2}\right)}{\Gamma\left(1-\frac{q_\mathcal{R}}{2}+ih_{\mathcal{R}}(r\sigma)-\frac{h_{\mathcal{R}}(\mathfrak{m})}{2}\right)} \, r^{1-q_{\mathcal{R}}+2ih_{\mathcal{R}}(r\sigma)} \,,
\end{equation}
where $q_\mathcal{R}$ is the 2d R-charge of the multiplet, and $\{h_\mathcal{R}\}$ denotes the set of weights of $\mathcal{R}$. Counting the factors of $r$ that appear in \eq{free-hyper-1-loop-matter-generic}, we see that $Z^{\rm matter}_{\rm 1-loop}$ transforms with weight $\sum_\mathcal{R} (1-q_\mathcal{R})\text{dim}~\mathcal{R}$ under Weyl re-scaling. For $\mathfrak{g}_{\text{2d}}$ a direct sum of semi-simple Lie algebras and $\mathfrak{u}(1)$'s, the remaining term in the Weyl re-scaling $\sum_\mathcal{R}\sum_{\{h_\mathcal{R}\}}2ih_\mathcal{R}(r\sigma)$ reduces to a sum over the charges under the $U(1)$ factors of $G_{\text{2d}}$, which vanishes.

Combining the weights from the gauge sector, the matter sector, and the measure, one finds that the partition function is scale-invariant up to an overall factor of $r^{c_\text{2d}/3}$, with $c_{\rm 2d}$ given in~\eq{c2d}. Hence, $b=c_\text{2d}$ in the round sphere limit $\epsilon_1=\epsilon_2=\frac{1}{r}$, as argued above.

Even though the 4d theory is very simple, considering the case of $N^2$ free massless hypers with a surface defect illustrates the important point that the non-trivial contribution to the central charge comes from the scaling of the 1-loop partition function. The simplicity of the above example stems from the factorisation in \eq{free-hyper-Z-generic}, which immediately led to identifying $b=c_\text{2d}$.

For a more generic 2d-4d system, one might suspect that matter charged under both 2d and 4d gauge groups would spoil the factorisation of $Z_{\Sigma\hookrightarrow S^4}$ and possibly alter $b$. However, that is not the case, if the system enjoys enough SUSY. It is now understood that a $1/2$-BPS surface defect engineered in a generic 4d $\CN=2$ gauge theory by gauging symmetries~\cite{Gorsky:2017hro} or through Higgsing~\cite{Pan:2016fbl} mixes ambient and defect degrees of freedom in only two ways. Firstly, any 4d adjoint hypermultiplet scalars frozen at their VEVs enter as twisted mass parameters in $Z_{\Sigma}$, while keeping its functional form unchanged. Secondly, any coupling of 2d and 4d degrees of freedom leads to an extra factor in the partition function that is entirely non-perturbative: it arises from the interactions of instantons and vortices. The 1-loop part of the partition function receives no modifications. Hence, central charges extracted from Weyl re-scalings of the partition function are unchanged. In other words, we expect $b = c_\text{2d}$ to be the case always. Indeed, we will see examples of this below.

Let us point out that the scaling behaviour of the partition function can often be obtained in a more straightforward, yet ad hoc way by using three facts: only the 1-loop partition function (and the measure of the integration over the VEV of the adjoint scalar) contributes, factors of $1/r$ arise in the evaluation of 1-loop determinants, and special functions are the result of zeta-function regularisation.

The explicit dependence on the scale $r$ is often left implicit. If one was given the partition function $Z_{\Sigma}$ without any scale factors, one could still deduce the scaling behaviour by re-instating the correct $r$-dependence, dealing with special functions appropriately and  accounting for the measure. For example, if one encounters the Euler Gamma-function $\Gamma\left(\frac{q_\mathcal{R}}{2}-ih_{\mathcal{R}}(\sigma)-\frac{h_{\mathcal{R}}(\mathfrak{m})}{2}\right)$, one first needs to insert appropriate factors of $r$ to make its argument dimensionless, i.e. $\sigma \rightarrow r\sigma$. The natural function that appears in the zeta-function regularisation of the matter sector 1-loop partition function is the Barnes single Gamma-function $\Gamma_1(z|a,b)$ defined in eq.~\eqref{eq:multipleGamma}. To obtain the scaling behaviour one should interpret the Euler Gamma-function as $\Gamma_1\left(\left.\frac{1}{r}\left(\frac{q_\mathcal{R}}{2}-ih_{\mathcal{R}}(r\sigma)-\frac{h_{\mathcal{R}}(\mathfrak{m})}{2}\right)\right| \frac{1}{r}\right)$. Using the properties
\begin{equation}
\Gamma_1\left(\frac{z}{r} \left| \frac{1}{r}\right.\right)= r^{\frac{1}{2}-z} \, \Gamma_1\left(\left.z\right| 1\right)\,, \qquad
\Gamma_1(z|1)=\dfrac{1}{\sqrt{2\pi}}\Gamma(z) \, ,  \label{eq:gamma_r_scaling}
\end{equation}
one correctly recovers the partition function $Z_{\Sigma}$ with appropriate scale factors. We refer to appendix \ref{ap:special_fn} for more details and definitions of these special functions. 

\subsubsection*{Examples of Defects Coupled to Free Massless Hypermultiplets}

Having determined $b=c_\text{2d}$ for superconformal surface defects coupled to 4d free massless hypers, we can now consider some specific defect models. All that needs to be done to compute $b$ is to determine the 2d R-charges $q_\mathcal{R}$ of the matter fields. 

Due to the $\mathfrak{su}(1|1)$-invariant coupling between the 2d $\mathcal{N}=(2,2)$ GLSM and the hypermultiplets, the 2d R-symmetry generators are linear combinations of the $U(1)_{N}$ generator of rotations in the normal bundle to $\Sigma$ and $U(1)_R\subset SU(2)_R$  of the ambient R-symmetry \cite{Gomis:2016ljm}. The coefficients determining the exact 2d R-symmetry depend on $\mathfrak{b}$. The 2d R-charges of the 4d hypermultiplet scalars restricted to $\Sigma$ can be found in terms of their 4d charges under $U(1)_{N} \times U(1)_R$. Requiring that the 2d-4d superpotentials have 2d R-charge $q=2$ together with constraints from the identified flavour symmetry then fixes the R-charges of the 2d fields and sets their twisted masses to zero.

The precise superpotential terms depend on the particular quiver diagram, and were found in refs.~\cite{Gomis:2014eya, Gomis:2016ljm}. In particular, they depend on whether the $j^{\textrm{th}}$ node has an adjoint chiral $X_j$. If it does, we define $\eta_j \equiv +1$, and if it does not, $\eta_j \equiv -1$.  Further let us define $\varepsilon_i \equiv \prod_{j=i}^n \eta_j$. One finds that the hypermultiplet scalars restricted to $\Sigma$ have 2d R-charge $q^{\text{\tiny{hyper}}}=1+\mathfrak{b}^2$, the fundamental and anti-fundamental chirals have $q^{\text{\tiny{fund}}}_n+q^{\text{\tiny{anti-fund}}}_n = 1-\mathfrak{b}^2$, the adjoint chirals have
\begin{align}
\label{R-charge_adjoint}
q_{X_j}=\begin{cases}
2+2\mathfrak{b}^2 \qquad &\text{if} \; \varepsilon_{j+1}=\varepsilon_j=-1 \\
-2 \mathfrak{b}^2 \qquad &\text{if} \; \varepsilon_{j+1}= \varepsilon_{j}=+1\, ,
\end{cases}
\end{align}
and the bifundamentals have R-charges
\begin{align}
\label{R-charge_bifund}
q^{\text{\tiny{bif}}}_{j(j-1)}+q^{\text{\tiny{bif}}}_{(j-1)j}=\begin{cases}
-2\mathfrak{b}^2 \qquad &\text{if} \; \varepsilon_j=-1 \\
2+ 2 \mathfrak{b}^2 \qquad &\text{if} \; \varepsilon_{j}=+1 \, ,
\end{cases}
\end{align}
where we have a total of $n$ nodes and $\varepsilon_{n+1} \equiv +1$. Notice that in a 2d SCFT, with a conserved stress tensor, unitarity and the BPS bound require positive R-charges. In contrast, our 2d defect fields do not have a conserved 2d stress tensor, and so can have negative R-charges.

\paragraph{Example 1: $\mathcal{N}=(2,2)$ SQCD.} As a first example consider $\mathcal{N}=(2,2)$ SQCD with gauge group $G_{\text{2d}}=U(K)$ and $N$ fundamental and $N$ anti-fundamental chiral multiplets coupled to $N^2$ ambient free massless hypers. Note that $q^{\text{\tiny{fund}}}_n+q^{\text{\tiny{anti-fund}}}_n = 0$ in the round sphere limit $\mf{b}=1$. Thus, using eq.~\eqref{eq:c2d} we find
\begin{equation}
\dfrac{b}{3}= 2NK - K^2 = K(2N-K) \, . \label{eq:SQCD}
\end{equation}

\paragraph{Example 2: $\mathcal{N}=(2,2)$ SQCDA.} We now add an adjoint chiral to the previous example, where $q_X = -2$ in the limit $\mf{b}=1$. Using eq.~\eqref{eq:c2d}, this ``extra'' field thus contributes an additional $(1-q_X) \text{dim} \,\mathcal{R} = 3 K^2$ to the value of $b$ of the previous example,
\begin{equation}
\dfrac{b}{3} = 2NK-K^2 +3K^2= 2K(N+K)\, . \label{eq:SQCDA}
\end{equation}
These two examples clearly obey the $b$-theorem~\cite{Jensen:2015swa}. If we start in the UV with SQCDA, with $b$ in eq.~\eqref{eq:SQCDA}, and deform the theory by a mass term for the adjoint chiral, then in the IR we will find SQCD~\cite{Gomis:2014eya}, with $b$ in eq.~\eqref{eq:SQCD}. In this case, $b_{\textrm{UV}}-b_{\textrm{IR}} = 9K^2 \geq 0$.

\paragraph{Example 3: $\mathcal{N}=(2,2)$ quiver with $n$ adjoint chirals.} We can also consider more general quiver gauge theories. For example, consider the $n$-node quiver depicted in figure~\ref{quiver1} with gauge group $G_{\text{2d}}=U(K_1)\times \ldots \times U(K_n)$, $N$ fundamental and $N$ anti-fundamental chirals of $U(K_n)$ and adjoint chirals on each node, coupled to $N^2$ free hypers. Using $q_{X_j}=-2$, $q^{\text{\tiny{bif}}}_{j(j-1)}+q^{\text{\tiny{bif}}}_{(j-1)j}=4$ and $q^{\text{\tiny{fund}}}_n+q^{\text{\tiny{anti-fund}}}_n=0$, and eq.~\eqref{eq:c2d} we find after a bit of algebra that
\begin{equation}
\begin{split}
\dfrac{b}{3}
&=2\sum_{i=1}^n (K_i-K_{i-1}) K_i + 2K_n N \, ,  \label{eq:b_n_adj_chirals}
\end{split}
\end{equation}
where we have defined $K_0 \equiv 0$. 

\paragraph{Example 4: $\mathcal{N}=(2,2)$ quiver with $(n-1)$ adjoint chirals.} Consider the same quiver as the previous example, but with adjoint chirals on all nodes but the $n^{\textrm{th}}$ one. In this case, $q_{X_j}=4$, $q^{\text{\tiny{bif}}}_{j(j-1)}+q^{\text{\tiny{bif}}}_{(j-1)j}=-2$, and $q^{\text{\tiny{fund}}}_n+q^{\text{\tiny{anti-fund}}}_n=0$, so that
\begin{equation}
\dfrac{b}{3}=-4\sum_{i=1}^n (K_i-K_{i-1}) K_i + 2K_n N + 3 K_n^2\,.
\label{eq:b_n-1_adj_chirals}
\end{equation}
These two examples also obey the $b$-theorem~\cite{Jensen:2015swa}. If we start in the UV with an $\mathcal{N}=(2,2)$ quiver with $n$ adjoint chirals, with $b$ in eq.~\eqref{eq:b_n_adj_chirals} and deform by a mass term for the $n^{\textrm{th}}$ adjoint chiral, then in the IR we will find an $\mathcal{N}=(2,2)$ quiver with $n-1$ adjoint chirals~\cite{Gomis:2014eya}, with $b$ in eq.~\eqref{eq:b_n-1_adj_chirals}. We thus have
\beq
\dfrac{b_{\textrm{UV}}}{3} - \dfrac{b_{\textrm{IR}}}{3} = - 3 K_n^2+6 \sum_{i=1}^n (K_i-K_{i-1}) K_i  = 3\sum_{i=1}^n (K_i - K_{i-1})^2,
\eeq
where the final equality holds because $K_0\equiv0$. Clearly in this case $b_{\textrm{UV}}-b_{\textrm{IR}} \geq 0$, and so the $b$-theorem is satisfied.

To our knowledge all four of the examples above, and indeed the general statement $b = c_\text{2d}$, are novel results for $b$ of 2d superconformal defects. Notice that in all of our examples $b \geq 0$: for eqs.~\eqref{eq:SQCD},~\eqref{eq:SQCDA}, and~\eqref{eq:b_n_adj_chirals} this is manifest, while for eq.~\eqref{eq:b_n-1_adj_chirals} this can be checked straightforwardly, for example by considering limiting cases.

\subsubsection{$\mathcal{N}=4$ SYM with a Generic Surface Defect}

To construct a $1/2$-BPS superconformal surface defect in $\mathcal{N}=4$ SYM, one can couple a 2d $\CN=(4,4)$ GLSM to the ambient theory~\cite{Gadde:2013ftv}. $\CN=(4,4)$ SUSY requires the $i^{\textrm{th}}$ node in figure~\ref{quiver1} to have an adjoint chiral multiplet $X_i$ for all $i$. The $\CN=(2,2)$ adjoint chiral recombines with the $\CN=(2,2)$ vector multiplet into an $\CN=(4,4)$ vector multiplet. Similarly, the bifundamentals $\phi^{\text{\tiny bif}}_{i(i+1)}$ and $\phi^{\text{\tiny bif}}_{(i+1)i}$ regroup into bifundamental hypers, and the $N$ (anti-)fundamental chirals $\phi^{\text{\tiny fund}}_n$ and $\tilde \phi^{\text{\tiny anti-fund}}_n$ recombine into $N$ fundamental hypermultiplets. The $N$ hypers enjoy $SU(N)$ flavour symmetry such that the GLSM can be coupled to 4d $\CN=4$ $SU(N)$ SYM theory by gauging the 2d flavour group. As argued in the previous subsection, $b=c_\text{2d}$ as there is no perturbative 2d-4d contribution to the partition function. We may thus calculate $b$ through yet another counting exercise.\footnote{We thank B.~Le Floch for pointing this out to us, and for discussions directly related to this computation.}

Assuming the GLSM flows to an IR fixed point, the central charge of the 2d SCFT is given by \eq{c2d}. To determine the 2d R-charges, one considers the allowed superpotential terms which schematically look like $W = \phi X \tilde{\phi}$ in $\CN=(2,2)$ language. The R-charge assignments are easily deduced by looking at the $U(1)_R$ action on the mesons built from fundamental chirals, which combine into non-compact scalars at the IR fixed point.  The exact low-energy $U(1)_R$ symmetry cannot act as a rotation on the mesons due to chiral factorisation of the R-symmetry in a CFT. This gives the assignment that matter sector chiral multiplets in the (anti-)fundamental and bifundamental representations have $q =0$, while the adjoint chiral multiplets carry $q=2$. Thus, \eq{c2d} gives
\beq\label{eq:bSYM}
\frac{b}{3} = \frac{c_\text{2d}}{3}= 2\sum_{i=1}^{n} K_i(K_{i+1} - K_i) = N^2 - \sum_{i=1}^{n+1} N_i^2\,,
\eeq
as quoted in \eq{c2dSYM}, which is in agreement with the complex dimension of the moduli space of the Levi type-$\mathds{L}$ defect in \eq{Gadde-Gukov-cpx-dim}, and with the holographic result in eq.~\eqref{eq:4d-Levi-surface-b}, thus proving that the latter is not merely the large-$N$ limiting value.

\subsubsection{$\mathcal{N}=4$ SYM with a Full Levi Defect}

A useful check of the previous result~\eq{bSYM} can be performed in special cases. In refs.~\cite{Kanno:2011fw, Nawata:2014nca}, the authors consider $\mathcal{N}=2^*$ SYM with gauge group $G=SU(N)$ and a full surface defect ($\mds{L}=\mds{T}$) engineered by putting the theory on the orbifold $\mds{C}\times \mds{C}/\mds{Z}_N$.\footnote{In ref.~\cite{Kanno:2011fw}, the authors found the following equivalence: instanton moduli space of a 4d $\mathcal{N}=2$ gauge theory with full surface defect $\iff$ instanton moduli space without defect but on the orbifold $\mds{C}\times \mds{C}/\mds{Z}_N$. This allows one to compute the instanton partition function of the coupled 2d-4d system by instead working on the orbifold. It was then conjectured in ref.~\cite{Nawata:2014nca} that this equivalence should hold more generally for the full partition function. The author of ref.~\cite{Nawata:2014nca} computes the 1-loop determinants on the orbifold and goes on to check that the partition function obtained in this way correctly encodes the coupled 2d-4d and 2d degrees of freedom.} By taking the mass of the 4d adjoint hyper to zero, the $\mathcal{N}=2^*$ SUSY enhances to $\mathcal{N}=4$. 

Let us now compute $b$ for this system. The non-trivial contribution comes from the 1-loop partition function,
\begin{equation}
Z^{\mathcal{N}=4}_{\text{1-loop}}[1^N] = \prod_{\substack{i,j = 1\\i\neq j}}^N \dfrac{\Upsilon\left(a_i-a_j + \ceil*{\frac{j-i}{N}}\epsilon_2|\epsilon_1,\epsilon_2\right)}{\Upsilon\left(a_i-a_j + \frac{\epsilon_1 + \epsilon_2}{2} + \ceil*{\frac{j-i}{N}}\epsilon_2|\epsilon_1,\epsilon_2\right)} \, , \label{eq:N=4[1^N]}
\end{equation}
where $a_i$ are the components of the VEV of the 4d adjoint scalar $\langle \Phi\rangle = \text{diag}(a_1, \ldots , a_N)$, and $\ceil{x}$ denotes the ceiling of $x$.

The arguments of the Upsilon-functions in \eq{N=4[1^N]} have mass dimension one. To make the overall scale factor explicit, we should factor out $1/r$ from their arguments. Define the dimensionless quantities $\tilde{\epsilon}_{1,2} \equiv r\epsilon_{1,2}$ and $Q \equiv \tilde{\epsilon}_1 +\tilde{\epsilon}_2 $. Under a re-scaling, the Upsilon-function transforms according to eq.~\eqref{eq:upsilonscaling}, which means \eq{N=4[1^N]} becomes
\begin{equation}
Z^{\mathcal{N}=4}_{\text{1-loop}}[1^N] = \prod_{\substack{i,j = 1\\i\neq j}}^N \dfrac{\Upsilon\left(x_{ij}|\tilde{\epsilon}_1,\tilde{\epsilon}_2\right)}{\Upsilon\left( x_{ij}+Q/2|\tilde{\epsilon}_1,\tilde{\epsilon}_2\right)} \, r^{\kappa_{ij}}\, ,
\end{equation}
where 
\begin{equation}
\begin{split}
x_{ij}&=r\left(a_i-a_j + \ceil*{\frac{j-i}{N}}\epsilon_2\right)\, ,\\
\kappa_{ij} &= -2\,\zeta_2 (0;x_{ij}|\tilde{\epsilon}_1, \tilde{\epsilon}_2)+2\,\zeta_2 (0;x_{ij} + Q/2|\tilde{\epsilon}_1, \tilde{\epsilon}_2)\, ,
\end{split}
\end{equation}
and $\zeta_2(s;z|a,b)$ is the Barnes double zeta-function defined in eq. \eqref{eq:Barnes}. Using eq.~\eqref{eq:Barneszero} and taking $\mf{b}\to1$, one finds
\begin{equation}
\sum_{\substack{i,j = 1\\i\neq j}}^N\kappa_{ij} = \sum_{\substack{i,j = 1\\i\neq j}}^N (2x_{ij}-1) = 2 \, \frac{N^2-N}{2} - (N^2-N) = 0 \, .
\end{equation}
In other words, the 1-loop determinant in the presence of the full Levi defect is scale-invariant. Hence,
\begin{equation}
-4a_\text{4d}+\frac{b}{3}=-(N-1) \, ,
\end{equation}
where the right-hand side is the contribution of the measure in eq.~\eqref{eq:Z4d}.  

The central charge $a_\text{4d}$ for $\CN=4$ $SU(N)$ SYM theory is well-known, $4a_\text{4d}=N^2-1$, and so we find
\begin{equation}
\frac{b}{3}=N^2-N\,, \label{eq:bN=4}
\end{equation}
which agrees with~\eq{bSYM} in this special case, as advertised.

This agreement may seem surprising, given the different M-theory origins of this surface defect and the defects that lead to eq.~\eqref{eq:bSYM}. This full surface defect comes from the compactification on a torus of the 6d $\mathcal{N}=(2,0)$ theory with a codimension-two defect~\cite{Kanno:2011fw, Nawata:2014nca}. This arises in M-theory from a stack of coincident M5-branes wrapping the torus and a second stack of M5-branes intersecting the first stack and also wrapping the torus, thus producing the orbifold in 4d, i.e. a codimension-two singular surface. On the other hand, the surface defects that lead to~\eq{bSYM}, namely the GLSM quivers reviewed in section~\ref{sec:review}, come from a codimension-four defect in the 6d theory. This arises in M-theory from M2-branes ending on M5-branes, localised at a point on the torus. In these two descriptions $b$ agrees because of the duality of refs.~\cite{Gukov:2006jk,Gukov:2008sn,Frenkel:2015rda,Ashok:2017odt}, mentioned in section~\ref{sec:review}, which leaves invariant the $S^4$, the $\Sigma = S^2$ wrapped by the defect, and the stress tensor, and hence leaves invariant $b$. Of course, also crucial is the fact that $b$ depends only on the Levi subgroup of each defect: if $b$ depended on more detailed information, then the equivalence would not be possible.

\subsubsection{$\mathcal{N}=2$ SQCD with $2N$ Flavours and a Full Levi Defect} 

A large class of SCFTs are the theories of class $\mathcal{S}$ introduced in ref.~\cite{Gaiotto:2009we}. One of the simplest, yet non-trivial examples of such theories is massless $\mathcal{N}=2$ SQCD with $2N$ flavours. A full surface defect in this theory is considered in ref.~\cite{Nawata:2014nca}.

The 1-loop determinant with a full surface defect is 
\begin{align}\label{eq:SQCD-1-loop-full-Levi}
&Z^{\text{SQCD}}_{\text{1-loop}} [1^N]\\
&=\dfrac{\prod_{\alpha \in \Delta^+} \Upsilon (\alpha(a)+\epsilon_2|\epsilon_1, \epsilon_2) \Upsilon (-\alpha(a)|\epsilon_1, \epsilon_2)}{\prod_{i,j=1}^N \Upsilon \left(h_i(a)+\frac{\epsilon_1+ \epsilon_2}{2}+\ceil*{\frac{N-i-j+1}{N}}\epsilon_2|\epsilon_1, \epsilon_2\right)\Upsilon \left(-h_i(a)+\frac{\epsilon_1+ \epsilon_2}{2}+\ceil*{\frac{i-j}{N}}\epsilon_2|\epsilon_1, \epsilon_2\right)} \, ,\nonumber
\end{align}
where $h_i$ are the weights of the fundamental representation of $SU(N)$. 

Following the same strategy as above, we factor out $1/r$ to write the arguments of the special functions in terms of dimensionless quantities $\tilde{\e}_{1,2}$. Let us consider the numerator first. The scaling behaviour of the Upsilon-functions in eq.~\eqref{eq:upsilonscaling} in the $\mf{b}\to1$ limit gives a scaling weight of the numerator of the form
\begin{equation}
-\frac{(N^2-N)}{3} -4\rho(ra) -\sum_{\alpha \in \Delta} (\alpha(ra))^2\,,
\end{equation}
where $\rho$ is the Weyl vector, and $\Delta$ is the set of all (positive and negative) roots. The denominator contributes to the overall scaling weight a factor
\begin{equation}
\frac{2}{3}N^2-N \, + 4\rho(ra) + 2N\sum_{i=1}^N (h_i(ra))^2 \, .
\end{equation}
A vanishing beta function implies (see e.g. ref.~\cite{Pestun:2007rz})
\begin{equation}
\sum_{\alpha \in \Delta} (\alpha(a))^2=2N\sum_{i=1}^N (h_i(a))^2~.\label{eq:vanishingbeta}
\end{equation}
Hence, upon summing the contributions of the numerator and denominator one finds that all terms that depend on the VEV $a$ cancel, giving an overall scaling weight for~\eq{SQCD-1-loop-full-Levi} of the form
\begin{equation}\label{eq:SQCD-1-loop-full-Levi-scaling}
\frac{1}{3}N^2-\frac{2}{3}N \, .
\end{equation}
Finally, to account for $a_{4d}$, we normalise by the partition function without the defect:
\begin{equation}
Z^{\text{SQCD}}_{\text{1-loop}} = \dfrac{\prod_{\alpha \in \Delta^+} \Upsilon (\alpha(a)|\epsilon_1, \epsilon_2) \Upsilon (-\alpha(a)|\epsilon_1, \epsilon_2)}{\prod_{i,j=1}^N \Upsilon \left(h_i(a)+\frac{\epsilon_1+ \epsilon_2}{2}|\epsilon_1, \epsilon_2\right)\Upsilon \left(-h_i(a)+\frac{\epsilon_1+ \epsilon_2}{2}|\epsilon_1, \epsilon_2\right)} \, ,
\end{equation}
which scales with weight
\begin{equation}\label{eq:SQCD-1-loop-scaling}
-\frac{7}{6}N^2+\frac{5}{6}N \, ,
\end{equation}
where we have again used eq.~\eqref{eq:vanishingbeta}. Thus subtracting \eq{SQCD-1-loop-scaling} from \eq{SQCD-1-loop-full-Levi-scaling}, we find that $b$ for a full Levi type-$\mds{L}$ defect in $\N=2$ conformal SQCD is given by
\begin{equation}
\frac{b}{3} =\frac{3}{2}(N^2-N)\, .
\end{equation}

\subsection{AGT Correspondence}
\label{sec:AGT}

In the seminal work \cite{Alday:2009aq}, Alday, Gaiotto and Tachikawa (AGT) proposed a remarkable correspondence between the partition function of a class of 4d asymptotically conformal $\mathcal{N}=2$ SUSY $SU(2)$ quiver gauge theories on $S^4_{\mathfrak{b}}$ and Liouville theory on a genus-$g$ Riemann surface with $n$ punctures, $\mc{C}_{g,n}$. This AGT correspondence has been further extended to $SU(N)$ quiver gauge theories and $A_{N-1}$ Toda field theories~\cite{Wyllard:2009hg}.

In this section, we will employ an extension of the methods originally used in ref.~\cite{Balasubramanian:2013kva} to compute the type A anomaly coefficients in a number of 4d class $\mc{S}$ theories in order to extract the central charge $b$ of a certain type of surface operators, via the AGT correspondence.  To begin with, we give a very brief review of the AGT correspondence, and then discuss its modification for taking into account the insertion of a certain class of surface operators in the 4d theory. In appendix~\ref{appendix_Toda}, we establish our notation, briefly review $A_{N-1}$ Toda field theories, and report all the formulae that we will need in this section.

The object of interest on the 4d side of the AGT correspondence is the SUSY localised partition function $Z_{S^4_{\mf{b}}}$ of an $\CN=2$ gauge theory on $S^4_{\mathfrak{b}}$ discussed in some detail above in section~\ref{sec:SUSYloc} and displayed in~\eq{Z4d}. In general, $Z_{S^4_{\mf{b}}}$ depends on some complexified couplings denoted by $\{q\}$, some masses $\{m\}$ and VEVs of the adjoint scalars $\{a\}$ in the vector multiplets. Hence, in most of the examples considered below the ambient theories are not strictly 4d SCFTs but rather are SUSY gauge theories that are conformal in certain limiting regimes. However, we will explicitly compute $b$ only for 4d massless free hypermultiplets, which are a CFT. We will of course reproduce the results of section~\ref{sec:fmh}, but now from the perspective of the Liouville/Toda theory.
\begin{figure}[t] 
	\begin{center}
		\includegraphics[scale=0.7]{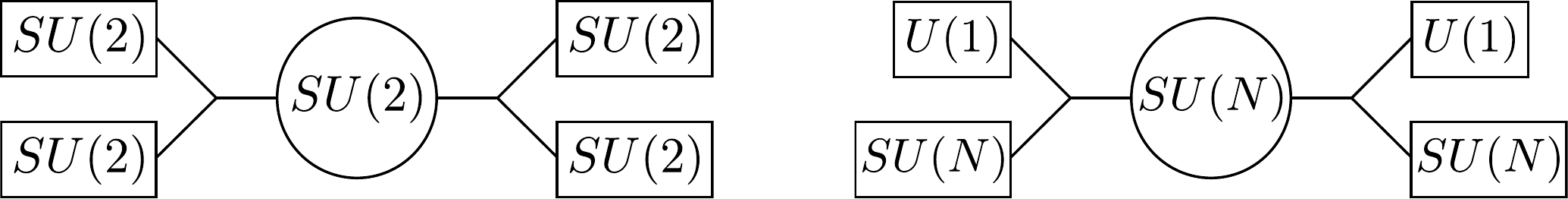}
	\end{center}
	\vspace{-.5cm}
	\caption{\label{quiver_4d}
		\small
		Left: Quiver diagram corresponding to the Liouville four-point function on the sphere. Right: Quiver diagram corresponding to the $A_{N-1}$ Toda four-point function on the sphere. The two $SU(N)$ flavour groups correspond to full punctures on the Riemann surface, while the two $U(1)$ factors correspond to simple punctures.  
	}
\end{figure}

In order to discuss the 2d side of the AGT correspondence, consider the class $\mc{S}$ theory constructed by an $n$-node linear quiver \cite{Gaiotto:2009we} where the gauge sector of the $i^{\rm th}$ node is described by the gauge group $SU(2)_i$, the coupling $\tau_i$, the ``sewing parameter'' $q_i=e^{2\pi i \tau_i}$, and adjoint scalar VEVs $a_i$. The matter sector of the quiver is described by two antifundamental hypermultiplets of mass $\mu_{1,2}$ at the first node, bifundamental hypermultiplets of mass $m_i$ between the $i^{\rm th}$ and $(i+1)^{\rm th}$ nodes, and two fundamental hypermultiplets of mass $\mu_{3,4}$ on the $n^{\rm th}$ and final node (see e.g. the left side of figure~\ref{quiver_4d} for the case $n=1$). The UV curve of the kind of quiver just described is given by the $(n+3)$-punctured sphere $\mc{C}_{0,n+3}$. Through the AGT correspondence, the partition function eq.~\eqref{eq:Z4d} is equivalent to the $(n+3)$-point correlation function of the Liouville field theory on $S^2$, namely
\begin{align}\label{eq:n-point-Toda-AGT}
Z_{S^4_\mathfrak{b}}(\{q\},\{m\};\epsilon_1,\epsilon_2)=\braketbis{\widehat V_{\alpha_0}(\infty)\widehat V_{r m_0}(1) \widehat V_{r m_1}(q_1)\dots \widehat V_{r m_n}(q_1\dots q_n)\widehat V_{\alpha_{n+1}}(0)}\,,
\end{align}
where
\begin{align}
\begin{split}
&\alpha_0=Q/2+ r(\mu_1-\mu_2)/2\,, \qquad \alpha_{n+1}=Q/2+r(\mu_3-\mu_4)/2\,, \\
& m_0=\mu_1+\mu_2\,, \qquad m_n=\mu_3+\mu_4\,,
\end{split}
\end{align} 
and $\widehat{V}_\alpha$ is the suitably normalised Liouville exponential (see eq.~\eqref{norm_1}). Since all Liouville/Toda correlation functions going forward will be of the form $\langle\widehat{V}_{\alpha_a}(\infty)\widehat{V}_{r m_0}(1)\widehat{V}_{\alpha_{b}}(0)\ldots\rangle$, we will omit position unless clarity is needed. Expanding the right hand side of \eq{n-point-Toda-AGT} using OPEs, the resulting conformal blocks correspond to the classical and instanton partition functions in \eq{Z4d}, while the one-loop contribution corresponds to the coefficient of the three-point function and structure constants.

As an example, let us consider the  four-punctured sphere in Liouville theory corresponding to the $SU(2)$ quiver diagram in figure~\ref{quiver_4d} (and to \eq{n-point-Toda-AGT} with $n=1$). The correlator can be decomposed in terms of $s$-channel conformal blocks $\mathcal{F}^{(s)}$ as \cite{Belavin:1984vu,Zamolodchikov:1995aa}
\begin{equation}
\braketbis{\widehat V_{\alpha_0 }(\infty) \widehat V_{r m_0}(1) \widehat V_{r m_1}(q) \widehat V_{\alpha_2}(0)}=\int \frac{d \alpha}{2\pi} \, \widehat C(\alpha_0,r m_0,\alpha) \widehat C^{\alpha}_{r m_1,\alpha_2} \left| \mathcal{F}^{(s)}_\alpha(q) \right|^2\,,
\end{equation}
where the integral is taken along the $Q/2+i \mathds{R}$ line, $\widehat C(\cdot,\cdot,\cdot)$ denotes the coefficient of the Liouville three-point function and $ \widehat C^{\cdot}_{\cdot,\cdot}$ the structure constants which correspond to the fusion of two primaries. Both $\widehat C(\cdot,\cdot,\cdot)$ and $ \widehat C^{\cdot}_{\cdot,\cdot}$ are defined in terms of the primary operators normalised as in eqs.~\eqref{norm_1} and \eqref{norm_2}. In ref.~\cite{Alday:2009aq}, the authors showed by explicit computations that the combination $\widehat C(\cdot,\cdot,\cdot) \widehat C^{\cdot}_{\cdot,\cdot}$ provides the one-loop part of the partition function together with the Vandermonde determinant, while the conformal blocks $\mathcal{F}_\alpha$ give the instanton and classical contributions. Finally, the integration over the internal Liouville/Toda momentum $\alpha$ corresponds to the integration over the VEV of the adjoint scalar of the $SU(2)$ gauge group.

When the rank of the gauge group $N>2$ the situation is slightly different. In particular, many types of punctures are possible on  $\mc{C}_{g,n}$, each of which corresponds to a hypermultiplet with some flavour symmetry: see for example ref.~\cite{Chacaltana:2010ks}. We will only consider punctures corresponding to $SU(N)$ flavour nodes, called \textit{full punctures}, and $U(1)$ flavour nodes, called \textit{simple punctures}~\cite{Wyllard:2009hg}. In the Toda theory picture, a full puncture corresponds to the insertion of a Toda primary operator with unconstrained momentum $\alpha$. A simple puncture corresponds to a \textit{semi-degenerate} Toda primary operator, i.e. with  momentum either of the form  $\alpha=\varkappa h_1$ or $\alpha=-\varkappa h_{N}$ with $h_1$ ($-h_{N}$) being the highest weight of the (anti-) fundamental representation of $SU(N)$ and $\varkappa$ a numerical factor. The standard example is 4d $\mathcal{N}=2$ conformal SQCD whose quiver is depicted in figure~\ref{quiver_4d}. The partition function corresponding to this quiver can be expressed as the Toda four-point function on the sphere with two full and two simple punctures.

\subsubsection{Surface operators and Toda degenerate primaries}

In refs.~\cite{Alday:2009fs,Drukker:2009id,Drukker:2010jp} the Toda/gauge theory dictionary was enlarged to describe the addition of 1/2-BPS line and surface operators. It has been shown that $1/2$-BPS surface operators in 4d class $\mathcal{S}$ theories descending from 2d superconformal defect operators in the 6d $\CN=(2,0)$ SCFT correspond to the insertion of one or more \textit{degenerate} Toda primary operators in \eq{n-point-Toda-AGT} \cite{Alday:2009fs,Dimofte:2010tz,Taki:2010bj,Bonelli:2011fq,Bonelli:2011wx,Doroud:2012xw}.\footnote{For surface operators descending from 4d defect operators in 6d, the correspondence is a bit different in that the legs of the 4d defect wrapping the Riemann surface deform the Toda theory on $\mc{C}_{g,n}$ to another 2d CFT, e.g. a WZNW model.  We will not consider such operators in this section.}  As opposed to the punctures giving rise to semi-degenerate primary insertions described above, a generic degenerate Toda primary operator has momentum $\alpha=-\mathfrak{b} \omega_1 -1/\mathfrak{b} \omega_2$ where $\omega_1$ and $\omega_2$ are highest weight vectors of two representations $\mc{R}_1$ and $\mc{R}_2$ of $A_{N-1}$. Note that the parameter $\mf{b}$ is mapped to the squashing parameter of $S^4_{\mf{b}}$ in the AGT correspondence.

In ref.~\cite{Gomis:2014eya} the authors found that inserting degenerate operators of the type $\alpha=-\mathfrak{b} \omega$, i.e. $\mc{R}_2$ is a trivial rep, corresponds to engineering an $\CN=(2,2)$ surface operator whose field content is described by the quiver in figure~\ref{quiver1}.\footnote{Even though in the present work we will focus on degenerate insertions parametrised by only one representation of $A_{N-1}$, we mention that a generic degenerate operator, with $\alpha=-\mathfrak{b} \omega_1 -1/\mathfrak{b} \omega_2$, corresponds to two surface defects supported on two (squashed) $2$-spheres that intersect at the north and south pole of $S^4_{\mathfrak{b}}$. These can be engineered by intersecting M2-branes ending on $N$ M5-branes wrapping the Riemann surface $\mathcal{C}_{g,n}$~\cite{Gomis:2016ljm}. } The representation $\mc{R}$ is described by a Young tableau consisting of $n$ columns with $N_j$ boxes for $1 \le j\le n$, which corresponds to an $n$-node quiver with adjoint chiral multiplets on every node except the $n^{\textrm{th}}$ node. The number of boxes of each column is related to the ranks of the 2d gauge nodes as $N_j\equiv K_j-K_{j-1}$ with $K_0=0$. The surface defect in the 4d theory sits at a marked point $(x,\,\overline{x})$ on $\mc{C}_{g,n}$, which corresponds to a non-trivial FI parameter and theta angle only in the $n^{\textrm{th}}$ node. This kind of surface operator has an M-theory realisation in terms of M2-branes labelled by the representation $\mc{R}$ of $A_{N-1}$ ending on the wrapped M5-branes. Considering multiple degenerate primary insertions with appropriate representations, one could turn on additional FI parameters or, alternatively, add an adjoint chiral on the $n^{\textrm{th}}$ node or remove them from other nodes \cite{Gomis:2014eya}.\footnote{As shown in ref.~\cite{Gomis:2014eya}, the insertion of an arbitrary degenerate operator $\widehat V_{\mathfrak{b}\omega}$ can be recovered from multiple degenerate insertions of the anti-symmetric type with fine-tuned FI parameters and theta angles.  }

In the following, we will focus only on rank-$k$ totally (anti-)symmetric representations. We will denote by $\mc{R}_{N_j,\varepsilon_j}$ the rank-$N_j$ totally symmetric ($\eps_j=+$) or anti-symmetric (${\eps_j=-}$) representation of $A_{N-1}$. The type of representation determines if an adjoint chiral occurs on a given node or not: For $\mc{R}_{N_n,+}$ there is an adjoint chiral on $U(K_n)$, while for $\mc{R}_{N_n,-}$ there is not. Then, if $\varepsilon_{j}=\varepsilon_{j+1}$ there is an adjoint chiral on $U(K_j)$, otherwise not.

Consider coupling $N^2$ free 4d hypermultiplets to a surface operator through a cubic superpotential involving the ambient hypermultiplets and the chiral multiplets in the (anti-) fundamental representation of $U(K_n)$. As discussed in section~\ref{sec:SUSYloc}, the partition function for this 2d-4d system factorises as $Z_{\Sigma \hookrightarrow S^4}= Z_{S^4} Z_{\Sigma}$, and the precise correspondence with the Toda theory reads \cite{Gomis:2014eya}
\begin{align}
\label{Toda/part_func}
Z_{\Sigma \hookrightarrow S^4}^{\{\mc{R}_{N_j,\e_j}\}}= \braketbis{\widehat V_{\alpha_\infty}\widehat V_{\widehat{m}}\widehat V_{\alpha_0}\prod_{j=1}^n \widehat V_{-\mathfrak{b} \omega_{(N_j,\varepsilon_j)}} (x_j,\bar x_j)}\,,
\end{align}   
where $\omega_{(N_j,\eps_j)}$ is the highest weight of the representation $\mc{R}_{N_j,\eps_j}$, and
\begin{align}
\label{masses_1}
\alpha_0 &=Q-\frac{1}{\mathfrak{b}}\sum_{s=1}^{N}i m_s h_s \,, &  \alpha_\infty&=Q-\frac{1}{\mathfrak{b}}\sum_{s=1}^{N} i \tilde m_s h_s\,, \\
\widehat{m} &=(\varkappa+N_n \mathfrak{b})h_1\,, & \varkappa&=\frac{1}{\mathfrak{b}}\sum_{s=1}^{N}(1+i m_s+i \tilde m_s)\,.   \label{masses_2}
\end{align}

Let $r_1=1/\epsilon_1$ be the equatorial radius of $\Sigma = S^2_{\epsilon_1}$, and denote the complexified twisted masses of the fundamental chiral multiplets $m^{\text{\tiny fund}} \equiv m=r_1 \bar{m}+iq/2$ where $\bar{m}$ is the (real) twisted mass and $q$ is its 2d R-charge. Similarly, let $m^{\text{\tiny anti-fund}}\equiv \tilde m$ be the complexified twisted masses of the anti-fundamental chirals. The R-charges are constrained by the superpotentials and are given in eqs.~\eqref{R-charge_adjoint} and~\eqref{R-charge_bifund} and the text above them. Similarly, the superpotential and the global symmetry $SU(N)\times SU(N)\times U(1)$ relate the hypermultiplet complexified twisted masses of the 4d theory to the (anti-)fundamental chiral masses as $M_{st}=i(1-\mathfrak{b}^2)/2\mathfrak{b}-1/\mathfrak{b}\,(m_s+\tilde m_t)$ for $s,t = 1, \ldots, N$ \cite{Gomis:2014eya,Gomis:2016ljm}. Finally, the normalisation of the semi-degenerate and degenerate primaries is given in eq.~\eqref{norm_2}.

\subsubsection{$b$ from Toda correlators}

Before considering specific examples, let us explain the strategy for computing $b$ from the 2d CFT correlators through the AGT correspondence. First of all, as observed above, $b$ is computed by the scaling weight of $Z_{\text{1-loop}}$ and the measure of the integral over the locus. This means that, via the AGT dictionary, we should be able to extract $b$ solely from the scaling behaviour of the three-point function coefficient and structure constants appearing in the Toda correlators and, in principle, the integration measure over internal Toda momenta in the conformal block expansion. In particular, we will not need the explicit form of the conformal blocks since, as summarised above, they correspond to the instanton part of the partition function, which does not contribute to the Weyl anomaly. 

To isolate $b$ from the scaling behaviour of the Toda correlator, we remove the contribution from the 4d ambient theory by dividing the Toda correlator with degenerate insertions by the same correlator without them. Crucially, since the fusion rule eq.~\eqref{deg_fusion} prevents the degenerate insertion from adding more integrals, the normalisation by the correlator without the degenerate operator insertion eliminates the contribution from the integration measure over internal Toda momenta. Thus, the only contributions to $b$ should come from the scaling behaviour of three-point function coefficients and structure constants themselves.

From the gauge theory side we know that the special functions in the structure constants arise from the zeta-function regularisation of some infinite product. In particular, the structure constants can be expressed in terms of $\gamma_1(x|1/\mathfrak{b})$, which is defined in eq.~\eqref{eq:gamma_1}, and $\Upsilon(x)\equiv \Upsilon(x|\mathfrak{b},1/\mathfrak{b})$. Following the same logic as above eq.~\eqref{eq:gamma_r_scaling}, one re-introduces the factors of $r$ by interpreting $\gamma_1(x | 1/\mathfrak{b})$ as $\gamma_1(x/r|1/r)$ where we set $\mathfrak{b}=1$. From eqs.~\eqref{eq:scale_gamma1} and \eqref{eq:scaleUpsilon}, we arrive at the following substitution rules under Weyl re-scaling:
\begin{align}
\label{r_dep_subs}
\gamma_1(x|1/\mathfrak{b}) \rightarrow r^{1-2x} \gamma_1(x|1/\mathfrak{b})\,, \qquad \Upsilon(x) \rightarrow r^{-2 \zeta_2(0;x|\mathfrak{b},1/\mathfrak{b})}\Upsilon( x)\,.
\end{align} 
We stress that this approach can be applied whenever the special functions come from the zeta-function regularisation of infinite products. 

Before moving on to more complicated examples, let us illustrate the method in the simplest non-trivial case: one degenerate insertion of the type $V_{-\mathfrak{b}\omega_{(N_1,\varepsilon)}}$ in the theory of $N^2$ free hypermultiplets, i.e. the correlator eq.~\eqref{Toda/part_func} with $n=1$. In the case of a single degenerate primary insertion, the fusion rule eq.~\eqref{deg_fusion} allows us to write the four-point function in the $s$-channel decomposition as \cite{Zamolodchikov:1995aa,Fateev:2007ab}
\begin{equation}
\begin{split}
\label{Toda_one_ins}
&\braketbis{\widehat V_{\alpha_\infty} \widehat V_{\widehat{m}}\widehat V_{-\mathfrak{b}\omega_{(N_1,\varepsilon_1)}}(x,\bar x)\widehat V_{\alpha_0} }\\
&=\sum_{\{h_{(N_1,\varepsilon_1)}\}} \widehat{C}(\alpha_\infty,\alpha_0-\mathfrak{b} h_{(N_1,\varepsilon_1)},\hat m)\widehat{C}^{\alpha_0-\mathfrak{b}h_{(N_1,\varepsilon_1)}}_{-\mathfrak{b}\omega_{(N_1,\varepsilon_1)},\alpha_0} |\mathcal{F}^{(s)}_{\alpha_0-\mathfrak{b}h_{(N_1,\varepsilon_1)}}|^2\,,
\end{split}
\end{equation}
where $|\mathcal{F}|^2 = \mc{F}(x)\mc{F}(\bar{x})$ denotes the conformal blocks, the sum runs over the weights $\{h_{(N_1,\varepsilon_1)}\}$ of the representation $\mc{R}_{N_1,\varepsilon}$, and $\widehat C(\boldsymbol{\cdot},\boldsymbol{\cdot},\boldsymbol{\cdot})$ and $\widehat C^{\boldsymbol{\cdot}}_{\boldsymbol{\cdot},\boldsymbol{\cdot}}$ are the appropriately normalised coefficient of the Toda three-point function eq.~\eqref{three-point_funct_bare} and the structure constants, respectively. Now, we need the structure constant part of the Toda correlator without any degenerate insertion. This is given by the three-point function~\cite{Gomis:2014eya}
\begin{align}
\label{three_point_trinion}
\begin{split}
\widehat{C}(\alpha_\infty, \alpha_0, \varkappa h_1)& = \frac{1}{\prod_{s,t=1}^{N} \Upsilon \left( \frac{\varkappa}{N}+\left(Q-\alpha_\infty,h_s \right)+ \left( Q-\alpha_0,h_t \right) \right)} \,,
\end{split}
\end{align}
which can easily be found by using eq.~\eqref{three-point_funct_bare} together with the normalisations eqs.~\eqref{norm_1} and \eqref{norm_2}. Notice that this corresponds to the partition function of the $N^2$ free hypermultiplets with the masses $M_{st}$ defined in the text below eq.~\eqref{masses_2}.    
Thus, the quantity to study is
\begin{align}
\label{eq:const_struc_part}
\frac{\widehat{C}(\alpha_\infty,\alpha_0-\mf{b} h_{(N_1,\varepsilon_1)},\widehat m)\widehat{C}^{\alpha_0-\mf{b}h_{(N_1,\varepsilon_1)}}_{-\mf{b}\omega_{(N_1,\varepsilon_1)},\alpha_0}}{\widehat{C}(\alpha_\infty, \alpha_0, \varkappa h_1)}\equiv \mf{C}_{N_1,\eps_1}(h_{(N_1,\varepsilon_1)})\,,
\end{align}
which in this case reduces to a combination of $\gamma$-functions. Note that the function $\mf{C}_{N_1,\eps_1}$ depends on the particular weight $h_{(N_1,\varepsilon_1)}$. However, its weight under Weyl re-scaling is independent of it. This is important because it ensures that the correlator \eqref{Toda_one_ins} has a well-defined scaling behaviour from which the charge $b$ can be extracted.  

\subsubsection*{Examples of Defects Coupled to Free Massless Hypermultiplets}

We will now compute $b$ for defects coupled to 4d free massless hypers, reproducing our SUSY localisation results from section~\ref{sec:fmh} and illustrating what those calculations look like in the corresponding Toda theory.

\paragraph{Example 1: $\mathcal{N}=(2,2)$ SQCD.} Let us consider the case of a single rank-$N_1$ anti-symmetric insertion corresponding to 2d $\mathcal{N}=(2,2)$ SQCD with gauge group $U(K_1)$ (${K_1=N_1}$). In  \eq{SQCD}, $b$ was found in the limit where the chiral multiplets are massless by using the formula \eq{c2d}.  Here, we are interested in finding $b$ by considering the structure constant part of the Toda correlator in eq.~\eqref{Toda_one_ins}, which corresponds to the quantity defined in \eq{const_struc_part}. 

$\mf{C}_{N_1,\eps_1}$ in this case is easily constructed from eqs.~\eqref{const_struc_anti} and \eqref{three-point_funct_bare} and by employing the property eq.~\eqref{eq:Upsilon_prop} to rewrite the $\Upsilon$-functions in terms of $\gamma_1$-functions. Using the explicit values of $\alpha_0$ and $\alpha_\infty$ given in eqs.~\eqref{masses_1} and \eqref{masses_2}, \eq{const_struc_part} takes the form 
\begin{align}
\begin{split}
\mf{C}_{N_1,-}\Big(h_{(N_1,-),{\{p\}}}\Big) = A_1 \prod_{t \in \{p\}} \left[ \frac{\prod^{N}_{s \notin \{p\}} \gamma_1 \left(\left. i m_t - i m_s \right| \frac{1}{\mathfrak{b}}\right)}{\prod^{N}_{s=1}\gamma_1 \left( \left. 1 +i \tilde{m}_s + i m_t \right| \frac{1}{\mathfrak{b}}\right)} \right]\,,  
\end{split}
\end{align}
where $\{p\}$ labels the weights of the representation as explained in appendix~\ref{appendix_Toda}. The constant $A_1$ contains factors of $\mathfrak{b}$ coming from the form of the structure constants eq.~\eqref{const_struc_anti}. We observe that those factors, being independent of the Toda momenta, are unaffected by the Weyl re-scaling, so they can be ignored.  

By performing the replacement eq.~\eqref{r_dep_subs} first, then setting the (anti-)fundamental chiral twisted masses to zero and leaving the R-charges generic, we find an overall prefactor expressible as a power of $r$. The exponent is given by
\begin{align}
\label{b_SQCD_AGT}
\begin{split}
\frac{b}{3}&=\sum_{t\in \{p\}} \left\{ \sum_{s \notin \{ p\}}^{N} \left[1-(q_s-q_t)\right] -\sum_{s=1}^{N} \left[ 1-2+(\tilde q_s + q_t) \right]  \right\}\\ 
&=K_1(2N-K_1)-K_1 \sum_{s=1}^{N}(q_s+\tilde q_s) \,.
\end{split}
\end{align}
Recall from above eq.~\eqref{R-charge_adjoint} that $q_s+\tilde q_s=0$ ($\tilde{q} \equiv q^\text{\tiny{anti-fund}}$) for $\mathfrak{b}=1$. Thus, the last term vanishes, and one finds agreement with \eq{SQCD}.

\paragraph{Example 2: $\mathcal{N}=(2,2)$ SQCDA.}
Now consider a single symmetric insertion that corresponds to 2d ${\mathcal{N}=(2,2)}$ SQCD with gauge group $U(K_1)$ ($K_1=N_1$) and an adjoint chiral on $U(K_1)$.
The structure constant part of the Toda correlator is~\cite{Gomis:2014eya}
\begin{align}
\begin{split}
\mf{C}_{N_1,+}\Big(h_{(N_1,+),[n]}\Big) =   A_1 \prod_{s,t=1}^{N} \prod_{\nu=0}^{n_t-1} \left[ \frac{\gamma_1\left( \left. i m_t + \nu \mf{b} - i m_s-n_s \mf{b} \right| \frac{1}{\mathfrak{b}}\right)}{\gamma_1 \left( \left. 1+ i \tilde m_s+i m_t + \nu \mf{b} \right| \frac{1}{\mathfrak{b}}\right)} \right] \, ,
\end{split}
\end{align}
where $n_s$ are non-negative integers that label the weights $h_{(N_1,+),[n]}$ as explained in appendix~\ref{appendix_Toda}. Also in this case, the constant $A_1$ contains factors independent of the momenta, and it does not play any role in the scaling behaviour of the structure constants. Thus, by applying again the substitution eq.~\eqref{r_dep_subs} and setting the twisted masses to zero, we find that the exponent of the overall factor of $r$ in the structure constants is
\begin{align}
\label{b_charge_anti}
\frac{b}{3} &= \sum_{s,t=1}^{N} \sum_{\nu=0}^{n_t-1} \left[ 2 + 2 n_s-(q_s + \tilde q_s) \right]=2 K_1 (N+K_1)\,,
\end{align}
which agrees with $b$ given in~\eq{SQCDA}. 

\paragraph{Example 3: Multiple degenerate insertions.} Finally, we consider the insertion of $n$ degenerate primary fields in totally symmetric or totally anti-symmetric representations. In this case, we can use the fusion rule eq.~\eqref{deg_fusion} to determine the allowed momenta which run between the degenerate insertions. The non-trivial part in writing the $(n+3)$-point function in terms of three-point functions is finding the conformal blocks. This has been discussed in ref.~\cite{Gomis:2014eya} by employing the correspondence with the four-dimensional gauge theory and the localisation results. However, to obtain $b$ it suffices to consider the following quantity
\begin{align}
\label{total_const_struc}
\frac{\widehat{C}\left(\alpha_\infty, \widehat{m}, \alpha_0 - \mathfrak{b} h_{[n^n]}\right)}{\widehat{C} (\alpha_\infty, \varkappa h_1,\alpha_0)} \prod_{j=1}^n\widehat{C}^{\alpha_0 -\mathfrak{b} h_{[n^j]}}_{-\mathfrak{b} \omega_{(N_j,\varepsilon_j)},\alpha_0- \mathfrak{b} h_{[n^{j-1}]}} \, .
\end{align}
In the equation above, $h_{[n^j]}=\sum_{t=1}^{N}n^j_t h_t$ for integers $n_t^j \ge 0$ such that $h_{[n^j]}-h_{[n^{j-1}]}$ is a weight of the representation $\mc{R}_{N_j,\varepsilon_j}$. For $\mathcal{R}_{N_j,+}$ (symmetric), $n_t^j-n_t^{j-1}\in \mathds{Z}_{>0}$, while for $\mathcal{R}_{N_j,-}$ (anti-symmetric), $n_t^j-n_t^{j-1}\in \{0,1\}$.  In both cases, we have $\sum_{t=1}^{N} n_t^j - n_t^{j-1}=N_j$, from which it follows $\sum_{t=1}^{N}n^j_t=\sum_{i=1}^{j}N_i$. 

The form of the first contribution does not depend on the type of representation (symmetric or anti-symmetric), and it can be found from eq.~\eqref{three_point_trinion}:
\begin{align}
\label{contr_1}
\begin{split}
\hspace{-0.2cm}\frac{\widehat{C}\left(\alpha_\infty, \widehat{m}, \alpha_0 - \mathfrak{b} h_{[n^n]}\right)}{\widehat{C} (\alpha_\infty, \varkappa h_1,\alpha_0)} 
& =\prod_{t,s=1}^{N} \prod_{\nu=0}^{n_t^n-1} \frac{1}{\gamma_1\left(\left. \frac{\varkappa}{N}+\left( Q-\alpha_\infty,h_s \right)+ \left(Q-\alpha_0,h_t \right) +\nu \mathfrak{b}  \right|\frac{1}{\mathfrak{b}} \right)}\,,
\end{split}
\end{align}
where we employed the shift property \eq{Upsilon_prop}.

Let us start with the symmetric case where we need the structure constants in eq.~\eqref{const_struc_symm}. It is convenient to compute the scaling weight of the structure constants (denoted in the following with square brackets) involving only a single symmetric degenerate insertion. By setting $\mathfrak{b}=1$ and the (anti-)fundamental chiral masses to zero, we obtain
\begin{equation}
\label{symmetric_contribution}
\begin{split}
&\left[\widehat{C}^{\alpha_0 -\mathfrak{b} h_{[n^j]}}_{-\mathfrak{b} \omega_{(N_j,+)},\alpha_0- \mathfrak{b} h_{[n^{j-1}]}} \right]  = -  \sum_{s,t=1}^{N}\sum_{\nu=0}^{n_t^j-n_t^{j-1}-1} \left[ 2(\nu-(n_s^{j}-n_t^{j-1}))-1 + (q_s-q_t) \right] \\
& =  -  \sum_{s,t=1}^{N} \left[ (n_t^j)^2-(n_t^{j-1})^2 \right] + 2 \sum_{i=1}^j N_j N_i +2 N  N_j + N \sum_{t=1}^N q_t (n_t^j-n_t^{j-1})-N_j \sum_{t=1}^N q_t\,.
\end{split}
\end{equation}
Then, by adding  the contribution of eq.~\eqref{contr_1} to eq.~\eqref{symmetric_contribution} we get
\begin{align}
\label{b_multi_ins_symm}
\begin{split}
\frac{b_{\text{sym}}}{3} &= 2\sum_{j=1}^n \sum_{i=1}^j N_j N_i +2 N \sum_{j=1}^n N_j - \sum_{s=1}^N \left(\tilde q_s +q_s \right)\sum_{j=1}^n N_j\\
& = 2\sum_{j=1}^n (K_j-K_{j-1})K_j+2 K_N\,N,
\end{split}
\end{align}
where in the second step we used that $N_i=K_i-K_{i-1}$ and $\sum_{i=1}^j N_i=K_j$. Again, this result reproduces \eq{b_n_adj_chirals} found by employing the formula \eq{c2d}.

Before we move on to study a generic combination of symmetric and anti-symmetric degenerate insertions let us first obtain the contribution due to a single anti-symmetric insertion. The structure constants for this case are given by eq.~\eqref{const_struc_anti}. By using repetitively the fact that $n_t^{j}-n_t^{j-1} \in \{0,1\}$ (since the j$^{\rm th}$ rep is anti-symmetric), we obtain the corresponding contribution to the central charge $b$, namely 
\begin{align}
\begin{split}
\left[ \widehat{C}^{\alpha_0 -\mathfrak{b} h_{[n^j]}}_{-\mathfrak{b} \omega_{(N_j,-)},\alpha_0- \mathfrak{b} h_{[n^{j-1}]}}  \right] 
=&\;(2N-N_j)N_j + 2\sum_{i=1}^{j-1}N_j N_i - N\sum_{t=1}^{N}  \left[ (n_t^j)^2-(n_t^{j-1})^2 \right] \\
& +N \sum_{t=1}^N q_t (n_t^j-n_t^{j-1})-N_j \sum_{t=1}^N q_t \,.
\end{split}
\end{align}

By comparing the above with eq.~\eqref{symmetric_contribution}, we find that the difference between the contributions of a single symmetric and anti-symmetric insertion is simply given by
\begin{align}
\label{sym_minus_anti}
\text{sym}-\text{anti-sym}=3 N_j^2= 3 \left( K_j-K_{j-1} \right)^2\,.
\end{align}
With this result, we can compute $b$ for free massless hypermultiplets coupled to the GLSM of figure~\ref{quiver1} with any combination of $\varepsilon=\pm$ (and thus any number of adjoint chirals). In particular, we can compute $b$ when all the insertions are anti-symmetric. We find
\begin{align}
\begin{split}
\frac{b_\text{anti-sym}}{3} & = \frac{b_{\text{sym}}}{3} - 3{\sum_{j=1}^n} \left( K_j-K_{j-1} \right)^2 \\
& =  4{\sum_{j}^n} K_{j-1} K_j +2N K_n  -\sum_{j=1}^n K_j^2 -3 \sum_{j=1}^n K_{j-1}^2 \,,
\end{split}
\end{align}
which reduces to eq.~\eqref{b_SQCD_AGT} for $n=1$, and agrees with \eq{b_n-1_adj_chirals}.

\section{SUSY Partition Function on $S^1\times S^{d-1}$}
\label{sec:Casimirs}

In an SCFT with at least a single $U(1)_R$ symmetry, the twisted partition function $Z$ of the theory on $\CM \equiv S^1_R\times S^{d-1}$ is identified with the superconformal index~\cite{Kinney:2005ej}, up to a normalisation factor (that will be important in what follows!). Here $S^1_R$ is a temporal circle of radius $R$. The superconformal index is, like other indices, functionally a count of a certain set of degrees of freedom obeying specific shortening conditions. After affecting the sum over multiplets obeying the shortening condition, the remaining non-trivial contributions to the superconformal index come from multiplets that are protected under RG flow. Further, the superconformal index is itself invariant under continuous deformations that preserve the supercharge used to define the index. Thus, $Z$ is a likely candidate to capture putative defect central charges. Indeed, as we will demonstrate below in certain cases, an overall prefactor in $Z$ depends explicitly on defect Weyl anomaly coefficients.  

Specifically, in this section we will first argue for the appearance of the central charge $d_2$ in $Z$ in the presence of a 2d superconformal defect wrapping $\Sigma = S^1_R \times S^{1}$. After setting the general framework in section~\ref{sec:Casimirs-Anomalies}, we will examine two models where $d_2$ has been calculated holographically~\cite{Jensen:2018rxu}: Levi type-$\mathds{L}$ defects in $\CN=4$ $U(N)$ SYM theory (section~\ref{sec:Casimirs-2d}) and the Wilson surface operator in the $A_{N-1}$ 6d $\CN=(2,0)$ SCFT (section~\ref{sec:Casimirs-6d}). We will see that upon deformation of $Z$ by these specific 2d defects the exponent in the normalisation, i.e. the SUSY Casimir energy (SCE)\footnote{The first place to identify a similar quantity to the SCE was ref.~\cite{Kim:2012ava} referring to it as an ``index Casimir energy'', but we will refer to this quantity by the more frequently used term ``SUSY Casimir energy''.} \cite{Assel:2014paa, Assel:2015nca,Bobev:2015kza}, changes by a factor proportional to $d_2$.

Actually, in the 6d $\CN=(2,0)$ SCFT we obtain a more general result: we compute the change in the SCE due to a \textit{pair} of Wilson surfaces wrapping $\Sigma_1 = S^1_R \times S^1_1$ and $\Sigma_2 = S^1_R \times S^1_2$, where $S^1_1$ and $S^1_2$ intersect only at the poles of the $S^5$. Taking the limit of a single Wilson surface on, say, $\Sigma_1$, we will find that the change in the SCE reduces to a term $\propto d_2$. Interestingly, for two intersecting Wilson surfaces the change in the SCE is not merely the sum of results for two lone Wilson surfaces, but also contains contributions purely from their intersections at the poles. A detailed interpretation of this result would require knowledge of the algebra of intersecting surface operators and the degrees of freedom intrinsic to the 1d submanifold they share, which is beyond the scope of this paper.

In the final part of this section, we will propose the form of a type B anomaly coefficient (or possibly a linear combination thereof) of a 1/2-BPS codimension two defect wrapping $S^1_R\times S^{3}$ in the $A_{N-1}$ 6d $\CN=(2,0)$ SCFT, our~\eq{6d-Levi-surface-character} below.  Despite the lack of an explicit form of the defect Weyl anomaly for a 4d defect in a 6d CFT, the logic of the construction that computed $d_2$ in the two examples considered below is straightforward to extend to superconformal defects of arbitrary codimension. We thus claim that~\eq{6d-Levi-surface-character} must be proportional to type B defect central charges.

\subsection{Anomalies and SUSY Casimir Energy}
\label{sec:Casimirs-Anomalies}

In this subsection we are concerned with setting up a general framework for arguing that the change of the SCE due to a defect on $\Sigma$ is proportional to $d_2$. To begin, consider the twisted partition function of a SCFT, $Z(R,\,\mu_j)$, on $\CM=S^1_R \times S^{d-1}$ for even $d$, where $\mu_j$ are chemical potentials for superconformal Cartan generators that commute with the supercharge used to define the index.  The main argument in refs.~\cite{Assel:2014paa, DiPietro:2014bca, Bobev:2015kza} is that by utilising SUSY localisation to compute $Z(R,\,\mu_j)$, one finds a general form proportional to the superconformal index $\mathcal{I}$:
\begin{align}\label{eq:Z-index-SCE}
Z(R,\,\mu_j) = e^{-R\,E_c(\mu_j)} \mc{I}(R\mu_j)\,,
\end{align}
where $E_c$ is the SCE. The understanding here is that, as an object counting protected operators starting from the identity operator, $\mathcal{I}$ is an ascending polynomial in non-negative powers of fugacities, $q_j$, start at one, i.e. $\mathcal{I} = 1 + q_j^{\#}+\ldots$, i.e. $\#>0$.  

In the presence of a defect preserving the supercharge used to define the index, $\mathcal{I}$ will generically pick up negative powers in an expansion in $q$, which will need to be compensated in order to maintain the normalisation that the index begins counting with the identity operator \cite{Gaiotto:2012xa, Cordova:2017mhb}.  That is, the superconformal index in the presence of a surface defect is still counting states, in a similar sense as in the ambient theory, but now including defect states in radial quantisation around the defect.

A form of the SCE as an integrated anomaly has been conjectured, but to our knowledge not rigorously proven. Here we will briefly present the existing pieces of evidence for this conjecture, which we will then use to motivate the appearance of defect type B anomaly coefficients in the SCE. We will subsequently show in the examples of sections~\ref{sec:Casimirs-2d} and~\ref{sec:Casimirs-6d} that $d_2$ indeed appears in the SCE, providing compelling evidence for our arguments.

In ref.~\cite{Assel:2014paa}, for 4d SCFTs on $S^1_R \times S^3$, $E_c$ was computed  by SUSY localisation to be proportional to the Weyl anomaly coefficients $a$ and $c$ as
\begin{align}\label{eq:Assel-Casimir-energy}
E_c(\mu_j) \equiv -\lim_{R\to \infty} \partial_R \log Z(R,\,\mu_j) = \frac{4\pi}{3}(|\mu_1|+|\mu_2|)\left((a-c) +\frac{(|\mu_1|+|\mu_2|)^2}{|\mu_1||\mu_2|}(3c-2a)\right)\,,
\end{align}
where $\mu_{1,2}$ are chemical potentials for $SO(2)_{1,2}$ rotations preserved in squashing the $S^3$.

This connection between $E_c$ and Weyl anomaly coefficients was refined in ref.~\cite{Assel:2015nca}. The authors of ref.~\cite{Assel:2015nca} draw a direct relationship between the Weyl anomaly and the SCE explicitly by reducing $Z(R,\,\mu_i)$ on the squashed three-sphere $S^3_{\e_1, \e_2}$, with squashing parameters $\e_1$ and $\e_2$, to SUSY Quantum Mechanics (SQM) on $S^1_R$.  The expectation value of the 1d theory's Hamiltonian, $\langle H_{\rm SQM}\rangle$, in the limit $R\to\infty$ was identified with the SCE in a manifestly scheme independent way, unlike directly computing $\int_{S^{d-1}}\sqrt{g}\langle T_{\tau\tau}\rangle$, which is scheme dependent. In the end, $\langle H_{SQM}\rangle$ was found to be given by~\eq{Assel-Casimir-energy} where the role of the chemical potentials $\mu_{1,2}$ is played explicitly by the squashing parameters $\e_{1,2}$.

In ref.~\cite{Closset:2019ucb} the above arguments of ref.~\cite{Assel:2014paa, DiPietro:2014bca} were extended to more general backgrounds, of the form $\CM = S_R^1 \times \mc{M}_{d-1}$, where for example in $d=4$ $\mc{M}_{d-1}$ is a circle bundle over a Riemann surface. In $d=4$ examples, it was demonstrated that in terms of $U(1)_f$ ``flavour parameters'' collectively referred to as $\nu$ and the ``geometric parameter'' $\hat{\tau}$, which is related to complex structure moduli, the twisted partition function has a Casimir contribution of the form
\begin{align}\label{eq:Closset-Casimir-energy}
E_c(\nu,\tau) = \frac{1}{6\hat{\tau}^3} \mc{A}^{abc}\nu_a\nu_b\nu_c - \frac{1}{12\hat{\tau}}\mc{A}^a\nu_a\,,
\end{align}
where $\mc{A}^{abc}$ and $\mc{A}^a$ are cubic $U(1)_f$ and mixed gravitational anomalies respectively. 

The authors of ref.~\cite{Bobev:2015kza} made a more general conjecture, that the SCE in a SCFT is given by the equivariant integration of the anomaly polynomial,  $\mc{A}_{d+2}(\CM)$,
\begin{align}
E_c = \int \mc{A}_{d+2}(\CM)\,,\label{eq:Bobev-Casimir-energy}
\end{align}
which, if true, obviously means $E_c$ can depend on Weyl anomaly central charges. However, we reiterate that eq.~\eqref{eq:Bobev-Casimir-energy} remains a conjecture, albeit one strongly supported by evidence from a number of examples in various dimensions~\cite{Bobev:2015kza}.  

Now, we would like to outline how we conjecture a 2d surface defect wrapping $\Sigma\hookrightarrow\CM$ modifies $E_c$. One line of reasoning starts from \eq{Bobev-Casimir-energy}, and requires that we make two assumptions from the start: (i) the deformed anomaly polynomial factorises into ambient and defect localised contributions
\begin{align}
\mc{A}_{d+2}(\Sigma\hookrightarrow\CM) \rightarrow \mc{A}_{d+2}(\CM) + \delta_{\Sigma}\mc{A}_4(\Sigma)\,,
\end{align}
(ii) there is a sufficient amount of superconformal symmetry preserved by the defect such that the defect Weyl anomaly sits in a multiplet with other global defect localised anomalies, e.g. defect chiral anomalies. In addition to finding a general proof of $E_c$ being given by $\int \mc{A}_{d+2}(\CM)$, proving the validity of these assumptions is the focus of on-going work.

If both assumptions (i) and (ii) hold, then the result of the equivariant integration of $\mc{A}_4(\Sigma)$ is related to the integrated defect Weyl anomaly. That is, the anomaly coefficients that can appear in $\int \mc{A}_4(\Sigma)$ are controlled by coefficients appearing in the non-vanishing contributions to the integrated defect Weyl anomaly.

From the form of the defect Weyl anomaly reviewed in section~\ref{sec:intro}, it is immediately clear in \eq{defecttrace} that the type A term will not contribute to the integrated anomaly: the Euler character of $\Sigma = S^1_R \times S^1$, and its squashings, vanishes. However, the integrated type B contributions coming from $\mathring{\II}{}^2$ and $W_{ab}{}^{ab}$ do not necessarily vanish on a squashed sphere. Moreover, for our 2d superconformal defects $d_1 =d_2$ has been proven in $d=4$ and conjectured in other $d>4$~\cite{Bianchi:2019sxz}. Thus, if assumptions (i) and (ii) hold, then the change in $E_c$ due to the presence of a superconformal defect wrapping $\Sigma$ must be proportional to $d_2$ when $d=4$, and, supported by evidence in the following subsections, we conjecture that it is proportional to $d_2$ in other $d$ as well.

It should be mentioned that there could be another, possibly more direct, way to show that the defect induced change in $E_c$ is related to $d_2$ following the logic in ref.~\cite{Assel:2015nca}. Since $S^1_R\subset \Sigma$, the reduction of the defect on $S^{d-1}$ will change the SUSY Hamiltonian $H_{SQM}$ on $S^1_R$ by additional chiral and Fermi multiplets, as well as possible superpotential deformations due to the reduction of couplings between defect and ambient degrees of freedom. In the cases where there is an explicit defect action, the computation would amount to the regulated counting done in ref.~\cite{Assel:2015nca}.  However, this would not in general be a constructive proof of the connection between defect central charges and the SCE, but if demonstrated to hold in a number of examples, could provide a useful computational tool to try to predict $d_2$ in novel models.

Finally, while not directly related to anomalies, a different line of reasoning also suggests the appearance of $d_2$ in the SCE.  From the point of view of constructing VOAs from 4d SCFTs~\cite{Beem:2013sza,Beem:2014rza} and 6d SCFTs~\cite{Beem:2014kka}, the Schur limit of the SUSY partition function of an $\CN \geq 2$ SCFT on $S^1 \times S^{3}$ or $\CN\geq (1,0)$ SCFT on $S^1\times S^5$ is the character of the vacuum module of the VOA, see e.g. ref.~\cite{Beem:2017ooy}.  As shown in ref.~\cite{Cordova:2017mhb}, the SUSY partition function in the presence of a superconformal surface defect inserted normal to the VOA plane instead computes in the Schur limit the character of some non-vacuum module. Crucially, the dimension of the defect identity in the module is given by $-d_2$~\cite{Bianchi:2019sxz}. This is precisely the statement that introducing the defect shifts $E_c$ by a term $\propto d_2$. This VOA perspective will be especially useful in our 6d computation below.

\subsection{4d SUSY Casimir Energy}
\label{sec:Casimirs-2d}

In this subsection we consider the SUSY partition function of $\N=4$ $SU(N)$ SYM theory on $\CM=S^1_R \times S^3$ with a Levi type-$\mathds{L}$ defect along $\Sigma=S^1_R \times S^1$. Crucially, for the reasons mentioned in section~\ref{sec:review}, we need to set $\beta_i=\gamma_i=0$ in eq.~\eqref{eq:4d-Levi-surface-d2}, so we can only study cases where $b = d_2$. Nevertheless, from the arguments above and our evidence in 6d in section~\ref{sec:Casimirs-6d}, we believe the SCE obtained from the SUSY partition function is proportional to $d_2$ alone.

To compute the twisted partition function, $Z$, on $\CM=S^1_R \times S^3$, we will use the correspondence between its Schur limit and correlators in 2d $q$-deformed YM theory (qYM) on a genus-$g$ Riemann surface with $n$-punctures in the zero-area limit, $\mc{C}_{g,n}$~\cite{Gadde:2011ik}. Ref.~\cite{Alday:2013kda} showed that the deformation of the 4d twisted partition function by surface defects corresponds in the qYM theory to local operator insertions $\mc{O}_{\mc{R}_i}$, where $\mc{R}_i$ is a label descending from the representation data of the $i^{\rm{th}}$ defect. In the description of qYM as Chern-Simons theory on $S^1 \times \mc{C}_{g,n}$, these local operators arise from Wilson lines in representations $\mc{R}_i$ extended along the $S^1$, and hence localised at a point on $\mc{C}_{g,n}$.

While the authors of ref.~\cite{Alday:2013kda} consider an arbitrary number of defect insertions, for us it will suffice to consider the 2d qYM one-point function corresponding to a single surface operator in representation $\mc{R}$ of the 4d gauge symmetry. This takes the form
\begin{align}\label{eq:2dqYM-defect-correlator}
\langle \mc{O}_{\mc{R}}\rangle_{g,n}  = \sum_{\mc{S}}\mathds{S}^{2-2g-n}_{\mc{S},0}\frac{\mathds{S}_{\mc{S},\mc{R}}}{\mathds{S}_{\mc{S},0}}\prod_{i=1}^n \chi_{\mc{S}}(\vec{a}_{(i)})  \,,
\end{align} 
where the sum is over partitions of $N$ schematically of the form $\mc{S} = [s_1,s_2,\ldots,s_{N-1},0]$ and ``$0$'' labels the trivial representation.  Each $\vec{a}_{(i)}$ for $i=1, \ldots, n$ is the holonomy around one of the $n$ punctures, and each $\chi_\mc{R}(\vec{a})$ is the Schur polynomial for a partition $[\ell_1,\ldots,\ell_{N-1},0]$ defining the representation $\mc{R}$.  

Explicitly, the Schur polynomial is computed as a ratio of determinants 
\begin{align}
\chi_{\mc{R}}(\vec{a}) = \frac{\det A(\vec{a},\,\ell)}{\det A(\vec{a},\,0)}~,
\end{align}
where the matrix $A$ has components $A_{ij}(\vec{a},\,\ell) = a_{j}^{\ell_i +N -i}$. The modular-S matrix, $\mds{S}$, appearing in \eq{2dqYM-defect-correlator} is defined by
\begin{align}\label{eq:modular-S}
\mds{S}_{\mc{R},\mc{R}^\prime}  = \mds{S}_{0,0} \chi_{\bar{\mc{R}}}(q^{\rho+\kappa})\chi_{\mc{R}^\prime}(q^\rho)~.
\end{align}
Here, we are using the form of the Weyl vector $\rho = (\rho_1,\ldots,\,\rho_N)$  in a particular orthogonal basis
\begin{align}
\rho &= \frac{1}{2} \left( N-1,\,N-3,\,\ldots, 1-N\right)~.
\end{align}
The original partition data $\ell = [\ell_1,\ldots,\ell_{N-1},0]$ re-expressed in the orthogonal basis is denoted $\kappa = [\kappa_1,\ldots,\kappa_{N}]$, where
\begin{align}
\kappa_i = \ell_i -\frac{1}{N}\sum_{j=1}^{N-1}j(\ell_j-\ell_{j+1})~.
\end{align}
The notation adopted in the arguments of the Schur polynomials in \eq{modular-S} is then to be interpreted as, e.g. $q^{\rho+\kappa} = (q^{\rho_1+\kappa_1},\ldots, q^{\rho_N+\kappa_N})$.

The case that we are interested in is a single $\CN=(4,4)$ Levi type-$\mds{L}$ defect labelled by a representation $\mc{R}$, specified by a partition $\ell$ of $N$, inserted in 4d $\CN=4$ $SU(N)$ SYM theory.  This corresponds to computing the one-point function of $\mc{O}_{\mc{R}}$ on a torus with no punctures ($g=1$, $n=0$), in which case~\eq{2dqYM-defect-correlator} completely collapses to 
\begin{align}\label{eq:2d-qYM-torus-correlator}
\langle \CO_{\mc{R}}\rangle_{1,0} = \chi_{\mc{R}}(q^{\rho})~.
\end{align}
From~\eq{2d-qYM-torus-correlator}, we can read off the defect SCE from the overall power of $q$ that needs to be stripped off in order to match the ``start-at-one'' normalisation of the index. Explicitly, the Schur polynomial $\chi_{\mc{R}}(q^\rho)$ can be expanded in $q$ as an overall prefactor multiplying an ascending polynomial in non-negative powers of $q$,
\begin{align}
\chi_{\mc{R}} = q^{-\sum_i \rho_i \ell_i} \left(1+q^{\#}+\ldots\right)~.
\end{align}
Defining the transpose (or conjugate) partition $\tilde{\ell} = [\tilde{\ell}_1,\ldots,\,\tilde{\ell}_N]$,\footnote{Formally, one starts with a partition $\ell=[\ell_1,\ldots,\,\ell_N]$ and constructs the transpose partition as
	\begin{align}
	\tilde{\ell}_i = \#\{j\Big|\ell_j \geq i\}.
	\end{align} 
	Put more plainly, $\tilde{\ell}_i$ is given by the number of entries in $\ell$ that are greater than or equal to $i$. The Young tableau of $\tilde{\ell}$ is obtained from the tableau of $\ell$ by exchanging columns and rows. Thus, if $\ell$ is a partition of $N$, then so is $\tilde{\ell}$.} 
we can easily show for Levi type-$\mds{L}$ defects that
\begin{align}
\sum_{i=1}^N \rho_i \ell_i = \frac{1}{2}\sum_{i=1}^N(N+1-2i)\ell_i =\frac{1}{2}\left(N^2 - \sum_{i=1}^N\tilde{\ell}_i^2\right)\,.
\end{align}
Comparing to the holographic result for $d_2$ in eq.~\eqref{eq:4d-Levi-surface-d2}, with $\beta_i=\gamma_i=0$, we thus identify the Levi type-$\mds{L}$ defect contribution to the SCE encoded in the 2d qYM one-point function:
\begin{subequations}
	\begin{align}\label{eq:2d-qYM-SCE}
	\langle \CO_{\mc{R}}\rangle_{1,0} = q^{-d_2/6}\left(1 + q^\#+\ldots\right),
	\end{align}
	\beq
	d_2 = 3 \left(N^2 - \sum_{i=1}^N\tilde{\ell}_i^2\right).
	\eeq
\end{subequations}

Crucially, this calculation involved no approximations, relying only on SUSY and the equivalence of the twisted partition function with qYM correlators. This calculation thus strongly suggests that the holographic results in eq.~\eqref{eq:4d-Levi-surface-d2} are in fact exact, and not merely large-$N$ and/or strong-coupling limits.

To repeat once again, in these cases where $\beta_i = \gamma_i=0$ the Levi-type defect in $\N=4$ SYM has $b = d_2$, so that our identification of $d_2$ alone in eq.~\eqref{eq:2d-qYM-SCE} is ambiguous. In other words, how do we know we obtain $d_2$ alone, rather than $b$  alone, or a linear combination of $b$ and $d_2$? If our arguments in section~\ref{sec:Casimirs-Anomalies} for the connection between the integrated defect Weyl anomaly and the defect contribution to the SCE hold, then the exponent in~\eq{2d-qYM-SCE} is $d_2$ and not $b$. Moreover, in the following subsection we will study Wilson surface defect indices in 6d, and in that case $b$ and $d_2$ in eq.~\eqref{eq:6d-Wilson-surface-b} are distinct, allowing us to identify $d_2$ unambiguously, which will provide compelling evidence for our arguments more generally.

\subsection{6d SUSY Casimir Energy}
\label{sec:Casimirs-6d}

In this subsection, we are concerned with the twisted partition function of the 6d $\CN=(2,0)$ $A_{N-1}$ SCFT on the squashed $S^1_R\times S^5$ in the presence of 2d or 4d superconformal defects.\footnote{The metric on the squashed $S^1_R\times S^5$ can be found in, e.g. appendix B in ref.~\cite{Bullimore:2014upa}.  Our calculations, however, will not require specific details about the ambient geometry.} In M-theory this SCFT arises as the low-energy theory on the worldvolume of $N$ coincident M5-branes, and we are interested in the defects arising from either M2- or M5-branes that end on this initial stack of M5-branes. The M2-branes give rise to a 2d defect (codimension four), namely a Wilson surface operator, which we place along $\Sigma = S^1_R \times S^1$. The M5-branes give rise to a 4d defect (codimension two), which we place along $S^1_R \times S^3$. Ref.~\cite{Bullimore:2014upa} carried out a systematic study of the twisted partition function of this 6d SCFT with both types of defects. Using the results of ref.~\cite{Bullimore:2014upa} and our arguments from section~\ref{sec:Casimirs-Anomalies}, we will calculate central charges for both types of defects. For the Wilson surfaces, we will unambiguously find $d_2$ in the SCE. For the 4d defects, we do not yet know their contribution to the trace anomaly, so we cannot say exactly which central charge(s) we are computing. Our result serves as a prediction for such putative central charge(s).

Let us briefly review the 6d $\CN=(2,0)$ superconformal index and its unrefined limit.  Let $\e_i$ be the squashing parameters of the $S^5$. The bosonic part of the superconformal algebra of the theory is $\mf{so}(6,2)\oplus \mf{usp}(4)_R\subset\mf{osp}(8^*|4)$ with Cartan generators $(E,\,R_1,\,R_2,\,h_1,\,h_2,\,h_3)$.  The generators $h_i$ rotate the planes $\mathds{R}_{\epsilon_i}^2\subset \mathds{R}^6$ into which the squashed $S^5$ is embedded.  Among the SUSY generators $Q^{R_1R_2}_{h_1h_2h_3}$, where the indices are all $\pm \frac{1}{2}$, the privileged supercharge used to construct the index is $Q\equiv Q^{++}_{---}$.  The states contributing to the superconformal index obey the shortening condition in saturating the bound
\begin{align}\label{eq:6d-shortening-condition}
E\geq 2(R_1+R_2) +h_1+h_2+h_3\,.
\end{align}

Assuming saturation of \eq{6d-shortening-condition}, the index can be expressed as
\begin{align}
\mc{I}  = \Tr_{\mc{H}_Q}(-1)^F p^{R_1-R_2} \prod_{i=1}^3 q_i^{h_i +\frac{R_1+R_2}{2}}\,,
\end{align}
where $\mc{H}_Q$ is the subspace of the Hilbert space annihilated by $Q$ and $Q^\dagger$.  The fugacities are $q_i \equiv e^{-R \e_i}$ and $p\equiv e^{-R\mu}$, where $\mu$ is the chemical potential for the R-symmetry generator $R_1-R_2$.  The unrefined limit of $\mc{I}$, defined by $\mu \to \frac{1}{2}(\e_1+\e_2-\e_3)$, has an additional supercharge $Q^\prime \equiv Q^{+-}_{++-}$ that commutes with the Cartan generators, and so the unrefined index collapses to
\begin{align}
\mc{I}_{\rm{unref}} = \Tr_{\mc{H}_{Q,\,Q^\prime}} (-1)^F q^{E-R_1}s^{h_1+R_2}\,,
\end{align}
where $q\equiv q_3$ and $s\equiv q_1/q_2$.  Note the privileged status of rotations in the plane $\mathds{R}_{\epsilon_3}^2$, which is identified with the VOA plane of the 6d theory~\cite{Beem:2014kka}.  The index $\mc{I}_{\rm{unref}}$ is then interpreted as the character of the vacuum module of the VOA.  

Due to the lack of a Lagrangian description of this 6d SCFT, the authors of ref.~\cite{Bullimore:2014upa} compute its twisted partition function by dimensionally reducing on $S^1_R$ and computing the twisted partition function of the 5d $U(N)$ $\CN=2$ SYM theory with coupling $g^2 = 2\pi R$~\cite{Douglas:2010iu} on the squashed $S^5$, $Z_{S^5}$. The codimension four and two defects wrapping $S^1_R$ in 6d reduce to Wilson lines or certain 3d defects in the 5d SYM theory on the squashed $S^5$. Further, the authors of ref.~\cite{Bullimore:2014upa} argue that both the perturbative and non-perturbative contributions to the partition function of the localised theory on the squashed $S^5$ are sufficient to count the states contributing to the 6d index, and hence to defect indices. Although this is far from a proven fact about the dimensional reduction to 5d, we will adopt the same working assumption. The fact that for Wilson surfaces we will recover precisely the holographic result for $d_2$ provides some evidence for this assumption.

In the absence of defects, the localised partition function of the 5d $U(N)$ $\CN=2$ SYM theory on a squashed $S^5$ takes the form
\begin{align}\label{eq:S5-SYM-Z}
Z_{S^5} = \int \frac{d^{N-1}a}{N!}i^{N-1}  e^{\frac{2\pi^2}{\e_1\e_2\e_3}\Tr a^2}Z_{1}Z_{2}Z_{3}\,,
\end{align}
where $Z_1$ is the Nekrasov partition function \cite{Nekrasov:2002qd} on $S^1_1 \times \mathds{R}^4_{\e_2,\,\e_3}$, with $Z_2$ and $Z_3$ obtained from $Z_1$ by cyclic permutation of the labels $\{1,2,3\}$, and $a$ is a constant adjoint-valued scalar parametrising the locus. 

Without any defects, the localised partition function of the 6d $\CN=(2,0)$ $A_{N-1}$ theory in the unrefined limit computes the character of the vacuum module in the $W_N$ algebra. Defining $2\pi i \tau = -R \e_3$ so that $q=e^{2\pi i\tau}$, and defining $\e_1\e_2=1$ and $\mf{b}^2 = \e_1/\e_2$,\footnote{Note that in this section we adopt a different convention for $\mf{b}$ compared to section~\ref{sec:s4}. Moreover, in contrast to section~\ref{sec:s4}, the squashing parameters $\e_{1,2,3}$ of the $S^5$ are chosen to be dimensionless, i.e. they come with appropriate factors of the equatorial radius of the $S^5$, which we take to be the identity in this section.} the partition function sees contributions in the unrefined limit from the three fixed points on the $S^1_i$ of the form
\begin{align}\label{eq:Zi-no-defect}
Z_1 = \prod_{e\in \Delta^+} 2\sin \frac{\pi}{\mf{b}}(e,a)\,,\quad Z_2 = \prod_{e\in \Delta^+} 2\sin \mf{b}\pi (e,a)\,,\qquad Z_3 = \eta \left(-{\tau}^{-1}\right)^{1-N}\,,
\end{align}
where $\eta(\cdot)$ is the Dedekind $\eta$ function.  Let $Q = \rho(\mf{b} +\mf{b}^{-1})$ with $\rho$ being the Weyl vector of $\mf{su}(N)$, and let $\mc{W}_{\mf{g}}$ be the Weyl group of $\mf{g}=A_{N-1}$.  After integrating over $a$, \eq{S5-SYM-Z} becomes 
\begin{align}\label{eq:ZS5-unrefined-no-defect}
Z_{S^5} = \frac{q^{-\frac{1}{2}(Q,Q)}}{\eta(\tau)^{N-1}}\sum_{\sigma \in \mc{W}_{\mf{g}}}\varepsilon(\sigma) q^{-(\sigma(\rho),\rho)+(\rho,\rho)}\,,
\end{align}
where $\varepsilon(\sigma)= (-1)^{\ell(\sigma)}$ and $\ell(\sigma)$ is the length of the Weyl group element $\sigma$. The exponent of the prefactor is related to the central charge $c$ of the VOA as $q^{-\frac{c}{24}}$. Recalling that $\eta(\tau) \propto q^{1/24}$, we thus have
\begin{align}
c = (N-1) +12 (Q,Q) = (N-1) +N(N^2-1)(\mf{b}+\mf{b}^{-1})^2\,,
\end{align}
where we identify $c/24$ as the chiral limit of the 6d SCE found in ref.~\cite{Bobev:2015kza}.

\subsubsection{2d Defects}

Adding two surface operators wrapping $S^1_R \times S^1_1$ or $S^1_R \times S^1_2$ will deform the index to compute the character of degenerate modules of the associated $W_N$-algebra in the VOA plane. The reduction to 5d yields Wilson loop operators with winding $n_1$ and $n_2$ on $S^1_1$ and $S^1_2$, respectively, and carrying irreducible representations of $\mf{su}(N)$ with highest weights $\omega_1$ and $\omega_2$, respectively. The fixed point contributions on $S^1_1$ and $S^1_2$ are modified from those in eq.~\eqref{eq:Zi-no-defect} to
\begin{align}
\label{eq:Zi-defect}
Z_1 = \prod_{e\in\Delta^+} 2\sin \frac{\pi}{\mf{b}}(e,a)\Tr_{\omega_1}e^{\frac{2\pi i a}{\mf{b}}}\,,\qquad Z_2 = \prod_{e\in\Delta^+} 2\sin \pi \mf{b}(e,a)\Tr_{\omega_2}e^{2\pi i a \mf{b}}\,,
\end{align}
where $\text{Tr}_{\omega}$ is a trace over the representation specified by $\omega$. Again, the plane $\mathds{R}^2_{\epsilon_3}$ is designated as the VOA plane and so the Wilson lines cannot wrap $S^1_3$ and also preserve the necessary nilpotent charge needed to define the VOA,\footnote{It is also true for 2d $\CN=(4,4)$ defects in 4d $\CN=4$ SYM theory that the surface operators must be inserted orthogonal to the chiral algebra plane.  Note, though, that a 2d chiral (e.g. $\CN=(0,8)$ or $(0,4)$) superconformal defect could be inserted along the chiral algebra plane while preserving the nilpotent supercharge used to define the VOA.  We thank W. Peelaers for pointing this out.} hence $Z_3$ remains unchanged compared to \eq{Zi-no-defect}.  Plugging $Z_1$ and $Z_2$ from eq.~\eqref{eq:Zi-defect} into the partition function and integrating over $a$ gives
\begin{align}\label{eq:ZS5-unrefined-defect}
Z_{S^5}^{\omega_1,\,\omega_2} = q^{-C_{\omega_1,\,\omega_2}/24}\sum_{\sigma\in\mc{W}(\mf{g})}\varepsilon(\sigma)e^{-(\sigma(\rho+\omega_2),\rho+\omega_1) +(\rho+\omega_2,\rho+\omega_1)}\,,
\end{align}
where the new ``central charge'' is
\begin{align}
C_{\omega_1,\omega_2} = (N-1) +12(Q + \mf{b}^{-1}\omega_1+\mf{b}\omega_2,\, Q+\mf{b}^{-1}\omega_1+\mf{b}\omega_2)\,.
\end{align}
To isolate the defect contribution to the partition function, we divide~\eq{ZS5-unrefined-defect} by the ambient theory result in~\eq{ZS5-unrefined-no-defect}, which gives the change in the central charge,
\beq
\label{eq:2defectsc}
C_{\omega_1,\omega_2} - c = 24 (Q,\mf{b}^{-1}\omega_1+\mf{b}\omega_2) + 12 (\mf{b}^{-1}\omega_1+\mf{b}\omega_2,\mf{b}^{-1}\omega_1+\mf{b}\omega_2)\,.
\eeq
Eq.~\eqref{eq:2defectsc} is our most general result for 2d defects in the $\N=(2,0)$ 6d SCFT, for two intersecting Wilson surfaces.

However, to compare to the holographic result for a single Wilson surface in section~\ref{sec:review}, we restrict to a single defect wrapping, say, $S^1_1$, in which case eq.~\eqref{eq:2defectsc} becomes
\beq
C_{\omega_1} - c = \frac{24}{\mf{b}} \,(Q,\omega_1) + \frac{12}{\mf{b}^2} \,(\omega_1,\omega_1)\,.
\eeq
Taking $\mf{b} \to 1$, so that $Q = \rho(\mf{b} +\mf{b}^{-1}) \to 2 \rho$, and using $d_2 = 24(\rho,\,\omega) + 6(\omega,\,\omega)$ from eq.~\eqref{eq:6d-Wilson-surface-d2}, we find
\beq
\label{eq:1defectc}
C_{\omega_1} - c = 48 \,(\rho,\omega_1) + 12 \,(\omega_1,\omega_1) = 2 \, d_2\,.
\eeq
We have thus shown that a single defect changes the normalisation factor from $q^{-c/24}$ to $q^{-C_{\omega_1}/24}$, or recalling that $q = e^{2 \pi i \tau} = e^{-R \epsilon_3}$, the defect shifts the SCE from $E_c = - \frac{c}{24} \,\epsilon_3$ to $E_c = - \frac{C_{\omega_1}}{24} \,\epsilon_3$. Our result eq.~\eqref{eq:1defectc} then shows that the change in $E_c$ is $\propto d_2$, as advertised.

Crucially, as mentioned above, for a Wilson surface in the 6d $\CN=(2,0)$ theory at large $N$ we can distinguish $d_2$ and $b$, namely $d_2 = b + 3(\omega,\,\omega)$, as opposed to the Levi defect in 4d. The comparison thus leaves no doubt: $d_2$ controls the defect contribution to the SCE.

However, similar to the Levi defect of section~\ref{sec:Casimirs-2d}, the calculation here involved no approximations, relying only on SUSY and the assumptions about the reduction on $S^1_R$ mentioned above. Our result eq.~\eqref{eq:1defectc} thus provides strong evidence that the holographic result for $d_2$ in eq.~\eqref{eq:6d-Wilson-surface-d2} is in fact exact, and not just the leading large-$N$ limiting value.

Our result for two intersecting Wilson surfaces, eq.~\eqref{eq:2defectsc}, is not merely the sum of two copies of the result for the single Wilson surface, eq.~\eqref{eq:1defectc}, due to a cross term $2 (\omega_1,\omega_2)$. This difference could potentially arise for various reasons: some special contributions to the Weyl anomaly from the intersections, some 1d degrees of freedom at the intersection points, and so on. We leave this as an important question for future research.

\subsubsection{4d Defects}

In 6d $\CN=(2,0)$ $A_{N-1}$ SCFTs, there is another class of superconformal defects that one could construct: 4d defects. In the M-theory description, these types of defects arise from $1/2$-BPS M5-M5-brane intersections. The authors of ref.~\cite{Bullimore:2014upa} also constructed the index for these 4d defects, using arguments similar to the 2d case.  

Codimension 2 operators in 6d $\CN=(2,0)$ $A_{N-1}$ SCFTs, in particular, are in one-to-one correspondence with homomorphisms $\varrho:\,\mf{su}(2)\to A_{N-1}$, and in the unrefined limit correspond to a deformation of the VOA by the insertion of a semi-degenerate operator labelled by a partition of N, i.e. $[N_1,\ldots, N_{n+1}]$ where $\sum_{i=1}^{n+1} N_i = N$.  That is, 4d superconformal operators preserve the Levi subalgebra $\mf{l} = \mf{s}\left[\bigoplus_{i=1}^{n+1}\mf{u}(N_i)\right]$.  

In the reduction along $S^1_R$, which the intersecting M5-branes wrap, such a codimension 2 defect has an equivalent description as a prescribed singularity in the gauge field of the resulting 5d $\CN=2$ SYM theory. Given a Levi subalgebra $\mf{l}$, the monodromy parameters are $\vec{m}= \oplus_{i=1}^{n+1} \vec{m}_i$ with each $\vec{m}_i$ being a rank $N_i$ vector whose components are all identically $m_i$, and the Weyl vector of $\mf{l}$ is $\rho_\mf{l} = \oplus_{i=1}^{n+1}\rho_{N_i}$ with each $\rho_{N_i}$ being the Weyl vector of $\mf{su}(N_i)$. The SUSY vacua of the localised theory are labelled by $\sigma\in \mc{W}_{\mf{g}}/\mc{W}_{\mf{l}}$ --- where $\mc{W}_{\mf{g}}$ and $\mc{W}_{\mf{l}}$ are the Weyl groups of $\mf{g}=A_{N-1}$ and $\mf{l}$, respectively ---  which also labels a permutation of the monodromies, i.e. different inequivalent choices of embeddings of $\mf{l}$ in $A_{N-1}$.

To compute the index in the presence of the defect, we need to use the form of the localised partition function in \eq{S5-SYM-Z} supplemented by the classical action from the monodromies given by $e^{-2\pi i(\sigma(\vec{m}),a)}$ and the Nekrasov partition functions corresponding to the particular $\vrho$ and choice of $\sigma$
\begin{align}
Z_1^{\vrho,\sigma} = \prod_{i=1}^{n+1} \prod_{e\in \Delta^+_i} 2\sin \frac{\pi}{\mf{b}}(e,\sigma(a))\,,\quad Z_2^{\vrho,\sigma} = \prod_{i=1}^{n+1} \prod_{e\in \Delta^+_i} 2\sin \pi\mf{b}(e,\sigma(a))~,
\end{align}
where $\Delta_{i}^+$ is the space of positives roots of the $i^{\rm th}$ summand of $\mf{l}$ and as above $Z_3 = \eta(-\tau^{-1})^{1-N}$. Summing over all $\sigma$ and integrating over the locus parametrised by $a$ gives
\begin{align}
Z^\vrho_{S^5} = q^{-C_{\vrho}/24}\sum_{\sigma} \e(\sigma) q^{-(\sigma(\rho_{\mf{l}})-\rho_{\mf{l}},\,\rho_{\mf{l}})}~.
\end{align}
Dividing $Z^\vrho_{S^5}$ by the ambient theory partition function changes the normalisation factor to $q^{-(C_\vrho -c)/24}$, where
\beq
C_{{\varrho}} - c = -24 (Q,\mu_\varrho) +12(\mu_\varrho,\mu_\varrho)\,,
\eeq
and
\beq
\mu_\varrho = Q+\vec{m} -(\mf{b}+\mf{b}^{-1})\rho_{\mf{l}}\,.
\eeq
Using $Q = \rho(\mf{b} +\mf{b}^{-1})$ and $(\vec{m},\rho_{\mf{l}})=0$, we find
\beq
C_{{\varrho}} - c =  12(\mf{b}+\mf{b}^{-1})^2\left [ (\rho_{\mf{l}},\rho_{\mf{l}})-(\rho,\,\rho) \right]+12 (\vec{m},\vec{m})\,.
\eeq
We can easily compute $(\rho_\mf{l},\,\rho_\mf{l}) = \frac{1}{12}(\sum_{i=1}^{n+1} (N_i^3 -N_i))$ by considering each individual $\mf{su}(N_i)$ summand in $\mf{l}$. In the limit $\mf{b}\to 1$ we thus find
\begin{align}
\label{eq:6d-Levi-surface-character}
C_{{\varrho}} - c = -4\left(N^3-\sum_{i=1}^{n+1} N_i^3 -3 (\vec{m},\,\vec{m})\right)\,.
\end{align}
As mentioned above, we do not have a sufficient understanding of the form of the Weyl anomaly of a 4d defect in a 6d CFT to state definitively which central charge(s) the above expression might be. For now, eq.~\eqref{eq:6d-Levi-surface-character} serves as a prediction for 4d superconformal defects in the 6d $\N=(2,0)$ SCFT.

Our result in~\eq{6d-Levi-surface-character} bears a resemblance, modulo overall sign and powers of $N$ and $N_i$, to $d_2$ for the $\CN=(4,4)$ Levi type-$\mds{L}$ surface operator in 4d $\CN=4$ SYM theory in eq.~\eqref{eq:4d-Levi-surface-d2}. Given the connection between the two constructions via dimensional reduction, this superficial resemblance is perhaps not surprising. Beyond the scope of the current work, but the focus of on-going investigation, is finding the behaviour of the defect Weyl anomaly of the 4d Levi type-$\mds{L}$ defect in 6d under dimensional reduction to a 2d Levi type-$\mds{L}$ defect in 4d $\CN\geq 2$ SCFTs.

\section{Summary and Outlook}
\label{sec:discussion}

We have illustrated a variety of techniques for computing the central charges $b$ and $d_2$ of 2d superconformal defects in SCFTs. These techniques rely only on a sufficient amount of SUSY, with no approximations. In particular, we used existing results for SUSY localisation, the AGT correspondence, and superconformal indices to extract new results for $b$ and $d_2$. Some of these results agreed perfectly with existing holographic results, proving that the latter were not merely large-$N$ or strong-coupling limits, but were in fact exact.

Our results pave the way for many fruitful generalisations. Obviously, a variety of other existing results for SUSY partition functions on $S^d$ and $S^1 \times S^{d-1}$ could be mined for further novel results for $b$ and $d_2$. This includes twist field defects relevant for calculations of SUSY R\'{e}nyi entropy~\cite{Nishioka:2013haa,Huang:2014pda,Yankielowicz:2017xkf}, where information theoretic constraints may imply bounds on the defect's central charges~\cite{Zhou:2016kcz}. Additionally, to our knowledge a variety of 2d superconformal defects have yet to be described using any of the SUSY methods we have discussed. A prominent example is chiral defects, such as defects with 2d $\N=(0,4)$ SUSY. Chiral defects break parity, producing parity-odd terms in the trace anomaly that define two parity-odd central charges~\cite{Cvitan:2015ana,Jensen:2018rxu}. These could in principle be calculated using the methods we have described. Furthermore, as deformations of the superconformal index, 2d $\CN=(0,4)$ defects can preserve the nilpotent supercharge used in the cohomological construction of chiral algebras from 4d SCFTs \cite{Beem:2013sza}, and so their central charges may appear in the vacuum character of a deformed chiral algebra.

Other approaches to computing SUSY partition functions on $S^d$ and $S^1 \times S^{d-1}$ could also be developed along similar lines as useful tools to extract defect central charges in novel systems. Examples include geometric engineering~\cite{Ooguri:1999bv,Dimofte:2010tz}, or computing a 5d SUSY partition function on $S^1 \times S^4$ with a 3d SUSY defect along $S^1 \times S^2$~\cite{Bullimore:2014awa} and then reducing on the common $S^1$ to obtain a 4d SUSY partition function on $S^4$ with a 2d defect along $S^2 \hookrightarrow S^4$~\cite{Gukov:2014gja}. More importantly, studying how the defect trace anomalies and associated central charges behave under dimensional reduction could provide a new window into how defect physics changes under RG flows across dimensions~\cite{Bobev:2017uzs}.

All the above methods could also be straightforwardly generalised to defects of other dimensions. For example, in 4d SCFTs various 1/2-BPS interfaces and domain walls have been studied using holography~\cite{DeWolfe:2001pq, DHoker:2007zhm,Bobev:2013yra}, SUSY localisation~\cite{Drukker:2010jp}, and other methods~\cite{Dimofte:2011ju}. In these cases the interface contribution to the trace anomaly defines two central charges~\cite{Melmed:1988hm,Dowker:1989ue,Herzog:2015ioa,Fursaev:2015wpa,Solodukhin:2015eca} that could in principle be calculated from existing results. In 5d and 6d SCFTs, higher-dimensional defects are possible, such as the 4d defect in the M5-brane theory that we discussed at the end of section~\ref{sec:Casimirs}. However, in these cases the defect contribution to the trace anomaly, and in fact many other quantities are unknown, so what (linear combination) of central charges the SUSY methods could compute is unclear.

Indeed, more generally the contributions of defects to trace anomalies as in \eq{defecttrace}, entanglement entropy \cite{Jensen:2013lxa}, and other quantities are clearly crucial for characterising and classifying defects, including for proving $c$-theorems \cite{Jensen:2015swa,Kobayashi:2018lil}, positivity \cite{Jensen:2018rxu} or other lower bounds \cite{Herzog:2017kkj}, and other constraints on defect central charges. We hope our methods provide useful tools for addressing such issues.


\acknowledgments

We would like to thank Nikolay Bobev, Matthew Buican, Nadav Drukker, John Estes, Davide Gaiotto, Christopher Herzog, Darya Krym, and Wolfger Peelaers for useful discussions and correspondence. We especially thank Lorenzo Bianchi, Bruno Le Floch, Madalena Lemos, Kristan Jensen, and Ronnie Rodgers for reading and commenting on a draft of the paper. A.~C. is supported by the Royal Society award RGF/EA/180098. A.~O'B. is a Royal Society University Research Fellow. B.~R. acknowledges partial support from STFC through Consolidated Grant ST/L000296/1 and by the KU Leuven C1 grant ZKD1118 C16/16/005 during the project. J.~S. is supported by the Royal Society award RGF/EA/181020.

\appendix

\section{Special Functions and Zeta-Function Regularisation} \label{ap:special_fn}

In this appendix, we give a quick overview of the special functions appearing in this paper and explain how they arise from zeta-function regularisation of infinite products. For more details on some of these functions we refer to refs.~\cite{Ruj00, Spr09}.

By meromorphic continuation to the complex $s$-plane, the Barnes multiple-zeta function $\zeta_N (s;z| a_1,\ldots , a_N)$ and the multiple Gamma-function are defined as follows
\begin{align}
\zeta_N (s;z| a_1,\ldots , a_N)&\equiv \sum_{n_1, \ldots, n_N \geq 0} \left(z+n_1 a_1 + \ldots n_N a_N\right)^{-s} \, , \label{eq:Barnes}\\
\Gamma_N (z|a_1, \ldots a_N)&\equiv \exp \left(\left.\frac{d}{ds}\zeta_N (s;z| a_1,\ldots , a_N)\right|_{s=0}\right) \, . \label{eq:multipleGamma}
\end{align}
The cases of particular interest to us are $N=1,2$, which include the single $\zeta_1(s;z|a)$ and double zeta-function $\zeta_2(s;z|a_1,a_2)$.  In particular, we will need their values at $s=0$
\begin{equation}
\begin{split}
\zeta_1(0;z|a)&= \dfrac{1}{2}-\dfrac{z}{a}\, ,\\
\zeta_2(0;z|a_1,a_2)&=\dfrac{1}{4}+\dfrac{1}{12}\left(\dfrac{a_1}{a_2}+\dfrac{a_2}{a_1}\right)-\dfrac{z}{2}\left(\dfrac{1}{a_1}+\dfrac{1}{a_2}\right) + \dfrac{z^2}{2a_1a_2}\, .\label{eq:Barneszero}
\end{split}
\end{equation}
From the definition above, the single Gamma-function $\Gamma_1 (z|a)$ is related to the ordinary Euler Gamma-function $\Gamma(z)$ via 
\begin{equation}
\Gamma_1(z|a^{-1})=\dfrac{a^{\frac{1}{2}-az}}{\sqrt{2\pi}}\Gamma(az)\, . \label{eq:Gamma_1_Euler}
\end{equation}
Further, the double Gamma-function is used in defining the special function
\begin{equation}
\Upsilon (z|a_1,a_2) \equiv \dfrac{1}{\Gamma_2(z|a_1,a_2) \Gamma_2(a_1+a_2-z|a_1,a_2)}\, , 
\end{equation}
which frequently appears in Liouville/Toda theory. This Upsilon-function obeys 
\begin{equation}
\Upsilon (z+a_2|a_1,a_2)=\gamma_1(z|a_1)\Upsilon(z|a_1,a_2)\, ,
\end{equation}
where
\begin{equation}
\gamma_1(z|a)\equiv \dfrac{\Gamma_1(z|a)}{\Gamma_1(a-z|a)}\,, \label{eq:gamma_1}
\end{equation}
and a similar relation for the shift $\Upsilon(z+a_1|a_1,a_2)$ replacing $a_1\to a_2$. This can be recast in the more familiar form
\begin{equation}
\Upsilon (z+a_2|a_1,a_2)=a_1^{2z/a_1-1} \gamma(z/a_1)\Upsilon(z|a_1,a_2)\, , \label{eq:Upsilon_prop}
\end{equation}
where 
\begin{equation}
\gamma(z)\equiv\dfrac{\Gamma(z)}{\Gamma(1-z)}\,.  \label{eq:gamma}
\end{equation}

The special functions above appear in the evaluation of 1-loop determinants as alluded to in section~\ref{sec:s4}. One usually encounters infinite products that diverge and require regularisation. Zeta-function regularisation instructs us to replace a diverging product
\begin{equation}
\prod_{k=0}^\infty \lambda_k \rightarrow \exp\left(-\left.\frac{d}{ds}\mathfrak{Z}(s)\right|_{s=0}\right) \, ,
\end{equation}
where $\mathfrak{Z}(s)$ is the associated zeta-function defined as the meromorphic continuation to the complex $s$-plane of the series
\begin{equation}
\sum_{k=0}^\infty \lambda_k{}^{-s}\, .
\end{equation}
For the divergent products of the form
\begin{equation}
\prod_{k=0}^\infty \left(\frac{k}{r}+z\right)\, ,
\end{equation}
the associated zeta function is $\zeta_1\left(s;z\left|r^{-1}\right.\right)$, and hence zeta-function regularisation gives
\begin{equation}
\prod_{k=0}^\infty \left(\frac{k}{r}+z\right)\rightarrow \dfrac{1}{\Gamma_1(z|r^{-1})} =\dfrac{\sqrt{2\pi}\,r^{rz-\frac{1}{2}}}{\Gamma(rz)} \, .  \label{eq:gamma_1_scaling}
\end{equation}

Most importantly for our analysis, we need to understand the behaviour of the multiple Gamma-function appearing in 1-loop determinants under a constant Weyl re-scaling.  Generically,
\begin{equation}
\Gamma_N \left(\left.\frac{z}{r}\right|\frac{a_1}{r}, \ldots , \frac{a_N}{r}\right) = r^{\zeta_N (0;z|a_1, \ldots, a_N)}\, \Gamma_N(z|a_1, \ldots , a_N)\, .
\end{equation}
For $N=1$ this reduces to
\begin{equation}
\Gamma_1 \left(\left.\frac{z}{r}\right|\frac{a}{r}\right) = r^{\frac{1}{2}-	\frac{z}{a}}\, \Gamma_1(z|a)\, ,
\end{equation}
such that
\begin{equation}
\gamma_1\left(\left.\frac{z}{r}\right|\frac{a}{r}\right)=r^{1-\frac{2z}{a}}\, \gamma_1(z|a)\,.\label{eq:scale_gamma1}
\end{equation}

For $N=2$ one finds
\begin{equation}
\Upsilon\left(\frac{z}{r}\left|\frac{a_1}{r},\frac{a_2}{r}\right.\right)=r^{-2\zeta_2 (0;z|a_1, a_2)}\Upsilon(z|a_1,a_2) \, .\label{eq:scaleUpsilon} 
\end{equation}

\section{$A_{N-1}$ Toda field theory}
\label{appendix_Toda}

In this appendix we review the essential features of the $A_{N-1}$ Toda field theory needed for the computations in section~\ref{sec:AGT}. For more details see for example refs.~\cite{Fateev:2007ab,Fateev:2008bm}.

The action for $A_{N-1}$ Toda field theory is given by 
\begin{align}
\label{Toda_action}
S=\int d^2 \sigma \left[ \frac{1}{8\pi} g^{\mu\nu} (\partial_\mu \phi, \partial_\nu \phi)+ \frac{(Q,\phi)}{4\pi} \mathcal{R}+\mu \sum_{i=1}^{{N-1}}e^{\mathfrak{b} (e_i,\phi)} \right]\,,
\end{align}  
where $g_{\mu\nu}$ is the metric of the Riemann surface, $\mathcal{R}$ the corresponding scalar curvature, $\mathfrak{b}$ is the dimensionless coupling constant, $e_i$ and $\rho$ are respectively the simple roots and Weyl vector of the $A_{N-1}$ Lie algebra, and $(\cdot,\cdot)$ denotes the scalar product on the weight space. The requirement of conformal invariance fixes $Q=(\mathfrak{b}+1/\mathfrak{b})\rho$. In terms of $Q$, the central charge is given by
\begin{align}
c= N-1+12 (Q,Q)\,.
\end{align}
Besides the conformal symmetry, the theory of eq.~\eqref{Toda_action} is invariant under higher-spin symmetry transformations generated by the $(n-1)$ holomorphic currents $\boldsymbol W^k(z)$ with spins $k=2,\dots,N$, whose algebra is called $W_N$-algebra. These currents can be written in terms of the field $\phi$ as
\begin{align}
\label{W-currents}
\prod^{N-1}_{i=0}\left( q \partial + (h_{N-i},\partial \phi) \right)= \sum_{k=0}^N \boldsymbol{W}^{N-\mathfrak{b}}(z)\left(q \partial \right)^k\,,
\end{align} 
where $h_i=h_1-e_1-\dots -e_k$ with $k=1,\dots,i$ are the weights of the fundamental representation of the $A_{N-1}$ with highest weight $h_1$. Their scalar product reads $(h_i,h_j)=\delta_{ij}-1/N$. We observe that $\boldsymbol{W}^2(z)=T(z)$. It is not difficult to see that for $N=2$, the Toda field theory reduces to the Liouville theory.

Primary fields with respect to the $W_N$-algebra are the vertex operators
\begin{align}
V_{\alpha}= e^{(\alpha,\phi)}\,,
\end{align}
with quantum numbers $\omega^{(k)}(\alpha)$ and conformal dimension  $\Delta(\alpha)=\omega^{(2)}(\alpha)=\frac{(\alpha,2Q-\alpha)}{2}$.

The three-point functions of $W_N$ primaries can generically be expressed as
\begin{align}
\label{three-point_general}
\braketbis{V_{\alpha_1}(z_1,\bar z_1)V_{\alpha_1}(z_2,\bar z_2)V_{\alpha_3}(z_3,\bar z_3)}=\frac{C(\alpha_1,\alpha_2,\alpha_3)}{|z_{12}|^{2(\Delta_1+\Delta_2-\Delta_3)}|z_{13}|^{2(\Delta_1+\Delta_3-\Delta_2)}|z_{23}|^{2(\Delta_2+\Delta_3-\Delta_1)}}\,.
\end{align}
While the $z$-dependence is fixed by conformal symmetry, all the non-trivial information about the three-point function is encoded in the coefficient $C(\alpha_1,\alpha_2,\alpha_3)$. In the Liouville case, this coefficient has been found for generic values of the momenta $\alpha$~\cite{Dorn:1994xn,Zamolodchikov:1995aa}. On the other hand, for $N>2$ the structure of the three-point function is more complicated and a general expression is not available.  However, there are useful limiting cases where analytic results can be obtained. For example, a simplification occurs if one of the primaries is semi-degenerate, i.e. it satisfies the special condition $\alpha=\varkappa h_1$ where $\varkappa$ is a real number. In this case, the coefficient in eq.~\eqref{three-point_general} can be expressed in a closed form
\begin{align}
\label{three-point_funct_bare}
\begin{split}
C(\alpha_1,\alpha_2,\varkappa h_1)= &\left[ \pi \mu \gamma(\mathfrak{b}^2)\mathfrak{b}^{2-2\mathfrak{b}^2} \right]^{\frac{(2Q-\sum_i \alpha_i,\rho)}{\mathfrak{b}}}  \\
&\times  (\Upsilon(\mathfrak{b}))^{N-1}\Upsilon(\varkappa) \frac{\prod_{e>0} \Upsilon\left( (Q-\alpha_1,e) \right)\Upsilon\left( (Q-\alpha_2,e) \right)}{\prod_{ij} \Upsilon \left( \frac{\varkappa}{N}+(Q-\alpha_1,h_i)+(Q-\alpha_2,h_j) \right)}\,,
\end{split}
\end{align}
where the function $\Upsilon(x)\equiv \Upsilon(x|\mathfrak{b},1/\mathfrak{b})$ (see eq.~\eqref{eq:Upsilon}). 

Another remarkable case is when the field is fully degenerate. Degenerate fields are parametrised by $\alpha=-\mathfrak{b} \omega_1 - 1/\mathfrak{b} \omega_2$ where $\omega_1$ and $\omega_2$ are highest weights of two representations $\mathcal{R}_1$ and $\mathcal{R}_2$ of $A_{N-1}$.  The operator product expansion of a degenerate primary with a generic primary field $V_\alpha$ consists of a finite number of primaries. Precisely, we have
\begin{align}
\label{deg_fusion}
V_{-\mf{b} \omega_1-1/\mf{b} \omega_2}V_\alpha = \sum_{s,p} C^{\alpha'_{s,p}}_{-\mf{b}\omega_1-1/\mf{b} \omega_2,\alpha} \left[ V_{\alpha'_{s,p}} \right]\,,
\end{align}
where the square brackets denotes all the descendants and $\alpha'_{s,p}=\alpha-\mathfrak{b} h_{(\omega_1),s}-1/\mathfrak{b} h_{(\omega_2),p}$, $h_{(\omega_1),s}$ and $h_{(\omega_2),s}$ being the weights of the representation $\mathcal{R}_1$ and $\mathcal{R}_2$, respectively. In this work we only consider the case $\alpha=-\mathfrak{b} \omega_1$.

In the computations performed in section~\ref{sec:AGT}, we will employ the following normalisation for the generic $W_N$-primary fields \cite{Gomis:2014eya}:
\begin{align}
\label{norm_1}
\widehat V_\alpha=  \frac{\left[ \pi \mu \gamma(\mathfrak{b}^2)\mathfrak{b}^{2-2\mathfrak{b}^2} \right]^{(\alpha-Q,\rho)/\mathfrak{b}}}{\prod_{s<t}^{N}\Upsilon((Q-\alpha,h_s-h_t))}  V_{\alpha}\,,
\end{align}
while for semi-degenerate and degenerate vertex operators we define
\begin{align}
\label{norm_2}
\widehat V_{\varkappa h_{1}}=\frac{\left[ \pi \mu \gamma(\mathfrak{b}^2)\mathfrak{b}^{2-2\mathfrak{b}^2} \right] ^{\frac{(\varkappa h_{1},\rho)}{\mathfrak{b}}}}{\Upsilon(\mathfrak{b})^{N-1}\Upsilon(\varkappa)} V_{\varkappa h_{1}}\,,  \qquad \widehat V_{-\mathfrak{b} \omega}= \left[ \pi  \mu \gamma(\mathfrak{b}^2) \mathfrak{b}^4 \right]^{\frac{(-\mathfrak{b}\omega,\rho)}{\mathfrak{b}}}V_{-\mathfrak{b}\omega}\,.
\end{align}
With these choices of normalisation, the three-point function eq.~\eqref{three-point_funct_bare} simplifies to eq.~\eqref{three_point_trinion} of the main text with $\alpha_1=\alpha_\infty$, $\alpha_2=\alpha_0$.

For reference and use in section~\ref{sec:AGT}, we list here the results for the structure constants in the cases that $\mc{R}_1$ is the rank-$k$ totally antisymmetric or totally symmetric representation of $A_{N-1}$.

The rank-$k$ antisymmetric representation denoted as $\mathcal{R}_{k,-}$ has highest weight given by $\omega_{(k,-)}=\sum_{s=1}^k h_s$ while all the other weights can be expressed in terms of the weights of the fundamental representation as $h_{{(k,-)},{\{p\}}}=\sum_{\{p\}}h_p$ where the set $\{p\}$ consists of $k$ numbers such that $1\le p_1  < \dots <p_k \le N$. The number of distinct weights is given by the number of ways in which one can choose such a set $\{p\}$. 
Using the normalisations in eqs.~\eqref{norm_1} and~\eqref{norm_2}, the structure constants take the form \cite{Fateev:2007ab} 
\begin{align}
\label{const_struc_anti}
\begin{split}
\widehat C ^{\alpha-\mathfrak{b}h}_{-\mathfrak{b} \omega_{(k,-)},\alpha} & = \mathfrak{b}^{-N (2(Q-\alpha)+\mathfrak{b}h,\mathfrak{b}h)}\prod_{s\notin\{p\}}^N \prod_{t\in\{p\}}^N \gamma\left(\mathfrak{b} (Q-\alpha,h_t-h_s) \right) \\
&=  \mathfrak{b}^{-(h,h)(1+\mathfrak{b}^2)}\prod_{s \notin \{p\}}^N \prod_{t \in \{p\}}^N \gamma_1 \left(  (Q-\alpha,h_t-h_s) \left| \frac{1}{\mathfrak{b}}\right. \right)\,,
\end{split}
\end{align}
where in the last step we employed that $\gamma(\mathfrak{b}x)=\mathfrak{b}^{2\mathfrak{b}x-1}\gamma_1(x|1/\mathfrak{b})$, which can be deduced from the definitions in eqs.~\eqref{eq:gamma_1}, \eqref{eq:gamma} and the property in eq.~\eqref{eq:Gamma_1_Euler}.

The rank-$k$ symmetric representation is denoted $\mathcal{R}_{k,+}$, its highest weight is $\omega_{(k,+)}=k h_1$, and all of its other weights are $h_{{(k,+)},[n]}=\sum_s^{N} n_s h_s$ with $\sum_{s=1}^{N} n_s=k$. The corresponding structure constants have been found in refs.~\cite{Fateev:2007ab,Gomis:2014eya}:
\begin{align}
\label{const_struc_symm}
\begin{split}
\widehat C ^{\alpha-\mathfrak{b}h}_{-\mathfrak{b}k h_1 ,\alpha} & = \frac{\mathfrak{b}^{-N (2(Q-\alpha)+\mathfrak{b}h,\mathfrak{b}h)}}{\prod_{\nu=1}^k \gamma(-\nu \mathfrak{b}^2)}\prod_{s,t=1}^N \prod_{\nu=0}^{n_t-1} \gamma\left( \mathfrak{b}(Q-\alpha,h_t-h_s)+(\nu-n_s)\mathfrak{b}^2 \right)
\\
& = \frac{\mathfrak{b}^{- k(N+k)(1+\mathfrak{b}^2)+k^2}}{\prod_{\nu=1}^k \gamma(-\nu \mathfrak{b}^2)}\prod_{s,t=1}^N \prod_{\nu=0}^{n_t-1} \gamma_1\left( (Q-\alpha,h_t-h_s)+(\nu-n_s)\mathfrak{b} \left|\frac{1}{\mathfrak{b}} \right. \right) \,.
\end{split}
\end{align}
%


\bibliographystyle{JHEP}
\bibliography{centralchargesv2}

\providecommand{\href}[2]{#2}\begingroup\raggedright\begin{thebibliography}{100}

\bibitem{Douglas:2010ic}
M.~R. Douglas, \emph{{Spaces of Quantum Field Theories}},
  \href{https://doi.org/10.1088/1742-6596/462/1/012011}{\emph{J. Phys. Conf.
  Ser.} {\bfseries 462} (2013) 012011},
  [\href{https://arxiv.org/abs/1005.2779}{{\ttfamily 1005.2779}}].

\bibitem{Aharony:2013hda}
O.~Aharony, N.~Seiberg and Y.~Tachikawa, \emph{{Reading between the lines of
  four-dimensional gauge theories}},
  \href{https://doi.org/10.1007/JHEP08(2013)115}{\emph{JHEP} {\bfseries 08}
  (2013) 115}, [\href{https://arxiv.org/abs/1305.0318}{{\ttfamily 1305.0318}}].

\bibitem{Gukov:2013zka}
S.~Gukov and A.~Kapustin, \emph{{Topological Quantum Field Theory, Nonlocal
  Operators, and Gapped Phases of Gauge Theories}},
  \href{https://arxiv.org/abs/1307.4793}{{\ttfamily 1307.4793}}.

\bibitem{Gukov:2014gja}
S.~Gukov, \emph{{Surface Operators}},  in \emph{New Dualities of Supersymmetric
  Gauge Theories} (J.~Teschner, ed.), pp.~223--259.
\newblock 2016.
\newblock \href{https://arxiv.org/abs/1412.7127}{{\ttfamily 1412.7127}}.
\newblock \href{https://doi.org/10.1007/978-3-319-18769-3_8}{DOI}.

\bibitem{Zamolodchikov:1986gt}
A.~B. Zamolodchikov, \emph{{Irreversibility of the Flux of the Renormalization
  Group in a 2D Field Theory}}, {\emph{JETP Lett.} {\bfseries 43} (1986)
  730--732}.

\bibitem{Cappelli:1990yc}
A.~Cappelli, D.~Friedan and J.~I. Latorre, \emph{{C theorem and spectral
  representation}},
  \href{https://doi.org/10.1016/0550-3213(91)90102-4}{\emph{Nucl. Phys.}
  {\bfseries B352} (1991) 616--670}.

\bibitem{Casini:2004bw}
H.~Casini and M.~Huerta, \emph{{A Finite entanglement entropy and the
  c-theorem}},
  \href{https://doi.org/10.1016/j.physletb.2004.08.072}{\emph{Phys. Lett.}
  {\bfseries B600} (2004) 142--150},
  [\href{https://arxiv.org/abs/hep-th/0405111}{{\ttfamily hep-th/0405111}}].

\bibitem{Komargodski:2011xv}
Z.~Komargodski, \emph{{The Constraints of Conformal Symmetry on RG Flows}},
  \href{https://doi.org/10.1007/JHEP07(2012)069}{\emph{JHEP} {\bfseries 07}
  (2012) 069}, [\href{https://arxiv.org/abs/1112.4538}{{\ttfamily 1112.4538}}].

\bibitem{Casini:2012ei}
H.~Casini and M.~Huerta, \emph{{On the RG running of the entanglement entropy
  of a circle}}, \href{https://doi.org/10.1103/PhysRevD.85.125016}{\emph{Phys.
  Rev.} {\bfseries D85} (2012) 125016},
  [\href{https://arxiv.org/abs/1202.5650}{{\ttfamily 1202.5650}}].

\bibitem{Casini:2017vbe}
H.~Casini, E.~Testé and G.~Torroba, \emph{{Markov Property of the Conformal
  Field Theory Vacuum and the a Theorem}},
  \href{https://doi.org/10.1103/PhysRevLett.118.261602}{\emph{Phys. Rev. Lett.}
  {\bfseries 118} (2017) 261602},
  [\href{https://arxiv.org/abs/1704.01870}{{\ttfamily 1704.01870}}].

\bibitem{Cardy:1988cwa}
J.~L. Cardy, \emph{{Is There a c Theorem in Four-Dimensions?}},
  \href{https://doi.org/10.1016/0370-2693(88)90054-8}{\emph{Phys. Lett.}
  {\bfseries B215} (1988) 749--752}.

\bibitem{Osborn:1989td}
H.~Osborn, \emph{{Derivation of a Four-dimensional $c$ Theorem}},
  \href{https://doi.org/10.1016/0370-2693(89)90729-6}{\emph{Phys. Lett.}
  {\bfseries B222} (1989) 97--102}.

\bibitem{Jack:1990eb}
I.~Jack and H.~Osborn, \emph{{Analogs for the $c$ Theorem for Four-dimensional
  Renormalizable Field Theories}},
  \href{https://doi.org/10.1016/0550-3213(90)90584-Z}{\emph{Nucl. Phys.}
  {\bfseries B343} (1990) 647--688}.

\bibitem{Osborn:1991gm}
H.~Osborn, \emph{{Weyl consistency conditions and a local renormalization group
  equation for general renormalizable field theories}},
  \href{https://doi.org/10.1016/0550-3213(91)80030-P}{\emph{Nucl. Phys.}
  {\bfseries B363} (1991) 486--526}.

\bibitem{Komargodski:2011vj}
Z.~Komargodski and A.~Schwimmer, \emph{{On Renormalization Group Flows in Four
  Dimensions}}, \href{https://doi.org/10.1007/JHEP12(2011)099}{\emph{JHEP}
  {\bfseries 12} (2011) 099},
  [\href{https://arxiv.org/abs/1107.3987}{{\ttfamily 1107.3987}}].

\bibitem{Affleck:1991tk}
I.~Affleck and A.~W.~W. Ludwig, \emph{{Universal noninteger 'ground state
  degeneracy' in critical quantum systems}},
  \href{https://doi.org/10.1103/PhysRevLett.67.161}{\emph{Phys. Rev. Lett.}
  {\bfseries 67} (1991) 161--164}.

\bibitem{Friedan:2003yc}
D.~Friedan and A.~Konechny, \emph{{On the boundary entropy of one-dimensional
  quantum systems at low temperature}},
  \href{https://doi.org/10.1103/PhysRevLett.93.030402}{\emph{Phys. Rev. Lett.}
  {\bfseries 93} (2004) 030402},
  [\href{https://arxiv.org/abs/hep-th/0312197}{{\ttfamily hep-th/0312197}}].

\bibitem{Casini:2016fgb}
H.~Casini, I.~S. Landea and G.~Torroba, \emph{{The g-theorem and quantum
  information theory}},
  \href{https://doi.org/10.1007/JHEP10(2016)140}{\emph{JHEP} {\bfseries 10}
  (2016) 140}, [\href{https://arxiv.org/abs/1607.00390}{{\ttfamily
  1607.00390}}].

\bibitem{Jensen:2015swa}
K.~Jensen and A.~O'Bannon, \emph{{Constraint on Defect and Boundary
  Renormalization Group Flows}},
  \href{https://doi.org/10.1103/PhysRevLett.116.091601}{\emph{Phys. Rev. Lett.}
  {\bfseries 116} (2016) 091601},
  [\href{https://arxiv.org/abs/1509.02160}{{\ttfamily 1509.02160}}].

\bibitem{Casini:2018nym}
H.~Casini, I.~Salazar~Landea and G.~Torroba, \emph{{Irreversibility in quantum
  field theories with boundaries}},
  \href{https://doi.org/10.1007/JHEP04(2019)166}{\emph{JHEP} {\bfseries 04}
  (2019) 166}, [\href{https://arxiv.org/abs/1812.08183}{{\ttfamily
  1812.08183}}].

\bibitem{Graham:1999pm}
C.~R. Graham and E.~Witten, \emph{{Conformal anomaly of submanifold observables
  in AdS / CFT correspondence}},
  \href{https://doi.org/10.1016/S0550-3213(99)00055-3}{\emph{Nucl. Phys.}
  {\bfseries B546} (1999) 52--64},
  [\href{https://arxiv.org/abs/hep-th/9901021}{{\ttfamily hep-th/9901021}}].

\bibitem{Henningson:1999xi}
M.~Henningson and K.~Skenderis, \emph{{Weyl anomaly for Wilson surfaces}},
  \href{https://doi.org/10.1088/1126-6708/1999/06/012}{\emph{JHEP} {\bfseries
  06} (1999) 012}, [\href{https://arxiv.org/abs/hep-th/9905163}{{\ttfamily
  hep-th/9905163}}].

\bibitem{Asnin:2008ak}
V.~Asnin, \emph{{Analyticity Properties of Graham-Witten Anomalies}},
  \href{https://doi.org/10.1088/0264-9381/25/14/145013}{\emph{Class. Quant.
  Grav.} {\bfseries 25} (2008) 145013},
  [\href{https://arxiv.org/abs/0801.1469}{{\ttfamily 0801.1469}}].

\bibitem{Schwimmer:2008yh}
A.~Schwimmer and S.~Theisen, \emph{{Entanglement Entropy, Trace Anomalies and
  Holography}},
  \href{https://doi.org/10.1016/j.nuclphysb.2008.04.015}{\emph{Nucl. Phys.}
  {\bfseries B801} (2008) 1--24},
  [\href{https://arxiv.org/abs/0802.1017}{{\ttfamily 0802.1017}}].

\bibitem{Estes:2018tnu}
J.~Estes, D.~Krym, A.~O'Bannon, B.~Robinson and R.~Rodgers, \emph{{Wilson
  Surface Central Charge from Holographic Entanglement Entropy}},
  \href{https://doi.org/10.1007/JHEP05(2019)032}{\emph{JHEP} {\bfseries 05}
  (2019) 032}, [\href{https://arxiv.org/abs/1812.00923}{{\ttfamily
  1812.00923}}].

\bibitem{Jensen:2018rxu}
K.~Jensen, A.~O'Bannon, B.~Robinson and R.~Rodgers, \emph{{From the Weyl
  Anomaly to Entropy of Two-Dimensional Boundaries and Defects}},
  \href{https://doi.org/10.1103/PhysRevLett.122.241602}{\emph{Phys. Rev. Lett.}
  {\bfseries 122} (2019) 241602},
  [\href{https://arxiv.org/abs/1812.08745}{{\ttfamily 1812.08745}}].

\bibitem{Cvitan:2015ana}
M.~Cvitan, P.~Dominis~Prester, S.~Pallua, I.~Smolić and T.~Štemberga,
  \emph{{Parity-odd surface anomalies and correlation functions on conical
  defects}},  \href{https://arxiv.org/abs/1503.06196}{{\ttfamily 1503.06196}}.

\bibitem{Deser:1993yx}
S.~Deser and A.~Schwimmer, \emph{{Geometric classification of conformal
  anomalies in arbitrary dimensions}},
  \href{https://doi.org/10.1016/0370-2693(93)90934-A}{\emph{Phys. Lett.}
  {\bfseries B309} (1993) 279--284},
  [\href{https://arxiv.org/abs/hep-th/9302047}{{\ttfamily hep-th/9302047}}].

\bibitem{Herzog:2017xha}
C.~P. Herzog and K.-W. Huang, \emph{{Boundary Conformal Field Theory and a
  Boundary Central Charge}},
  \href{https://doi.org/10.1007/JHEP10(2017)189}{\emph{JHEP} {\bfseries 10}
  (2017) 189}, [\href{https://arxiv.org/abs/1707.06224}{{\ttfamily
  1707.06224}}].

\bibitem{Herzog:2019rke}
C.~P. Herzog and I.~Shamir, \emph{{How a-type anomalies can depend on marginal
  couplings}},  \href{https://arxiv.org/abs/1907.04952}{{\ttfamily
  1907.04952}}.

\bibitem{Bianchi:2019umv}
L.~Bianchi, \emph{{Marginal deformations and defect anomalies}},
  \href{https://doi.org/10.1103/PhysRevD.100.126018}{\emph{Phys. Rev.}
  {\bfseries D100} (2019) 126018},
  [\href{https://arxiv.org/abs/1907.06193}{{\ttfamily 1907.06193}}].

\bibitem{Herzog:2017kkj}
C.~Herzog, K.-W. Huang and K.~Jensen, \emph{{Displacement Operators and
  Constraints on Boundary Central Charges}},
  \href{https://doi.org/10.1103/PhysRevLett.120.021601}{\emph{Phys. Rev. Lett.}
  {\bfseries 120} (2018) 021601},
  [\href{https://arxiv.org/abs/1709.07431}{{\ttfamily 1709.07431}}].

\bibitem{Gustavsson:2003hn}
A.~Gustavsson, \emph{{On the Weyl anomaly of Wilson surfaces}},
  \href{https://doi.org/10.1088/1126-6708/2003/12/059}{\emph{JHEP} {\bfseries
  12} (2003) 059}, [\href{https://arxiv.org/abs/hep-th/0310037}{{\ttfamily
  hep-th/0310037}}].

\bibitem{Gustavsson:2004gj}
A.~Gustavsson, \emph{{Conformal anomaly of Wilson surface observables: A Field
  theoretical computation}},
  \href{https://doi.org/10.1088/1126-6708/2004/07/074}{\emph{JHEP} {\bfseries
  07} (2004) 074}, [\href{https://arxiv.org/abs/hep-th/0404150}{{\ttfamily
  hep-th/0404150}}].

\bibitem{Berenstein:1998ij}
D.~E. Berenstein, R.~Corrado, W.~Fischler and J.~M. Maldacena, \emph{{The
  Operator product expansion for Wilson loops and surfaces in the large N
  limit}}, \href{https://doi.org/10.1103/PhysRevD.59.105023}{\emph{Phys. Rev.}
  {\bfseries D59} (1999) 105023},
  [\href{https://arxiv.org/abs/hep-th/9809188}{{\ttfamily hep-th/9809188}}].

\bibitem{Drukker:2008wr}
N.~Drukker, J.~Gomis and S.~Matsuura, \emph{{Probing N=4 SYM With Surface
  Operators}}, \href{https://doi.org/10.1088/1126-6708/2008/10/048}{\emph{JHEP}
  {\bfseries 10} (2008) 048},
  [\href{https://arxiv.org/abs/0805.4199}{{\ttfamily 0805.4199}}].

\bibitem{Garriga:2008ks}
J.~Garriga and A.~Vilenkin, \emph{{Holographic Multiverse}},
  \href{https://doi.org/10.1088/1475-7516/2009/01/021}{\emph{JCAP} {\bfseries
  0901} (2009) 021}, [\href{https://arxiv.org/abs/0809.4257}{{\ttfamily
  0809.4257}}].

\bibitem{Garriga:2009hy}
J.~Garriga and A.~Vilenkin, \emph{{Holographic Multiverse and Conformal
  Invariance}},
  \href{https://doi.org/10.1088/1475-7516/2009/11/020}{\emph{JCAP} {\bfseries
  0911} (2009) 020}, [\href{https://arxiv.org/abs/0905.1509}{{\ttfamily
  0905.1509}}].

\bibitem{Fiol:2010un}
B.~Fiol, \emph{{Flavor from M5-branes}},
  \href{https://doi.org/10.1007/JHEP07(2010)046}{\emph{JHEP} {\bfseries 1007}
  (2010) 046}, [\href{https://arxiv.org/abs/1005.2133}{{\ttfamily 1005.2133}}].

\bibitem{Fiol:2010wf}
B.~Fiol, \emph{{Defect CFTs and Holographic Multiverse}},
  \href{https://doi.org/10.1088/1475-7516/2010/07/005}{\emph{JCAP} {\bfseries
  1007} (2010) 005}, [\href{https://arxiv.org/abs/1004.0618}{{\ttfamily
  1004.0618}}].

\bibitem{Jensen:2013lxa}
K.~Jensen and A.~O'Bannon, \emph{{Holography, Entanglement Entropy, and
  Conformal Field Theories with Boundaries or Defects}},
  \href{https://doi.org/10.1103/PhysRevD.88.106006}{\emph{Phys. Rev.}
  {\bfseries D88} (2013) 106006},
  [\href{https://arxiv.org/abs/1309.4523}{{\ttfamily 1309.4523}}].

\bibitem{Korovin:2013gha}
Y.~Korovin, \emph{{First Order Formalism for the Holographic Duals of Defect
  CFTs}}, \href{https://doi.org/10.1007/JHEP04(2014)152}{\emph{JHEP} {\bfseries
  1404} (2014) 152}, [\href{https://arxiv.org/abs/1312.0089}{{\ttfamily
  1312.0089}}].

\bibitem{Estes:2014hka}
J.~Estes, K.~Jensen, A.~O'Bannon, E.~Tsatis and T.~Wrase, \emph{{On Holographic
  Defect Entropy}}, \href{https://doi.org/10.1007/JHEP05(2014)084}{\emph{JHEP}
  {\bfseries 05} (2014) 084},
  [\href{https://arxiv.org/abs/1403.6475}{{\ttfamily 1403.6475}}].

\bibitem{Seminara:2017hhh}
D.~Seminara, J.~Sisti and E.~Tonni, \emph{{Corner contributions to holographic
  entanglement entropy in AdS$_{4}$/BCFT$_{3}$}},
  \href{https://doi.org/10.1007/JHEP11(2017)076}{\emph{JHEP} {\bfseries 11}
  (2017) 076}, [\href{https://arxiv.org/abs/1708.05080}{{\ttfamily
  1708.05080}}].

\bibitem{Seminara:2018pmr}
D.~Seminara, J.~Sisti and E.~Tonni, \emph{{Holographic entanglement entropy in
  AdS$_{4}$/BCFT$_{3}$ and the Willmore functional}},
  \href{https://doi.org/10.1007/JHEP08(2018)164}{\emph{JHEP} {\bfseries 08}
  (2018) 164}, [\href{https://arxiv.org/abs/1805.11551}{{\ttfamily
  1805.11551}}].

\bibitem{Gaiotto:2009hg}
D.~Gaiotto, G.~W. Moore and A.~Neitzke, \emph{{Wall-crossing, Hitchin Systems,
  and the WKB Approximation}},
  \href{https://arxiv.org/abs/0907.3987}{{\ttfamily 0907.3987}}.

\bibitem{Alday:2009fs}
L.~F. Alday, D.~Gaiotto, S.~Gukov, Y.~Tachikawa and H.~Verlinde, \emph{{Loop
  and surface operators in N=2 gauge theory and Liouville modular geometry}},
  \href{https://doi.org/10.1007/JHEP01(2010)113}{\emph{JHEP} {\bfseries 01}
  (2010) 113}, [\href{https://arxiv.org/abs/0909.0945}{{\ttfamily 0909.0945}}].

\bibitem{Gaiotto:2009fs}
D.~Gaiotto, \emph{{Surface Operators in N = 2 4d Gauge Theories}},
  \href{https://doi.org/10.1007/JHEP11(2012)090}{\emph{JHEP} {\bfseries 11}
  (2012) 090}, [\href{https://arxiv.org/abs/0911.1316}{{\ttfamily 0911.1316}}].

\bibitem{Kozcaz:2010af}
C.~Kozcaz, S.~Pasquetti and N.~Wyllard, \emph{{A \& B model approaches to
  surface operators and Toda theories}},
  \href{https://doi.org/10.1007/JHEP08(2010)042}{\emph{JHEP} {\bfseries 08}
  (2010) 042}, [\href{https://arxiv.org/abs/1004.2025}{{\ttfamily 1004.2025}}].

\bibitem{Alday:2010vg}
L.~F. Alday and Y.~Tachikawa, \emph{{Affine SL(2) conformal blocks from 4d
  gauge theories}},
  \href{https://doi.org/10.1007/s11005-010-0422-4}{\emph{Lett. Math. Phys.}
  {\bfseries 94} (2010) 87--114},
  [\href{https://arxiv.org/abs/1005.4469}{{\ttfamily 1005.4469}}].

\bibitem{Ganor:1996nf}
O.~J. Ganor, \emph{{Six-dimensional tensionless strings in the large N limit}},
  \href{https://doi.org/10.1016/S0550-3213(96)00702-X}{\emph{Nucl. Phys.}
  {\bfseries B489} (1997) 95--121},
  [\href{https://arxiv.org/abs/hep-th/9605201}{{\ttfamily hep-th/9605201}}].

\bibitem{Bianchi:2019sxz}
L.~Bianchi and M.~Lemos, \emph{{Superconformal surfaces in four dimensions}},
  \href{https://arxiv.org/abs/1911.05082}{{\ttfamily 1911.05082}}.

\bibitem{Maldacena:1997re}
J.~M. Maldacena, \emph{{The Large N limit of superconformal field theories and
  supergravity}}, \href{https://doi.org/10.1023/A:1026654312961,
  10.4310/ATMP.1998.v2.n2.a1}{\emph{Int. J. Theor. Phys.} {\bfseries 38} (1999)
  1113--1133}, [\href{https://arxiv.org/abs/hep-th/9711200}{{\ttfamily
  hep-th/9711200}}].

\bibitem{Gubser:1998bc}
S.~S. Gubser, I.~R. Klebanov and A.~M. Polyakov, \emph{{Gauge theory
  correlators from noncritical string theory}},
  \href{https://doi.org/10.1016/S0370-2693(98)00377-3}{\emph{Phys. Lett.}
  {\bfseries B428} (1998) 105--114},
  [\href{https://arxiv.org/abs/hep-th/9802109}{{\ttfamily hep-th/9802109}}].

\bibitem{Witten:1998qj}
E.~Witten, \emph{{Anti-de Sitter space and holography}},
  \href{https://doi.org/10.4310/ATMP.1998.v2.n2.a2}{\emph{Adv. Theor. Math.
  Phys.} {\bfseries 2} (1998) 253--291},
  [\href{https://arxiv.org/abs/hep-th/9802150}{{\ttfamily hep-th/9802150}}].

\bibitem{Nekrasov:2002qd}
N.~A. Nekrasov, \emph{{Seiberg-Witten prepotential from instanton counting}},
  \href{https://doi.org/10.4310/ATMP.2003.v7.n5.a4}{\emph{Adv. Theor. Math.
  Phys.} {\bfseries 7} (2003) 831--864},
  [\href{https://arxiv.org/abs/hep-th/0206161}{{\ttfamily hep-th/0206161}}].

\bibitem{Pestun:2007rz}
V.~Pestun, \emph{{Localization of gauge theory on a four-sphere and
  supersymmetric Wilson loops}},
  \href{https://doi.org/10.1007/s00220-012-1485-0}{\emph{Commun. Math. Phys.}
  {\bfseries 313} (2012) 71--129},
  [\href{https://arxiv.org/abs/0712.2824}{{\ttfamily 0712.2824}}].

\bibitem{Drukker:2010jp}
N.~Drukker, D.~Gaiotto and J.~Gomis, \emph{{The Virtue of Defects in 4D Gauge
  Theories and 2D CFTs}},
  \href{https://doi.org/10.1007/JHEP06(2011)025}{\emph{JHEP} {\bfseries 06}
  (2011) 025}, [\href{https://arxiv.org/abs/1003.1112}{{\ttfamily 1003.1112}}].

\bibitem{Kanno:2011fw}
H.~Kanno and Y.~Tachikawa, \emph{{Instanton counting with a surface operator
  and the chain-saw quiver}},
  \href{https://doi.org/10.1007/JHEP06(2011)119}{\emph{JHEP} {\bfseries 06}
  (2011) 119}, [\href{https://arxiv.org/abs/1105.0357}{{\ttfamily 1105.0357}}].

\bibitem{Benini:2012ui}
F.~Benini and S.~Cremonesi, \emph{{Partition Functions of ${\mathcal{N}=(2,2)}$
  Gauge Theories on S$^{2}$ and Vortices}},
  \href{https://doi.org/10.1007/s00220-014-2112-z}{\emph{Commun. Math. Phys.}
  {\bfseries 334} (2015) 1483--1527},
  [\href{https://arxiv.org/abs/1206.2356}{{\ttfamily 1206.2356}}].

\bibitem{Doroud:2012xw}
N.~Doroud, J.~Gomis, B.~Le~Floch and S.~Lee, \emph{{Exact Results in D=2
  Supersymmetric Gauge Theories}},
  \href{https://doi.org/10.1007/JHEP05(2013)093}{\emph{JHEP} {\bfseries 05}
  (2013) 093}, [\href{https://arxiv.org/abs/1206.2606}{{\ttfamily 1206.2606}}].

\bibitem{Gomis:2014eya}
J.~Gomis and B.~Le~Floch, \emph{{M2-brane surface operators and gauge theory
  dualities in Toda}},
  \href{https://doi.org/10.1007/JHEP04(2016)183}{\emph{JHEP} {\bfseries 04}
  (2016) 183}, [\href{https://arxiv.org/abs/1407.1852}{{\ttfamily 1407.1852}}].

\bibitem{Nawata:2014nca}
S.~Nawata, \emph{{Givental J-functions, Quantum integrable systems, AGT
  relation with surface operator}},
  \href{https://doi.org/10.4310/ATMP.2015.v19.n6.a4}{\emph{Adv. Theor. Math.
  Phys.} {\bfseries 19} (2015) 1277--1338},
  [\href{https://arxiv.org/abs/1408.4132}{{\ttfamily 1408.4132}}].

\bibitem{Lamy-Poirier:2014sea}
J.~Lamy-Poirier, \emph{{Localization of a supersymmetric gauge theory in the
  presence of a surface defect}},
  \href{https://arxiv.org/abs/1412.0530}{{\ttfamily 1412.0530}}.

\bibitem{Gomis:2016ljm}
J.~Gomis, B.~Le~Floch, Y.~Pan and W.~Peelaers, \emph{{Intersecting Surface
  Defects and Two-Dimensional CFT}},
  \href{https://doi.org/10.1103/PhysRevD.96.045003}{\emph{Phys. Rev.}
  {\bfseries D96} (2017) 045003},
  [\href{https://arxiv.org/abs/1610.03501}{{\ttfamily 1610.03501}}].

\bibitem{Pan:2016fbl}
Y.~Pan and W.~Peelaers, \emph{{Intersecting Surface Defects and Instanton
  Partition Functions}},
  \href{https://doi.org/10.1007/JHEP07(2017)073}{\emph{JHEP} {\bfseries 07}
  (2017) 073}, [\href{https://arxiv.org/abs/1612.04839}{{\ttfamily
  1612.04839}}].

\bibitem{Gorsky:2017hro}
A.~Gorsky, B.~Le~Floch, A.~Milekhin and N.~Sopenko, \emph{{Surface defects and
  instanton–vortex interaction}},
  \href{https://doi.org/10.1016/j.nuclphysb.2017.04.010}{\emph{Nucl. Phys.}
  {\bfseries B920} (2017) 122--156},
  [\href{https://arxiv.org/abs/1702.03330}{{\ttfamily 1702.03330}}].

\bibitem{Alday:2009aq}
L.~F. Alday, D.~Gaiotto and Y.~Tachikawa, \emph{{Liouville Correlation
  Functions from Four-dimensional Gauge Theories}},
  \href{https://doi.org/10.1007/s11005-010-0369-5}{\emph{Lett. Math. Phys.}
  {\bfseries 91} (2010) 167--197},
  [\href{https://arxiv.org/abs/0906.3219}{{\ttfamily 0906.3219}}].

\bibitem{Wyllard:2009hg}
N.~Wyllard, \emph{{A(N-1) conformal Toda field theory correlation functions
  from conformal N = 2 SU(N) quiver gauge theories}},
  \href{https://doi.org/10.1088/1126-6708/2009/11/002}{\emph{JHEP} {\bfseries
  11} (2009) 002}, [\href{https://arxiv.org/abs/0907.2189}{{\ttfamily
  0907.2189}}].

\bibitem{Dimofte:2010tz}
T.~Dimofte, S.~Gukov and L.~Hollands, \emph{{Vortex Counting and Lagrangian
  3-manifolds}}, \href{https://doi.org/10.1007/s11005-011-0531-8}{\emph{Lett.
  Math. Phys.} {\bfseries 98} (2011) 225--287},
  [\href{https://arxiv.org/abs/1006.0977}{{\ttfamily 1006.0977}}].

\bibitem{Kim:2012ava}
H.-C. Kim and S.~Kim, \emph{{M5-branes from gauge theories on the 5-sphere}},
  \href{https://doi.org/10.1007/JHEP05(2013)144}{\emph{JHEP} {\bfseries 05}
  (2013) 144}, [\href{https://arxiv.org/abs/1206.6339}{{\ttfamily 1206.6339}}].

\bibitem{Assel:2014paa}
B.~Assel, D.~Cassani and D.~Martelli, \emph{{Localization on Hopf surfaces}},
  \href{https://doi.org/10.1007/JHEP08(2014)123}{\emph{JHEP} {\bfseries 08}
  (2014) 123}, [\href{https://arxiv.org/abs/1405.5144}{{\ttfamily 1405.5144}}].

\bibitem{Bullimore:2014upa}
M.~Bullimore and H.-C. Kim, \emph{{The Superconformal Index of the (2,0) Theory
  with Defects}}, \href{https://doi.org/10.1007/JHEP05(2015)048}{\emph{JHEP}
  {\bfseries 05} (2015) 048},
  [\href{https://arxiv.org/abs/1412.3872}{{\ttfamily 1412.3872}}].

\bibitem{Bobev:2015kza}
N.~Bobev, M.~Bullimore and H.-C. Kim, \emph{{Supersymmetric Casimir Energy and
  the Anomaly Polynomial}},
  \href{https://doi.org/10.1007/JHEP09(2015)142}{\emph{JHEP} {\bfseries 09}
  (2015) 142}, [\href{https://arxiv.org/abs/1507.08553}{{\ttfamily
  1507.08553}}].

\bibitem{Gadde:2011ik}
A.~Gadde, L.~Rastelli, S.~S. Razamat and W.~Yan, \emph{{The 4d Superconformal
  Index from q-deformed 2d Yang-Mills}},
  \href{https://doi.org/10.1103/PhysRevLett.106.241602}{\emph{Phys. Rev. Lett.}
  {\bfseries 106} (2011) 241602},
  [\href{https://arxiv.org/abs/1104.3850}{{\ttfamily 1104.3850}}].

\bibitem{Bullimore:2014nla}
M.~Bullimore, M.~Fluder, L.~Hollands and P.~Richmond, \emph{{The superconformal
  index and an elliptic algebra of surface defects}},
  \href{https://doi.org/10.1007/JHEP10(2014)062}{\emph{JHEP} {\bfseries 10}
  (2014) 062}, [\href{https://arxiv.org/abs/1401.3379}{{\ttfamily 1401.3379}}].

\bibitem{Beem:2013sza}
C.~Beem, M.~Lemos, P.~Liendo, W.~Peelaers, L.~Rastelli and B.~C. van Rees,
  \emph{{Infinite Chiral Symmetry in Four Dimensions}},
  \href{https://doi.org/10.1007/s00220-014-2272-x}{\emph{Commun. Math. Phys.}
  {\bfseries 336} (2015) 1359--1433},
  [\href{https://arxiv.org/abs/1312.5344}{{\ttfamily 1312.5344}}].

\bibitem{Beem:2014kka}
C.~Beem, L.~Rastelli and B.~C. van Rees, \emph{{$ \mathcal{W} $ symmetry in six
  dimensions}}, \href{https://doi.org/10.1007/JHEP05(2015)017}{\emph{JHEP}
  {\bfseries 05} (2015) 017},
  [\href{https://arxiv.org/abs/1404.1079}{{\ttfamily 1404.1079}}].

\bibitem{Beem:2014rza}
C.~Beem, W.~Peelaers, L.~Rastelli and B.~C. van Rees, \emph{{Chiral algebras of
  class S}}, \href{https://doi.org/10.1007/JHEP05(2015)020}{\emph{JHEP}
  {\bfseries 05} (2015) 020},
  [\href{https://arxiv.org/abs/1408.6522}{{\ttfamily 1408.6522}}].

\bibitem{Cordova:2017mhb}
C.~Cordova, D.~Gaiotto and S.-H. Shao, \emph{{Surface Defects and Chiral
  Algebras}}, \href{https://doi.org/10.1007/JHEP05(2017)140}{\emph{JHEP}
  {\bfseries 05} (2017) 140},
  [\href{https://arxiv.org/abs/1704.01955}{{\ttfamily 1704.01955}}].

\bibitem{Kinney:2005ej}
J.~Kinney, J.~M. Maldacena, S.~Minwalla and S.~Raju, \emph{{An Index for 4
  dimensional super conformal theories}},
  \href{https://doi.org/10.1007/s00220-007-0258-7}{\emph{Commun. Math. Phys.}
  {\bfseries 275} (2007) 209--254},
  [\href{https://arxiv.org/abs/hep-th/0510251}{{\ttfamily hep-th/0510251}}].

\bibitem{Gadde:2011uv}
A.~Gadde, L.~Rastelli, S.~S. Razamat and W.~Yan, \emph{{Gauge Theories and
  Macdonald Polynomials}},
  \href{https://doi.org/10.1007/s00220-012-1607-8}{\emph{Commun. Math. Phys.}
  {\bfseries 319} (2013) 147--193},
  [\href{https://arxiv.org/abs/1110.3740}{{\ttfamily 1110.3740}}].

\bibitem{DiPietro:2014bca}
L.~Di~Pietro and Z.~Komargodski, \emph{{Cardy formulae for SUSY theories in $d
  =$ 4 and $d =$ 6}},
  \href{https://doi.org/10.1007/JHEP12(2014)031}{\emph{JHEP} {\bfseries 12}
  (2014) 031}, [\href{https://arxiv.org/abs/1407.6061}{{\ttfamily 1407.6061}}].

\bibitem{Assel:2015nca}
B.~Assel, D.~Cassani, L.~Di~Pietro, Z.~Komargodski, J.~Lorenzen and
  D.~Martelli, \emph{{The Casimir Energy in Curved Space and its Supersymmetric
  Counterpart}}, \href{https://doi.org/10.1007/JHEP07(2015)043}{\emph{JHEP}
  {\bfseries 07} (2015) 043},
  [\href{https://arxiv.org/abs/1503.05537}{{\ttfamily 1503.05537}}].

\bibitem{Closset:2019ucb}
C.~Closset, L.~Di~Pietro and H.~Kim, \emph{{'t Hooft anomalies and the
  holomorphy of supersymmetric partition functions}},
  \href{https://doi.org/10.1007/JHEP08(2019)035}{\emph{JHEP} {\bfseries 08}
  (2019) 035}, [\href{https://arxiv.org/abs/1905.05722}{{\ttfamily
  1905.05722}}].

\bibitem{Gaiotto:2012xa}
D.~Gaiotto, L.~Rastelli and S.~S. Razamat, \emph{{Bootstrapping the
  superconformal index with surface defects}},
  \href{https://doi.org/10.1007/JHEP01(2013)022}{\emph{JHEP} {\bfseries 01}
  (2013) 022}, [\href{https://arxiv.org/abs/1207.3577}{{\ttfamily 1207.3577}}].

\bibitem{Alday:2013kda}
L.~F. Alday, M.~Bullimore, M.~Fluder and L.~Hollands, \emph{{Surface defects,
  the superconformal index and q-deformed Yang-Mills}},
  \href{https://doi.org/10.1007/JHEP10(2013)018}{\emph{JHEP} {\bfseries 10}
  (2013) 018}, [\href{https://arxiv.org/abs/1303.4460}{{\ttfamily 1303.4460}}].

\bibitem{Gentle:2015ruo}
S.~A. Gentle, M.~Gutperle and C.~Marasinou, \emph{{Holographic entanglement
  entropy of surface defects}},
  \href{https://doi.org/10.1007/JHEP04(2016)067}{\emph{JHEP} {\bfseries 04}
  (2016) 067}, [\href{https://arxiv.org/abs/1512.04953}{{\ttfamily
  1512.04953}}].

\bibitem{DHoker:2006qeo}
E.~D'Hoker, J.~Estes and M.~Gutperle, \emph{{Interface Yang-Mills,
  supersymmetry, and Janus}},
  \href{https://doi.org/10.1016/j.nuclphysb.2006.07.001}{\emph{Nucl. Phys.}
  {\bfseries B753} (2006) 16--41},
  [\href{https://arxiv.org/abs/hep-th/0603013}{{\ttfamily hep-th/0603013}}].

\bibitem{Gukov:2006jk}
S.~Gukov and E.~Witten, \emph{{Gauge Theory, Ramification, And The Geometric
  Langlands Program}},  \href{https://arxiv.org/abs/hep-th/0612073}{{\ttfamily
  hep-th/0612073}}.

\bibitem{Gukov:2008sn}
S.~Gukov and E.~Witten, \emph{{Rigid Surface Operators}},
  \href{https://doi.org/10.4310/ATMP.2010.v14.n1.a3}{\emph{Adv. Theor. Math.
  Phys.} {\bfseries 14} (2010) 87--178},
  [\href{https://arxiv.org/abs/0804.1561}{{\ttfamily 0804.1561}}].

\bibitem{Frenkel:2015rda}
E.~Frenkel, S.~Gukov and J.~Teschner, \emph{{Surface Operators and Separation
  of Variables}}, \href{https://doi.org/10.1007/JHEP01(2016)179}{\emph{JHEP}
  {\bfseries 01} (2016) 179},
  [\href{https://arxiv.org/abs/1506.07508}{{\ttfamily 1506.07508}}].

\bibitem{Ashok:2017odt}
S.~K. Ashok, M.~Billo, E.~Dell'Aquila, M.~Frau, R.~R. John and A.~Lerda,
  \emph{{Modular and duality properties of surface operators in N=2* gauge
  theories}}, \href{https://doi.org/10.1007/JHEP07(2017)068}{\emph{JHEP}
  {\bfseries 07} (2017) 068},
  [\href{https://arxiv.org/abs/1702.02833}{{\ttfamily 1702.02833}}].

\bibitem{Ashok:2019rwa}
S.~K. Ashok, S.~Ballav, M.~Frau and R.~R. John, \emph{{Surface operators in
  $N=$ 2 SQCD and Seiberg Duality}},
  \href{https://doi.org/10.1140/epjc/s10052-019-6866-5}{\emph{Eur. Phys. J.}
  {\bfseries C79} (2019) 372},
  [\href{https://arxiv.org/abs/1901.09630}{{\ttfamily 1901.09630}}].

\bibitem{Kapustin:2006pk}
A.~Kapustin and E.~Witten, \emph{{Electric-Magnetic Duality And The Geometric
  Langlands Program}},
  \href{https://doi.org/10.4310/CNTP.2007.v1.n1.a1}{\emph{Commun. Num. Theor.
  Phys.} {\bfseries 1} (2007) 1--236},
  [\href{https://arxiv.org/abs/hep-th/0604151}{{\ttfamily hep-th/0604151}}].

\bibitem{Hitchin:1987mz}
N.~J. Hitchin, \emph{{Stable bundles and integrable systems}},
  \href{https://doi.org/10.1215/S0012-7094-87-05408-1}{\emph{Duke Math. J.}
  {\bfseries 54} (1987) 91--114}.

\bibitem{Hitchin:1986ea}
N.~J. Hitchin, A.~Karlhede, U.~Lindstrom and M.~Rocek, \emph{{Hyperkahler
  Metrics and Supersymmetry}},
  \href{https://doi.org/10.1007/BF01214418}{\emph{Commun. Math. Phys.}
  {\bfseries 108} (1987) 535}.

\bibitem{Gadde:2013ftv}
A.~Gadde and S.~Gukov, \emph{{2d Index and Surface operators}},
  \href{https://doi.org/10.1007/JHEP03(2014)080}{\emph{JHEP} {\bfseries 03}
  (2014) 080}, [\href{https://arxiv.org/abs/1305.0266}{{\ttfamily 1305.0266}}].

\bibitem{Benini:2016qnm}
F.~Benini and B.~Le~Floch, \emph{{Supersymmetric localization in two
  dimensions}}, \href{https://doi.org/10.1088/1751-8121/aa77bb}{\emph{J. Phys.}
  {\bfseries A50} (2017) 443003},
  [\href{https://arxiv.org/abs/1608.02955}{{\ttfamily 1608.02955}}].

\bibitem{Gaiotto:2009we}
D.~Gaiotto, \emph{{N=2 dualities}},
  \href{https://doi.org/10.1007/JHEP08(2012)034}{\emph{JHEP} {\bfseries 08}
  (2012) 034}, [\href{https://arxiv.org/abs/0904.2715}{{\ttfamily 0904.2715}}].

\bibitem{Kozcaz:2010yp}
C.~Kozcaz, S.~Pasquetti, F.~Passerini and N.~Wyllard, \emph{{Affine sl(N)
  conformal blocks from N=2 SU(N) gauge theories}},
  \href{https://doi.org/10.1007/JHEP01(2011)045}{\emph{JHEP} {\bfseries 01}
  (2011) 045}, [\href{https://arxiv.org/abs/1008.1412}{{\ttfamily 1008.1412}}].

\bibitem{Kobayashi:2018lil}
N.~Kobayashi, T.~Nishioka, Y.~Sato and K.~Watanabe, \emph{{Towards a
  $C$-theorem in defect CFT}},
  \href{https://doi.org/10.1007/JHEP01(2019)039}{\emph{JHEP} {\bfseries 01}
  (2019) 039}, [\href{https://arxiv.org/abs/1810.06995}{{\ttfamily
  1810.06995}}].

\bibitem{Gentle:2015jma}
S.~A. Gentle, M.~Gutperle and C.~Marasinou, \emph{{Entanglement entropy of
  Wilson surfaces from bubbling geometries in M-theory}},
  \href{https://doi.org/10.1007/JHEP08(2015)019}{\emph{JHEP} {\bfseries 08}
  (2015) 019}, [\href{https://arxiv.org/abs/1506.00052}{{\ttfamily
  1506.00052}}].

\bibitem{Rodgers:2018mvq}
R.~Rodgers, \emph{{Holographic entanglement entropy from probe M-theory
  branes}}, \href{https://doi.org/10.1007/JHEP03(2019)092}{\emph{JHEP}
  {\bfseries 03} (2019) 092},
  [\href{https://arxiv.org/abs/1811.12375}{{\ttfamily 1811.12375}}].

\bibitem{Balasubramanian:2013kva}
A.~K. Balasubramanian, \emph{{The Euler anomaly and scale factors in
  Liouville/Toda CFTs}},
  \href{https://doi.org/10.1007/JHEP04(2014)127}{\emph{JHEP} {\bfseries 04}
  (2014) 127}, [\href{https://arxiv.org/abs/1310.5033}{{\ttfamily 1310.5033}}].

\bibitem{Belavin:1984vu}
A.~A. Belavin, A.~M. Polyakov and A.~B. Zamolodchikov, \emph{{Infinite
  Conformal Symmetry in Two-Dimensional Quantum Field Theory}},
  \href{https://doi.org/10.1016/0550-3213(84)90052-X}{\emph{Nucl. Phys.}
  {\bfseries B241} (1984) 333--380}.

\bibitem{Zamolodchikov:1995aa}
A.~B. Zamolodchikov and A.~B. Zamolodchikov, \emph{{Structure constants and
  conformal bootstrap in Liouville field theory}},
  \href{https://doi.org/10.1016/0550-3213(96)00351-3}{\emph{Nucl. Phys.}
  {\bfseries B477} (1996) 577--605},
  [\href{https://arxiv.org/abs/hep-th/9506136}{{\ttfamily hep-th/9506136}}].

\bibitem{Chacaltana:2010ks}
O.~Chacaltana and J.~Distler, \emph{{Tinkertoys for Gaiotto Duality}},
  \href{https://doi.org/10.1007/JHEP11(2010)099}{\emph{JHEP} {\bfseries 11}
  (2010) 099}, [\href{https://arxiv.org/abs/1008.5203}{{\ttfamily 1008.5203}}].

\bibitem{Drukker:2009id}
N.~Drukker, J.~Gomis, T.~Okuda and J.~Teschner, \emph{{Gauge Theory Loop
  Operators and Liouville Theory}},
  \href{https://doi.org/10.1007/JHEP02(2010)057}{\emph{JHEP} {\bfseries 02}
  (2010) 057}, [\href{https://arxiv.org/abs/0909.1105}{{\ttfamily 0909.1105}}].

\bibitem{Taki:2010bj}
M.~Taki, \emph{{Surface Operator, Bubbling Calabi-Yau and AGT Relation}},
  \href{https://doi.org/10.1007/JHEP07(2011)047}{\emph{JHEP} {\bfseries 07}
  (2011) 047}, [\href{https://arxiv.org/abs/1007.2524}{{\ttfamily 1007.2524}}].

\bibitem{Bonelli:2011fq}
G.~Bonelli, A.~Tanzini and J.~Zhao, \emph{{Vertices, Vortices and Interacting
  Surface Operators}},
  \href{https://doi.org/10.1007/JHEP06(2012)178}{\emph{JHEP} {\bfseries 06}
  (2012) 178}, [\href{https://arxiv.org/abs/1102.0184}{{\ttfamily 1102.0184}}].

\bibitem{Bonelli:2011wx}
G.~Bonelli, A.~Tanzini and J.~Zhao, \emph{{The Liouville side of the Vortex}},
  \href{https://doi.org/10.1007/JHEP09(2011)096}{\emph{JHEP} {\bfseries 09}
  (2011) 096}, [\href{https://arxiv.org/abs/1107.2787}{{\ttfamily 1107.2787}}].

\bibitem{Fateev:2007ab}
V.~A. Fateev and A.~V. Litvinov, \emph{{Correlation functions in conformal Toda
  field theory. I.}},
  \href{https://doi.org/10.1088/1126-6708/2007/11/002}{\emph{JHEP} {\bfseries
  11} (2007) 002}, [\href{https://arxiv.org/abs/0709.3806}{{\ttfamily
  0709.3806}}].

\bibitem{Beem:2017ooy}
C.~Beem and L.~Rastelli, \emph{{Vertex operator algebras, Higgs branches, and
  modular differential equations}},
  \href{https://doi.org/10.1007/JHEP08(2018)114}{\emph{JHEP} {\bfseries 08}
  (2018) 114}, [\href{https://arxiv.org/abs/1707.07679}{{\ttfamily
  1707.07679}}].

\bibitem{Douglas:2010iu}
M.~R. Douglas, \emph{{On D=5 super Yang-Mills theory and (2,0) theory}},
  \href{https://doi.org/10.1007/JHEP02(2011)011}{\emph{JHEP} {\bfseries 02}
  (2011) 011}, [\href{https://arxiv.org/abs/1012.2880}{{\ttfamily 1012.2880}}].

\bibitem{Nishioka:2013haa}
T.~Nishioka and I.~Yaakov, \emph{{Supersymmetric Renyi Entropy}},
  \href{https://doi.org/10.1007/JHEP10(2013)155}{\emph{JHEP} {\bfseries 10}
  (2013) 155}, [\href{https://arxiv.org/abs/1306.2958}{{\ttfamily 1306.2958}}].

\bibitem{Huang:2014pda}
X.~Huang and Y.~Zhou, \emph{{$ \mathcal{N}=4 $ Super-Yang-Mills on conic space
  as hologram of STU topological black hole}},
  \href{https://doi.org/10.1007/JHEP02(2015)068}{\emph{JHEP} {\bfseries 02}
  (2015) 068}, [\href{https://arxiv.org/abs/1408.3393}{{\ttfamily 1408.3393}}].

\bibitem{Yankielowicz:2017xkf}
S.~Yankielowicz and Y.~Zhou, \emph{{Supersymmetric Rényi entropy and Anomalies
  in 6d (1,0) SCFTs}},
  \href{https://doi.org/10.1007/JHEP04(2017)128}{\emph{JHEP} {\bfseries 04}
  (2017) 128}, [\href{https://arxiv.org/abs/1702.03518}{{\ttfamily
  1702.03518}}].

\bibitem{Zhou:2016kcz}
Y.~Zhou, \emph{{Information Theoretic Inequalities as Bounds in Superconformal
  Field Theory}},  \href{https://arxiv.org/abs/1607.05401}{{\ttfamily
  1607.05401}}.

\bibitem{Ooguri:1999bv}
H.~Ooguri and C.~Vafa, \emph{{Knot invariants and topological strings}},
  \href{https://doi.org/10.1016/S0550-3213(00)00118-8}{\emph{Nucl. Phys.}
  {\bfseries B577} (2000) 419--438},
  [\href{https://arxiv.org/abs/hep-th/9912123}{{\ttfamily hep-th/9912123}}].

\bibitem{Bullimore:2014awa}
M.~Bullimore, H.-C. Kim and P.~Koroteev, \emph{{Defects and Quantum
  Seiberg-Witten Geometry}},
  \href{https://doi.org/10.1007/JHEP05(2015)095}{\emph{JHEP} {\bfseries 05}
  (2015) 095}, [\href{https://arxiv.org/abs/1412.6081}{{\ttfamily 1412.6081}}].

\bibitem{Bobev:2017uzs}
N.~Bobev and P.~M. Crichigno, \emph{{Universal RG Flows Across Dimensions and
  Holography}}, \href{https://doi.org/10.1007/JHEP12(2017)065}{\emph{JHEP}
  {\bfseries 12} (2017) 065},
  [\href{https://arxiv.org/abs/1708.05052}{{\ttfamily 1708.05052}}].

\bibitem{DeWolfe:2001pq}
O.~DeWolfe, D.~Z. Freedman and H.~Ooguri, \emph{{Holography and defect
  conformal field theories}},
  \href{https://doi.org/10.1103/PhysRevD.66.025009}{\emph{Phys. Rev.}
  {\bfseries D66} (2002) 025009},
  [\href{https://arxiv.org/abs/hep-th/0111135}{{\ttfamily hep-th/0111135}}].

\bibitem{DHoker:2007zhm}
E.~D'Hoker, J.~Estes and M.~Gutperle, \emph{{Exact half-BPS Type IIB interface
  solutions. I. Local solution and supersymmetric Janus}},
  \href{https://doi.org/10.1088/1126-6708/2007/06/021}{\emph{JHEP} {\bfseries
  06} (2007) 021}, [\href{https://arxiv.org/abs/0705.0022}{{\ttfamily
  0705.0022}}].

\bibitem{Bobev:2013yra}
N.~Bobev, K.~Pilch and N.~P. Warner, \emph{{Supersymmetric Janus Solutions in
  Four Dimensions}}, \href{https://doi.org/10.1007/JHEP06(2014)058}{\emph{JHEP}
  {\bfseries 06} (2014) 058},
  [\href{https://arxiv.org/abs/1311.4883}{{\ttfamily 1311.4883}}].

\bibitem{Dimofte:2011ju}
T.~Dimofte, D.~Gaiotto and S.~Gukov, \emph{{Gauge Theories Labelled by
  Three-Manifolds}},
  \href{https://doi.org/10.1007/s00220-013-1863-2}{\emph{Commun. Math. Phys.}
  {\bfseries 325} (2014) 367--419},
  [\href{https://arxiv.org/abs/1108.4389}{{\ttfamily 1108.4389}}].

\bibitem{Melmed:1988hm}
J.~Melmed, \emph{{Conformal Invariance and the Regularized One Loop Effective
  Action}}, \href{https://doi.org/10.1088/0305-4470/21/23/005}{\emph{J. Phys.}
  {\bfseries A21} (1988) L1131--L1134}.

\bibitem{Dowker:1989ue}
J.~S. Dowker and J.~P. Schofield, \emph{{Conformal Transformations and the
  Effective Action in the Presence of Boundaries}},
  \href{https://doi.org/10.1063/1.528814}{\emph{J. Math. Phys.} {\bfseries 31}
  (1990) 808}.

\bibitem{Herzog:2015ioa}
C.~P. Herzog, K.-W. Huang and K.~Jensen, \emph{{Universal Entanglement and
  Boundary Geometry in Conformal Field Theory}},
  \href{https://doi.org/10.1007/JHEP01(2016)162}{\emph{JHEP} {\bfseries 01}
  (2016) 162}, [\href{https://arxiv.org/abs/1510.00021}{{\ttfamily
  1510.00021}}].

\bibitem{Fursaev:2015wpa}
D.~Fursaev, \emph{{Conformal anomalies of CFT’s with boundaries}},
  \href{https://doi.org/10.1007/JHEP12(2015)112}{\emph{JHEP} {\bfseries 12}
  (2015) 112}, [\href{https://arxiv.org/abs/1510.01427}{{\ttfamily
  1510.01427}}].

\bibitem{Solodukhin:2015eca}
S.~N. Solodukhin, \emph{{Boundary terms of conformal anomaly}},
  \href{https://doi.org/10.1016/j.physletb.2015.11.036}{\emph{Phys. Lett.}
  {\bfseries B752} (2016) 131--134},
  [\href{https://arxiv.org/abs/1510.04566}{{\ttfamily 1510.04566}}].

\bibitem{Ruj00}
S.~Ruijsenaars, \emph{On barnes' multiple zeta and gamma functions},
  \href{https://doi.org/https://doi.org/10.1006/aima.2000.1946}{\emph{Advances
  in Mathematics} {\bfseries 156} (2000) 107 -- 132}.

\bibitem{Spr09}
M.~Spreafico, \emph{On the barnes double zeta and gamma functions},
  \href{https://doi.org/https://doi.org/10.1016/j.jnt.2009.03.005}{\emph{Journal
  of Number Theory} {\bfseries 129} (2009) 2035 -- 2063}.

\bibitem{Fateev:2008bm}
V.~A. Fateev and A.~V. Litvinov, \emph{{Correlation functions in conformal Toda
  field theory II}},
  \href{https://doi.org/10.1088/1126-6708/2009/01/033}{\emph{JHEP} {\bfseries
  01} (2009) 033}, [\href{https://arxiv.org/abs/0810.3020}{{\ttfamily
  0810.3020}}].

\bibitem{Dorn:1994xn}
H.~Dorn and H.~J. Otto, \emph{{Two and three point functions in Liouville
  theory}}, \href{https://doi.org/10.1016/0550-3213(94)00352-1}{\emph{Nucl.
  Phys.} {\bfseries B429} (1994) 375--388},
  [\href{https://arxiv.org/abs/hep-th/9403141}{{\ttfamily hep-th/9403141}}].

\end{thebibliography}\endgroup

\end{document}